\documentclass{aa} 

\usepackage{txfonts}
\usepackage{natbib}
\usepackage{soul}
\usepackage{xcolor}
\usepackage{multicol}
\usepackage{graphicx,graphics,rotating,amssymb,amsmath,amsfonts}
\usepackage{xcolor}
\usepackage{multirow}
\usepackage{longtable}
\usepackage{booktabs}
\usepackage{subfig}
\usepackage{float,capt-of}
\usepackage{natbib}
\usepackage[english]{babel}
\usepackage{morefloats}
\usepackage{color}
\usepackage{booktabs}
\usepackage{sidecap}
\usepackage{epsfig,color}
\usepackage{wrapfig}
\usepackage{lscape}
\usepackage{ulem}
\usepackage{url}
\usepackage{tablefootnote}
\usepackage{hyperref}
\newcommand{\herschel}{{\it Herschel}}
\newcommand{\nthptrans}{N$_2$H$^+$ (1-0)}
\newcommand{\hnctrans}{HNC (1-0)}
\newcommand{\hcntrans}{HCN (1-0)}
\newcommand{\nthp}{N$_2$H$^+$}
\newcommand{\hnc}{HNC}
\newcommand{\hctn}{HC$_3$N (10-9)}
\newcommand{\kms}{km~s$^{-1}$}
\newcommand{\kmspc}{km~s$^{-1}$~pc$^{-1}$}
\usepackage{graphicx}
\usepackage{stfloats} 
\usepackage{booktabs}

\begin{document}

   \title{Emergence of high-mass stars in complex fiber networks (EMERGE)}

   \subtitle{VI. Turbulence dissipation and the formation of dense fibers}

   \author{Francesca Bonanomi
          \inst{1}
          \and
          Alvaro Hacar\inst{1}
          \and
          Andrea Socci\inst{1} 
           \and
          Stefan Heigl\inst{2} 
          }

    \institute{Institute for Astronomy (IfA), University of Vienna,
              T\"urkenschanzstrasse 17, A-1180 Vienna\\
              \email{francesca.bonanomi@univie.ac.at}
             \and
             Universit\"ats-Sternwarte, Ludwig-Maximilians-Universit\"at M\"unchen,
             Scheinerstrasse 1, 81679 Munich, Germany
             }
   \date{Received 02.06.2025; accepted 02.07.2026}

\titlerunning{EMERGE: VI. Turbulence dissipation and the formation of dense fibers}

  \abstract
    {The turbulent cascade naturally generates a hierarchy of structures within molecular clouds. Filament networks arise from this turbulent cascade, with fibers suggested to be the first (tran-)sonic components formed out of it. Consequently, the properties of cores forming within fibers may be inherited from the larger spatial scale of the filaments, where the dissipation of turbulence occurs.
    }
    {We aim to investigate the diffuse gas kinematics and its interaction with the dense gas composing fibers. We will characterize the diffuse material via the \hnc~molecular tracer.   
    }
    {We use high-resolution (4.5~arcsec or $\sim$ 2000~au) large-scale ALMA+IRAM-30m mosaics to survey five star-forming regions in Orion, as part of the EMERGE Early ALMA Survey. Our sample includes low- (OMC-4 South, NGC 2023), intermediate (OMC-3, LDN 1641N), and high-mass (Flame nebula) star-forming regions, covering a wide range of stellar activity, cloud morphology, and evolutionary stages. We observe our targets in \hnctrans~as probe of diffuse gas in the regions and compare it to the \nthptrans~emission tracing the dense gas. We systematically investigate the emission of the diffuse gas in relation to the dense structures identified in \nthp, focusing on its kinematics and velocity field.
    }
    {The high resolution (2000~au) observations reveal that \hnc~traces lukewarm, diffuse ($\sim5\times10^{21}$ cm$^{-2}$) material around dense fibers.
    The properties of the diffuse gas appear to be similar across our sample, despite the wide range of different environments. Compared to the quiescent and subsonic gas inside fibers, the diffuse gas is, however, more turbulent ($\tilde{\mathcal{M}_\text{s}}=2.9$). We observed the \hnc~non-thermal velocity dispersion to be roughly constant and homogeneous where the dense gas is detected ($\mathcal{M}_\text{s}\sim2$), and growing toward its edges up to $\mathcal{M}_\text{s}\sim5$. 
    Understanding the dissipation process is crucial to mark the transition between the dense subsonic gas and diffuse turbulent material occurs. We investigated the turbulence dissipation through the statistical analysis of the \hnc~velocity gradients. While the majority of the gas shows $\nabla V_{lsr}\le5$~\kmspc~with roughly constant values, we identified high-shear regions showing higher gradients with $\nabla V_{lsr}\ge10$~\kmspc~concentrated in small features of 0.1-0.3~pc in size located near the dense gas. While small in size, these high-shear structures appear to be major contributors of the turbulence dissipation in our targets.
    }
    {Our results suggest that in Orion the transition to coherence occurs at the fiber level, as suggested by the turbulence being effectively dissipated before the formation of cores and during the formation of these first dense gas structures.
    }


\maketitle

\section{Introduction}\label{sec:intro}

The interstellar medium (ISM) has long been known to have a filamentary nature \citep{1907Barnard}. Filaments play a key role in the star formation process \citep[see][for recent reviews]{2014Andre,2023Pineda}, connecting molecular clouds to cores. These elongated structures are now observed in the ISM at all scales \citep{2023Hacar}. In the last decade, molecular observations revealed that parsec-size filaments identified in continuum can be resolved in a plethora of velocity-coherent, sub-filaments \citep{2013Hacar}, known as fibers, highlighting the intrinsic hierarchical nature of the ISM at all scales. 

How these fibers are formed is still under debate. A first hypothesis argues that gravitational instabilities are responsible for the formation of filaments in infinite sheets \citep{1983Tomisaka,1985Larson,1987Miyama,1987Miyamab,1998Nagai,2014VanLoo}. On the other hand, simulations of turbulent boxes showed that, in the absence of gravity, interactions between sheets can lead to the formation of filaments \citep[e.g.,][]{1994Porter,1999Padoan,2001Padoan,2001Heitsch,2013Pudritz,2013Hennebelle}. According to the \citet{1941Kolmogorov} model of turbulence, the energy injected at large scales is transferred to smaller scales producing a turbulent cascade. This cascade is interrupted at much smaller scales when the turbulence is dissipated by several mechanisms such as viscosity \citep[see][for a comprehensive review on turbulence]{2004Elmegreen,2004Scalo}. Turbulence naturally forms networks of filaments, since the turbulent cascade naturally generates an internal hierarchy of structures.
More recent numerical simulations support this turbulence-induced formation mechanism \citep{2016Smith,2017Clarke}, although through two different scenarios. \citet{2014Smith, 2016Smith} proposed a bottom-up approach, known as "fray and gather". The sub-filaments exist within the cloud formed by turbulent compression and energy dissipation, and later are swept together into a single filament by large-scale motions. 
The top-down picture is known as the "fray and fragment" scenario \citep{2015Tafalla,2017Clarke}, where the parental filament is formed first by two gas flows colliding. As mass is added to the filament, it fragments to form fibers due to turbulence, and eventually some sub-filaments fragment into cores.
Investigating the kinematics of the gas allows us to better constrain the initial condition for star formation, allowing us to understand the role played by turbulence. 

Non-thermal velocity dispersion is a common property used to describe the turbulence regime of the gas. Although velocity dispersion is known to be supersonic in molecular clouds \citep[e.g.,][]{1981Larson}, dense cores exhibit roughly constant subsonic non-thermal motions \citep{1998Goodman,2002Caselli}. Studying the lower-density material surrounding cores, \citet{1998Goodman} noticed that its supersonic velocity dispersion decreases towards smaller sizes. Later observations tracing both dense and diffuse gas in NH$_3$ \citep{2010Pineda} denoted this transition as sharp and characterized by high velocity gradients.
Thus, a transition between turbulent diffuse gas and the more quiescent dense gas should occur at core scales.  
\citet{2011Hacar} found these velocity-coherent regions to extend on scales significantly larger than cores, considering the small-scales filaments as the first subsonic structures formed out of the turbulent cascade. The connection between filaments and cores has been well established since first systematic studies of filaments~\citep{1979SchneiderElmegreen} and has been recently highlighted by recent \herschel~studies \citep{2010Andre}. The comprehensive view on filaments provided by \herschel~further confirmed that core formation would therefore not be directly related to turbulence dissipation, but instead influenced by their parental filaments.

The transition to coherence requires, however, turbulence dissipation. Dissipation is known to be intermittent at small scales \citep{1990Falgarone}, meaning that the spatio-temporal fluctuations of the density and velocity fields relative to their average values increase as the scale decreases \citep{1959Landau,1962Kolmogorov,1999Miesch,2002Ossenkopf}. Investigating how the dissipative process occurs is crucial to understand fiber formation, however identifying these regions of intermittent turbulence is still challenging since they have a low filling factor and therefore correspond to rare events in time and space.
The kinematic transitions between dense and diffuse gas will be tackled in this paper, investigating how the dissipation occurs, and its influence on star formation.

This study is part of the Emergence of high-mass stars in complex fiber networks (EMERGE) project \citep[see][hereafter Paper I]{2024Hacara}\footnote{EMERGE Project website: https://emerge.univie.ac.at/}. 
The aim of the EMERGE project is to survey the internal gas organization in a series of star-forming regions based on high-resolution ALMA observations combined with high-sensitivity single-dish (SD) data \citep[the data combination procedure is presented in][hereafter Paper II]{2024Bonanomi}. Previous papers (III-IV) of this series describe the dense gas substructure down to 2000~au resolution using \nthptrans~ observations instead \citep{2024Soccia,2024Soccib}.

In this paper, the sixth of its series (Paper VI), we investigate the behaviour of the diffuse gas, traced by \hnc, surrounding the dense structures, and its role in fiber formation. We focus on its kinematics to determine how the diffuse and dense gas interacts.
We first introduce the EMERGE Early ALMA Survey and the observations used in this work (Sect.~\ref{sec:survey}).
Among these, we present the \hnc~integrated intensity maps, the total column density maps obtained from high-resolution ALMA + IRAM-30m observations, and the gas kinetic temperature maps (Sect.~\ref{sec:diffuse_gas}). 
We analyse the kinematics of the diffuse gas focusing on its linewidth and non-thermal velocity dispersion (Sect.\ref{sec:kinematics}). We further evaluate the impact of feedback on the properties of the gas (Sect.\ref{sec:feedback}).
Finally, we discuss our results about the coherent scale in the literature context (Sect.\ref{sec:coherence}), and focus on turbulence dissipation and its intermittent behaviour through the analysis of velocity gradients across our high-resolution maps (Sect.~\ref{sec:velocity_gradients} and \ref{sec:dissipation_statistics}). 

\section{The EMERGE Early ALMA Survey}\label{sec:survey}

\begin{table*}[]
\caption{EMERGE Early ALMA Survey: properties of the targets}
    \centering
    \begin{tabular}{lcccccccr}
        \toprule
        Target &  Cloud    &   D   &   SF-regime   &   Evolutionary stage  &   O-stars 
        &   P($^\ast$)   &   P+D($^\ast$) &   P/D($^\ast$)\\
        && (pc)\\
        \midrule
         OMC-3  & Orion A & 400 & Intermediate- & Young & No & 26 & 102 & 0.34\\
         OMC-4 South    & Orion A & 400 & Low- & Evolved & ? & 11 & 59 & 0.23\\
         LDN 1641N  & Orion A & 400 & Intermediate- & Young & No & 13 & 51 & 0.34\\
         \midrule
         NGC 2023   & Orion B & 423 & Low- & Young & No & 6 & 8 & 0.75\\
         Flame Nebula   & Orion B & 423 & High- & Evolved &Yes & 21 & 141 & 0.18\\
         \bottomrule
        \\
         \multicolumn{9}{c}{\textbf{Notes.} ($^\ast$) The number of protostars (P) and disks (D) here listed are merely used as reference in this paper,}\\
         \multicolumn{9}{c}{as they refer to large footprints compared to our ALMA fields (see Paper I for a full discussion).}\\

    \end{tabular}
    \label{tab:sample_properties}
\end{table*}

The EMERGE Early ALMA Survey observed seven prototypical star-forming regions across Orion, namely OMC-1, 2, 3, 4 South, and LDN 1641N in Orion A, and NGC 2023 and the Flame Nebula (or NGC 2024) in Orion B (Paper I). 

Our survey covers a wide range of different environments, from high- (OMC-1 and the Flame Nebula) to intermediate- (OMC-2, 3, and LDN 1641N) and low-mass (OMC-4 South and NGC 2023) star-forming regions, characterized by different degrees of stellar activity, mass, and evolutionary stage. Stark is the contrast between OMC-1 and the Flame Nebula with the highest peak stellar densities, by hosting the two most massive embedded clusters in Orion, compared to OMC-4 South and NGC 2023 only forming a few low-mass stars in isolation \citep{2016Megeath}. 
Different degrees of feedback affect all targets in our sample. 
Considering the ratio between Class 0/I (P) and Class I/II (D) objects as a classical evolutionary tracer \citep{2009Evans}, we catalogued our targets as young (OMC-2, 3, LDN 1641N, and NGC 2023) and evolved (OMC-1, 4 South, and the Flame Nebula).
Table~\ref{tab:sample_properties} summarises the properties of our sample (see Paper I for a full discussion).

We observed OMC-3, 4 South, LDN 1641N, NGC 2023, and the Flame Nebula in Band 3 with the ALMA 12m array during Cycle 7 (Project ID: 2019.1.00641.S, PI: Hacar), reaching a final resolution of $\sim4.5$~arcsec \citep[$\sim$2000 au at 414~pc;][]{2007Menten}. 
In Band 3 observations we targeted mainly molecular emission, \nthptrans~ (93.17 GHz), \hnctrans~ (90.66 GHz), and \hctn~ (90.97 GHz), surveyed at high spectral resolution ($\delta v \le 0.233$~\kms), together with the 3mm-continuum. We complement our pilot survey with additional \nthptrans~($\delta v\leq0.15$~\kms) observations of OMC-1 and 2 from ALMA Cycle-3 (project ID: 2015.1.00669.S, PI: Hacar). Since \hnctrans~data are not available in these regions \citep[see][for a full description]{2018Hacar}, hereafter we do not include OMC-1 and 2 whenever referring to the EMERGE Early ALMA Survey. 
We combined our ALMA maps with additional IRAM-30m observations (project IDs: 032-13, 034-16, 120-20, 060-22, and 133-22), used as zero-spacing information (see Papers I and II for a full discussion).

The EMERGE Early ALMA Survey also includes temperature and column density maps of our dataset. We produced the gas kinetic temperature ($T_\mathrm{K}$) maps from the \hcntrans~to \hnctrans~lines ratio observed with IRAM-30m at 30~arcsec resolution \citep{2020Hacar}. 
We produced total gas column density $N(\text{H}_2$) maps of these regions as well, derived from \herschel~observations at 36~arcsec resolution \citep{2014Lombardi}.
We used ancillary catalogues of young stellar objects \citep[YSOs, including Class 0/I protostars and Class II disks;][]{2012Megeath, 2016Furlan, 2013Stutz}, and O-B stars \citep{2000Wenger} to explore the connection between diffuse gas and star formation feedback.
We refer the reader to Paper I for a full description of the data used in this work.

Among the suite of tracers included in our survey we selected \nthp~ and \hnc~ to sample different gas density regimes. \nthp~is a density-selective species for cold ($\le$20 K) and dense ($>10^4$ cm$^{-3}$) gas \citep{2002Tafalla} formed after the depletion of CO and usually found regions in regions of column density above $N(\text{H}_2)\ge10^{22}$ ~cm$^{-2}$ \citep[][]{2007Bergin,2017Pety,2017Kauffmann,2021Tafalla}. 
The properties of \nthp~as probe of the dense gas in our sample are presented in Papers III and IV. Similar to other active star-forming regions  \citep{2014Henshaw, Chen2019}, our \nthptrans~emission maps reveal large amounts of dense gas usually corresponding to those dense filaments and cores in our targets.

On the other hand, to sample the cloud gas from low to intermediate column density \citep[$\sim10^{21}-10^{23}$ cm$^{-2}$;][]{2021Tafalla, 2023Tafalla}, we picked the \hnctrans~as tracer of the diffuse gas. As shown Paper I, \hnc~is detected beyond the boundaries of our maps showing a bright emission at column densities A$_V\gtrsim$~3~mag.
The \hnctrans~line is effectively excited at densities of a few $10^3$ cm$^{-3}$ \citep{2015Shirley}, and its abundance is roughly invariant to temperature variation between 10 and 40 K. The \hnc~intensity shows an almost linear correlation with the total gas column density before being depleted at high density \citep{2017Pety, 2021Tafalla, 2023Tafalla} on timescales comparable to the free-fall time at densities above $n(\text{H}_2)>10^4$~cm$^{-3}$ \citep{2007Bergin} and the best Pearson correlation coefficient (p=0.90) among all traditional 3-mm line tracers of the diffuse gas in molecular clouds ($^{12}$CO, $^{13}$CO, C$^{18}$O, HCN, and HCO$^+$). 
The \hnc~(1-0) transitions also shows a roughly constant line ratio with respect to its optically thin $HN^{13}C$ isotopologue between column densities of $N(H_2)\sim10^{22}-10^{23}$~cm$^{-2}$ \citep[e.g., see Fig.D1 in ][]{2021Tafalla} suggesting 
\hnc~ to present a low opacity up to $\sim$100 A$_V$.
Additional RADEX-LVG calculations \citep{2007_vanderTak_RADEX} for typical values of the gas conditions traced by HNC, that is $n(H_2)=(0.5-5)\times10^4$~cm$^{-3}$, $T_K=20-40$~K \citep{2024Hacara}, $N(H_2)=10^{21}-10^{23}$~cm$^{-2}$ assuming a standard $X(HNC/H_2)=10^{-9}$ \citep{2021Tafalla}, and $\Delta V=$~1.8~\kms (Sect.~\ref{sec:kinematics}), retrieve  opacities of $\tau\sim0.2-15$ for the \hnc~(1-0) transition. The highest opacity values are likely upper limits as the expected increase of volume density with high column density \citep[i.e. N(H$_2$)~$\propto n(H_2)^{0.4-0.5}$][]{Gaches2025} would decrease the opacity of the J=1-0 line by populating higher J-states.
with most points in our maps within the range of low to moderate $\tau\sim0.5-5$ opacities.

A full discussion on the different regimes characterized by these tracers in our sample is reported in Paper I.
In this work, we investigate the environmental conditions in which fibers formed focusing on the analysis of \hnc~and its comparison with the structures identified in \nthp. 

\section{\hnc~as a tracer of lukewarm and diffuse gas}\label{sec:diffuse_gas}

\begin{figure*}[htbp]
	\centering
        \includegraphics[width=\textwidth]{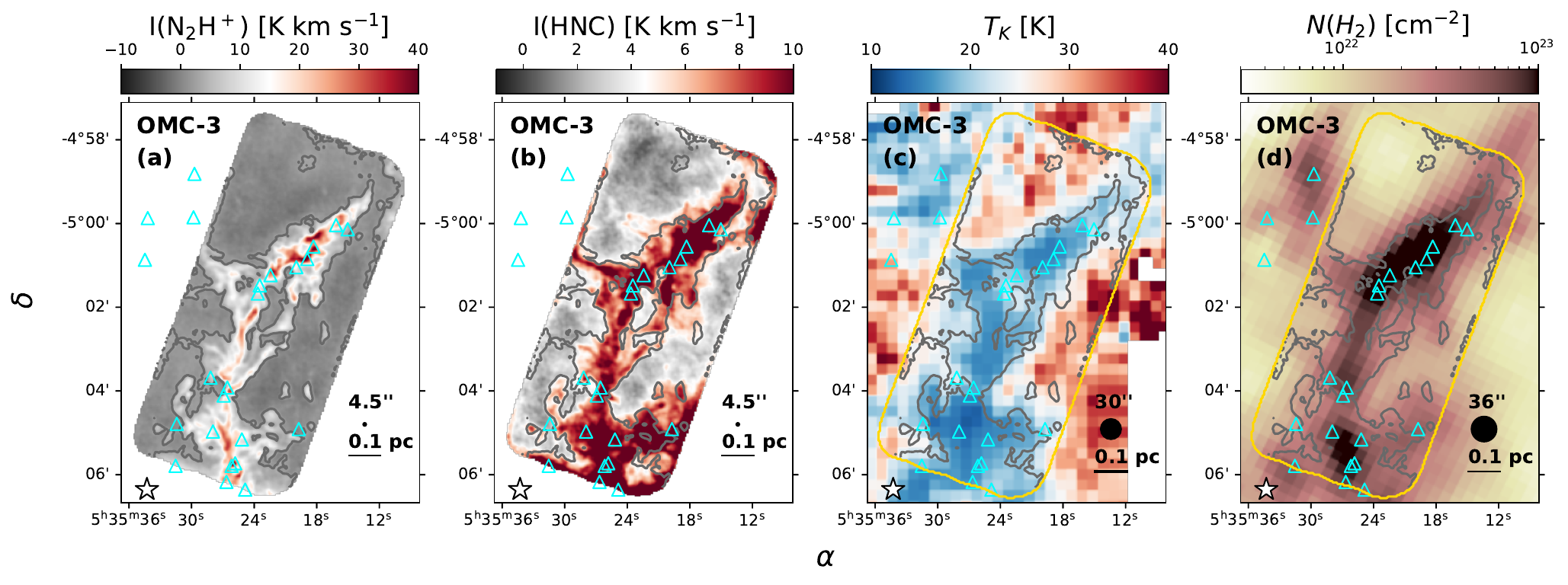}
	\caption{Maps in OMC-3, the representative showcase of our sample. From left to right we plot integrated intensity I(\nthp) and I(HNC) observed at 4.5~arcsec resolution with ALMA+IRAM-30m, gas kinetic temperature $T_K$  derived from the HCN-HNC ratio observed with IRAM-30m at 30~arcsec resolution (see Paper I), and total column density $N(\text{H}_2)$ map as observed by Herschel at 36~arcsec resolution \citep{2014Lombardi}. 
    The gray contours show the \nthptrans~integrated intensity above 3$\sigma$ (the \nthp~maps are visible in Paper III). In all panels, cyan triangles are the protostellar objects from \citet{2012Megeath,2013Stutz,2016Furlan}, while the yellow and white stars are the O and B stars collected from Simbad \citep{2000Wenger}, respectively. The maps of the remaining targets in the sample are shown in Fig.~\ref{fig:survey_maps}. Beam sizes and scale bars are placed in the bottom right corner.}
	\label{fig:OMC3_maps} 
\end{figure*}

We will address the characterization of the diffuse gas in the EMERGE Early ALMA Survey from two different points of view. On the one hand, we will investigate the morphology of the \hnc~emission in relation to the dense gas distribution traced by \nthp~ (Sect.~\ref{sec:intensitymaps}), the temperature regime covered by our observations (Sect.~\ref{sec:tempmaps}), and the total column density maps extracted from \hnc~and \nthp~(Sect.~\ref{sec:coldensmaps}).
On the other hand, we will probe the kinematics of the diffuse gas in these regions via the analysis of the \hnc~spectra (Sect.~\ref{sec:kinematics}).

\subsection{\hnc~integrated emission: morphology of the diffuse gas}\label{sec:intensitymaps}

Figure~\ref{fig:OMC3_maps} (panels a and b) shows the \nthptrans~ and \hnctrans~integrated intensity maps for OMC-3, as showcase of our survey. The \hnc~maps of the four remaining regions in our survey are displayed in Fig.~\ref{fig:survey_maps} (panels a and b). We selected the integration range in velocity case by case according to each individual cloud velocity. We plot a gray contour enclosing the \nthptrans~intensity at 3$\sigma$ (the dense gas maps are presented in Paper III) and mark the location of protostars (cyan triangles), B and O stars (white and yellow stars, respectively).

\hnc~emission has been detected above the noise level in all our maps (Fig.~\ref{fig:OMC3_maps}, panel b). We calculated the noise level in each map considering sub-regions free of emission, with a final estimate ranging between 0.7 and 1.3 K km s$^{-1}$. The comparison of the \nthp~ and \hnc~maps (panels a and b in Fig.~\ref{fig:OMC3_maps}) shows that \hnc~is much more extended than dense gas (gray contours). \hnc~emission is recovered almost everywhere in the OMC-3 map, with its brightest feature always overlying the dense gas. The brightest intensity peaks in \hnc~are found in the Flame Nebula (40 K km s$^{-1}$), LDN 1641N (23 K km s$^{-1}$), and OMC-3 (21 K km s$^{-1}$), located in proximity of several protostellar objects (cyan triangles in Figs.~\ref{fig:OMC3_maps} and \ref{fig:survey_maps}). The correlation between the location of the protostars and the bright line emission, already observed in \nthp~(see Paper I and III), is particularly clear in regions with a large stellar populations such as OMC-3 (OMC-3 MMS8-9) and LDN 1641N. This correlation is, instead, lost in the Flame Nebula possibly due to the strong UV-radiation. Low-mass regions, such as NGC 2023 and OMC-4 South, show weaker \hnc~ emission (intensity peaks $\sim$ 12 K km s$^{-1}$), yet still detected up to $>10\sigma$ level across our maps.

The morphology of the diffuse gas traced by \hnc~ varies from region to region. 
This behaviour was previously noted at low-resolution (30~arcsec) in the IRAM-30m maps (see Paper I) of both the dense and diffuse gas across our sample (see Paper III for a full discussion). The emission in OMC-3 is bright and elongated, roughly following the large-scale cloud structure; OMC-4 South and NGC 2023, show weaker and irregular emission, while LDN 1641N and the Flame Nebula a cometary shape.
The \hnc~emission appears, however, to be much more extended than \nthp~in all regions, with the brightest features always coinciding with the dense gas peaks (see Sect.~5 in Paper I for a full discussion). 

Thanks to its high-resolution (4.5~arcsec), our survey allows us to investigate the complex morphology of the dense and diffuse gas in high details. At $\sim$2000~au the \nthp~emission appears organised in a plethora of fibers. These fibers constitute networks of different complexity across our sample, but still driven by the total dense gas mass despite their star-formation regime, evolutionary stage, or stellar content (see Paper III for a full discussion). Figures~\ref{fig:OMC3_maps} and \ref{fig:survey_maps} show that the dense gas filamentary organization is reflected in the more diffuse gas which extends this complex ramification of the structures even at larger scales.

\subsection{Total column density maps}\label{sec:coldensmaps}

In order to describe the gas organization down to 2000 au, we need to consider the column density regimes sampled in our survey (see Sect.\ref{sec:survey}).
Figures~\ref{fig:OMC3_maps} (panel d) and \ref{fig:survey_maps} (right panels) show the total gas column density maps as observed by Herschel at 36~arcsec resolution \citep{2014Lombardi}. As expected, we can easily identify two different column density regimes probed by the two distinct tracers. The dense gas ($\ge10^{22}$ cm$^{-2}$) traced by \nthp~and enclosed in gray 3$\sigma$ contours, is organized in similar fiber-like structures in low- to high-mass star-forming regions (see Paper III). The diffuse gas ($\leq10^{22}$ cm$^{-2}$), traced by \hnc~, is characterized by a more complex organization, showing more irregular and chaotic shapes that extend outside the dense material. From the analysis of these maps and the distribution in Paper I comparing the density regimes sampled by our different molecules, we confirm that \hnc~traces the emission of the material surrounding the dense fibers at typical column densities of $\sim5\times10^{21}$ cm$^{-2}$.

Overall, despite the difference in morphology between different regions, the organization of the diffuse gas appears similar across our sample, which covers a wide range of different environments.

\subsection{Gas kinetic temperature: lukewarm gas}\label{sec:tempmaps}

We derived the gas kinetic temperature ($T_\mathrm{K}$) in our clouds from the ratio of the \hcntrans~and \hnctrans~lines observed with IRAM-30m at 30~arcsec resolution following \citet{2020Hacar} (see Paper I for more details).
This empirical method has been calibrated in the Orion Integral Shape Filament (ISF, including OMC-3 and OMC-4 South) against independent NH$_3$ data obtained by the GAS survey \citep{2017Friesen} within an optimal working range between $15\leq T_\text{K}\leq40$~K, and is limited to regions exposed to low or intermediate UV radiation. Recent observations \citep{2023Santa-Maria,2023Martinez,2024Nanase} indicate that the \hcntrans~to \hnctrans~lines ratio may be also sensitive to higher levels of UV-radiation in agreement to previous studies \citep{1992Schilke}. 
We have manually verified that the temperature values extracted from our IRAM-30m observations are in close agreement with previous NH$_3$ GAS measurements in the region of interest finding a good agreement (of few K difference) in those regions where both HCN/HNC and NH$_3$ molecules are simultaneously detected, typically at A$_V>$~15~mag \citep[see also Sect.~4.3.1 in][for a full discussion]{2020Hacar}.

Figures~\ref{fig:OMC3_maps} (panel c) and \ref{fig:survey_maps} (middle panels) show our large-scale $T_\mathrm{K}$ maps for OMC-3 and the whole sample, respectively.
The dense gas \nthp~distribution is always located in the coldest regions of the maps ($\leq$ 25 K, except in the Flame Nebula, where it is still detected at T$_K>$ 35 K,), \hnc~probes instead both cold and warmer gas (up to $\sim$ 40 K), heated by either feedback or the diffuse interstellar radiation field. All our maps show a temperature gradient, with the cold dense gas always surrounded by the lukewarm diffuse gas.

Feedback is one of the key players that affect the temperature distribution within the clouds. The gas in low-mass star-forming regions (OMC-3 and 4 South) is usually found at temperatures between 10 and 30 K, while in regions with embedded high-mass stars (Flame Nebula) the average values are $\geq$ 30 K and up to 45 K, see the histograms of $T_\mathrm{K}$ in Paper I for additional details). An interesting case in our sample is LDN 1641N: the region has a large fraction of diffuse gas above 40 K surrounding the dense gas. Although the origin of this warm material is unclear, it may be due to the interaction with the high-mass stars $\iota$ Ori and NGC 1980, as suggested by its cometary shape pointing towards the latter (Paper I).

\section{Kinematic analysis of the diffuse gas}\label{sec:kinematics}

\begin{table}[]
\caption{Kinematics of the diffuse gas: single component analysis}
    \centering
    \begin{tabular}{lcccr}
        \toprule
        \multicolumn{1}{l}{ Target} &\multicolumn{1}{c}{}& \multicolumn{2}{c}{\hnc} & \multicolumn{1}{r}{\nthp} \\
        && $\Delta$V & $\sigma_{nt}/c_s$ & $\sigma_{nt}/c_s^\ast$\\
        && (\kms)\\
        \midrule
        \bf
         OMC-3  && $1.7_{-0.4}^{+0.5}$ & $2.6_{-0.5}^{+0.7}$ & $0.64$\\
         OMC-4 South    && $2.0_{-0.6}^{+0.7}$ & $3.3_{-0.8}^{+1.1}$ & $0.66$\\
         LDN 1641N  && $1.8_{-0.4}^{+0.5}$ & $3.0_{-0.8}^{+0.9}$ & $1.09$\\
         NGC 2023   && $1.8_{-0.6}^{+0.6}$ & $2.9_{-0.9}^{+0.9}$ & $0.94$\\
         Flame Nebula   && $2.2_{-0.9}^{+0.9}$ & $3.0_{-0.9}^{+1.1}$ & $0.63$\\
         \midrule
         Sample && $1.8_{-0.5}^{+0.7}$ & $2.9_{-0.8}^{+1.0}$ & $0.74^\dagger$\\
         \bottomrule
         \multicolumn{5}{l}{($^\ast$) \small{Average fiber properties across the survey (see Paper III and IV).}}\\
         \multicolumn{5}{l}{($^\dagger$) \small{This value is computed including also the OMC-1 and 2.}}\\
    \end{tabular}
    \label{tab:single_fit_properties}
\end{table}

Our survey consists of five high-resolution \hnctrans~ maps across Orion. These maps are described by a total of $\sim130000$ spectra. We retrieved the main spectral line parameters, as centroid velocity ($V_\text{lsr}$), linewidth ($\Delta V$), and peak intensity ($I$), through a supervised fitting of these spectra using a customized GILDAS/CLASS \citep{2005Pety, 2013GildasTeam} routine \citep[see][for a detailed discussion of the fitting routine used in this work]{2013Hacar, 2018Hacar}. Among the total of $\sim130000$ spectra extracted from the five \hnctrans~ maps, we effectively fitted $\sim110000$ spectra ($\sim85\%$). To ensure the quality of the data, we applied two selection criteria: (a) the spectrum should contain at least one component with a S/N~$\ge$~5, and (b) $2\le V_{lsr}\le15$ \kms, the velocity range in Orion \citep{2008Bally}, leading to a final number of $\sim104000$ spectra considered in our analysis ($\sim93\%$ of the fitted ones).

Among all the fitted spectra, most ($\sim79\%$) present a single component in \hnc, while the residual multi-component spectra show two ($\sim20\%$) or three ($<1\%$) components. The relative number of single and multiple components recovered in our HNC spectra is similar to the one derived from C$^{18}$O observations in L1495/B213 by \citet[$\sim$70~\% single, $\sim$30\% multiple;][]{2016Hacarb}.
The distribution of single and multi-component spectra is regular throughout the sample (all values reported in Table~\ref{tab:fit_statistics}), with the Flame Nebula being the only outlier with $\sim45\%$ of multi-component spectra (with $\sim43\%$ and $\sim2\%$ of two and three component spectra, respectively).
We find typical median errors of 0.09~\kms~for the linewidths and of 0.04~\kms~for the velocity centroids obtained from our Gaussian fits, that is of less than one resolution channel of our spectra (Tab.~\ref{tab:fit_errors} summarises the errors in each region in the single and multi-component fitting).

The goal of this study is investigating the kinematics of the diffuse gas and its local behaviour across different locations in our maps (e.g., nearby the dense fibers and regions affected by feedback). A clearer description of the kinematics is achieved by selecting a single value for each pixel. Thus, we need to handle all the spectra fitted with multiple components either by taking (i) the brightest component among the
ones fitted, (ii) the broadest, or (iii) their average.

To avoid biases and considering that most of the fitted spectra have only a single component in \hnc~(see Table~\ref{tab:fit_statistics}), we decided to run a new fitting procedure with a single-component constraint. The linewidth estimated following the latter routine is to be considered as an upper limit, as multiple components may be fitted by a single broad Gaussian function. For completeness, in Appendix~\ref{sec:appendix_fit} we compare the linewidth values and the maps estimated from the single and multi-component fit. To conclude, we also produced velocity dispersion maps using moments (see Appendix~\ref{sec:appendix_fit} and Fig.~\ref{fig:method_comparison_linewidth.}). Overall, the linewidth values estimated using the latter methods appear to be lower in respect with the ones derived from the fit, especially when compared to the single component fit. This discrepancy may arise due to the moment map definition, i.e. summing the emission along the spectral axis: considering the positions with multiple Gaussian components, the velocity dispersion estimated by the second moment map will represent the separation between the components, not its linewidth. Following this interpretation, this difference is mostly visible in the Flame Nebula, as expected since is the region where the highest number of multiple components has been fitted. A comparison between the results of different methods is reported in Fig.~\ref{fig:method_comparison_linewidth.}. To further check if a single-component selection would critically affect our analysis, we compare the location of the fitted multi-components and the linewidth and velocity gradient maps for OMC-3 and the Flame Nebula (Fig.~\ref{fig:num_components}), not finding any spatial correlation.
Nevertheless, since none of the methods is flawless and since the results agree well, we report below the analysis based on the single component fit routine.
As already mentioned in sect.~\ref{sec:survey}, \hnc~emission is optically thin \citep{2023Tafalla}. However, even if optically thick, it would have not critically impacted our analysis: opacity would produce a broadening of the linewidth that may be responsible of blending multiple narrow velocity components \citep{2016Hacarb}. We removed this blending issue by considering a single-component fitting routine and, for this reason, we already considered our linewidth estimates as upper limits.

\subsection{Supersonic gas: linewidth and velocity dispersion}

\begin{figure*}[htbp]
	\centering
        \includegraphics[width=\textwidth]{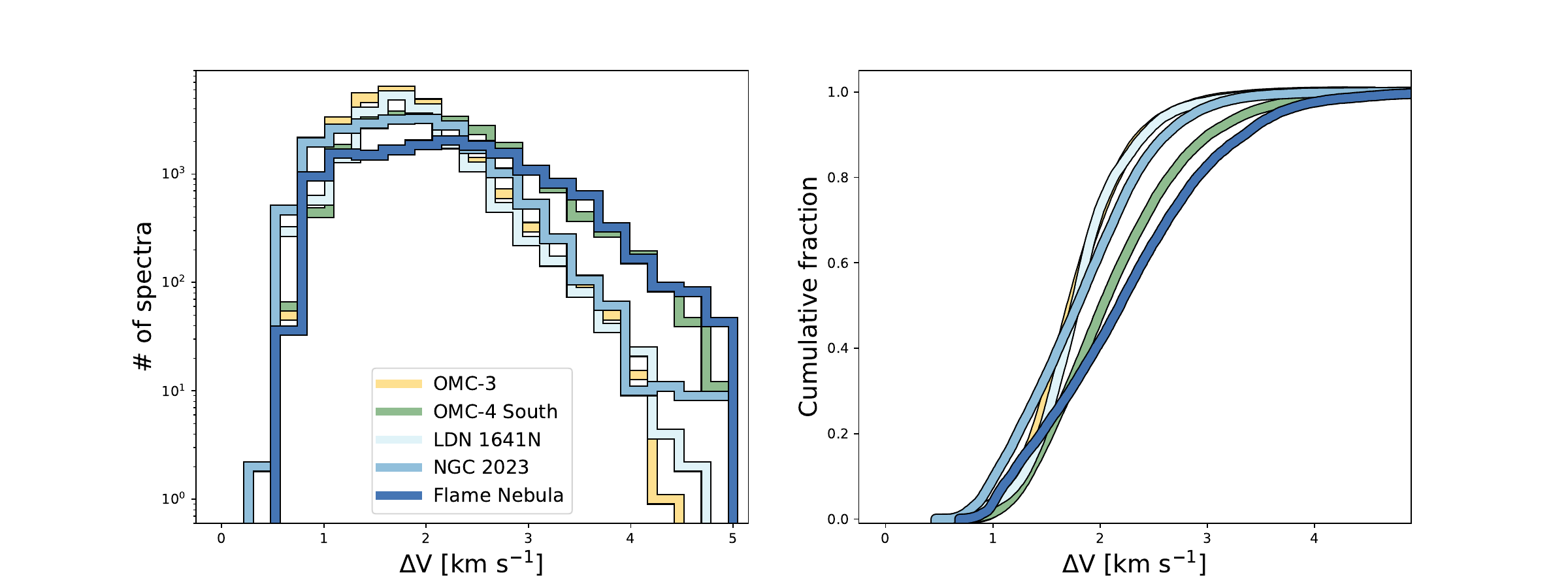}
	\caption{Distribution of the linewidth for all the five regions in our sample. Histogram (left; in logarithmic scale) and cumulative distribution (right) of $\Delta$V.}
	\label{fig:linewidth_statistics} 
\end{figure*}

\begin{figure*}[htbp]
	\centering
        \includegraphics[trim={0 2cm 0 1cm},clip,width=0.75\textwidth]{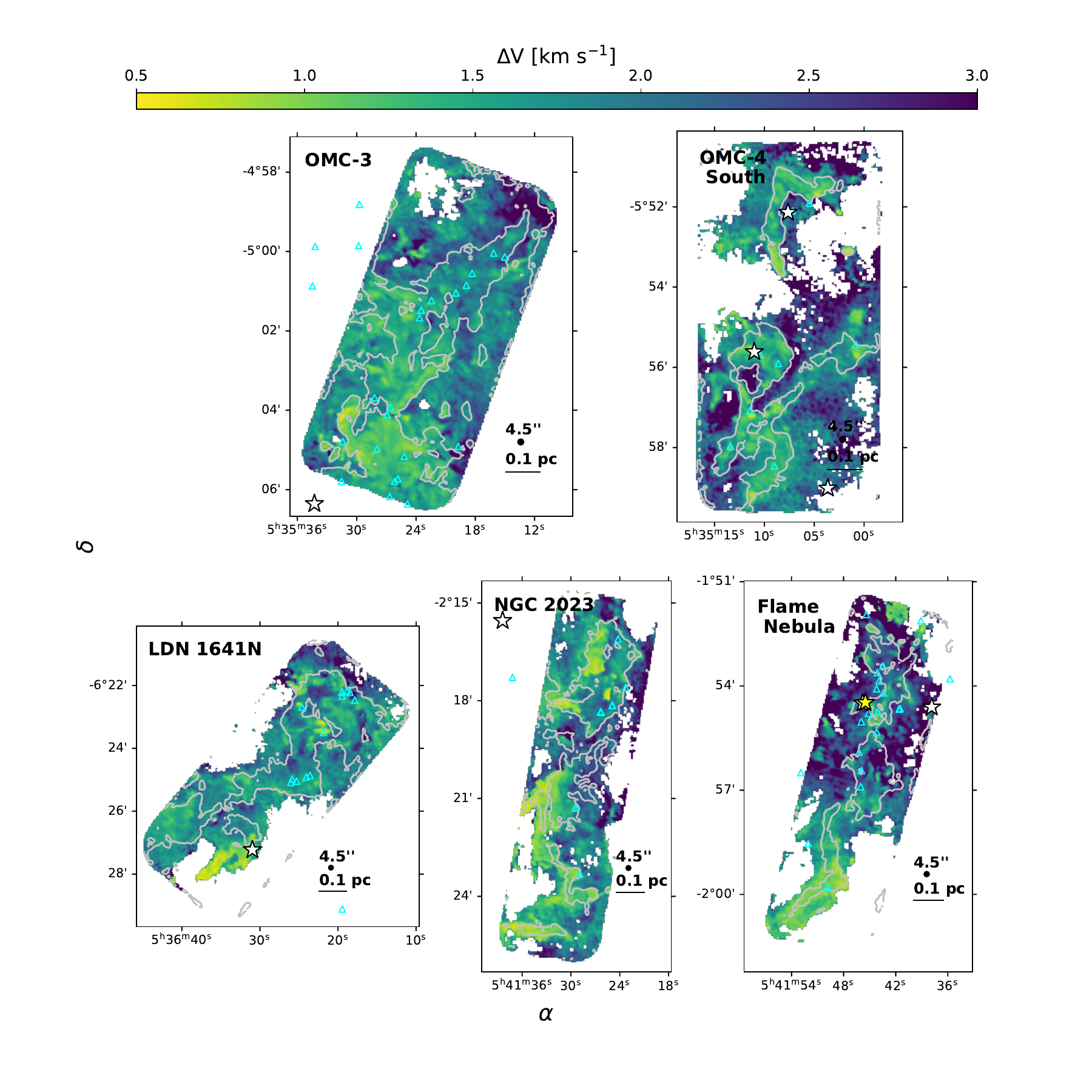}
	\caption{\hnctrans~ linewidth maps of the five star-forming regions in the EMERGE Early ALMA Survey at 4.5~arcsec resolution with ALMA+IRAM-30m. The gray contours show the \nthptrans~integrated intensity above 3$\sigma$ (the \nthp~maps are visible in Paper III). In all panels, cyan triangles are the protostellar objects from \citet{2012Megeath,2013Stutz,2016Furlan}, while the yellow and white stars are the O and B stars collected from Simbad \citep{2000Wenger}, respectively. Beam sizes and scale bars are placed in the bottom right corner.}
	\label{fig:linewidth_maps} 
\end{figure*}

\begin{figure*}[htbp]
	\centering
        \includegraphics[width=\textwidth]{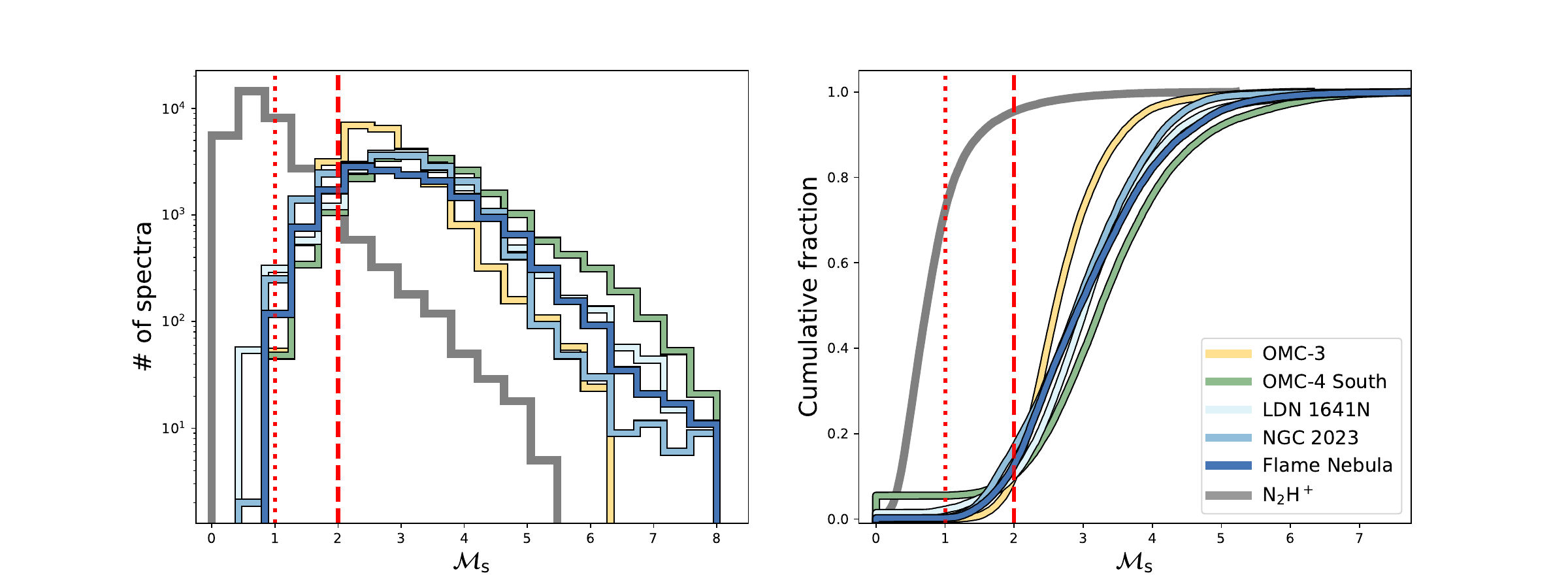}
	\caption{Distribution of the Mach number for all the five regions in our sample. Histogram (left) and cumulative distribution (right) of the Mach number. The gray line shows the distribution of the dense gas $\mathcal{M}_\text{s}$ values measured for the whole EMERGE Early ALMA Survey \citep[excluding OMC-1 and -2, see][for a full discussion]{2024Soccia,2024Soccib}. The red dotted and dashed lines at $\sigma_{nt}/c_s$=1 and 2 mark the transition between the sub-, tran-, and super-sonic regime, respectively.}
	\label{fig:mach_number_statistics} 
\end{figure*}

\begin{figure*}[htbp]
	\centering
        \includegraphics[trim={0 2cm 0 1cm},clip,width=0.75\textwidth]{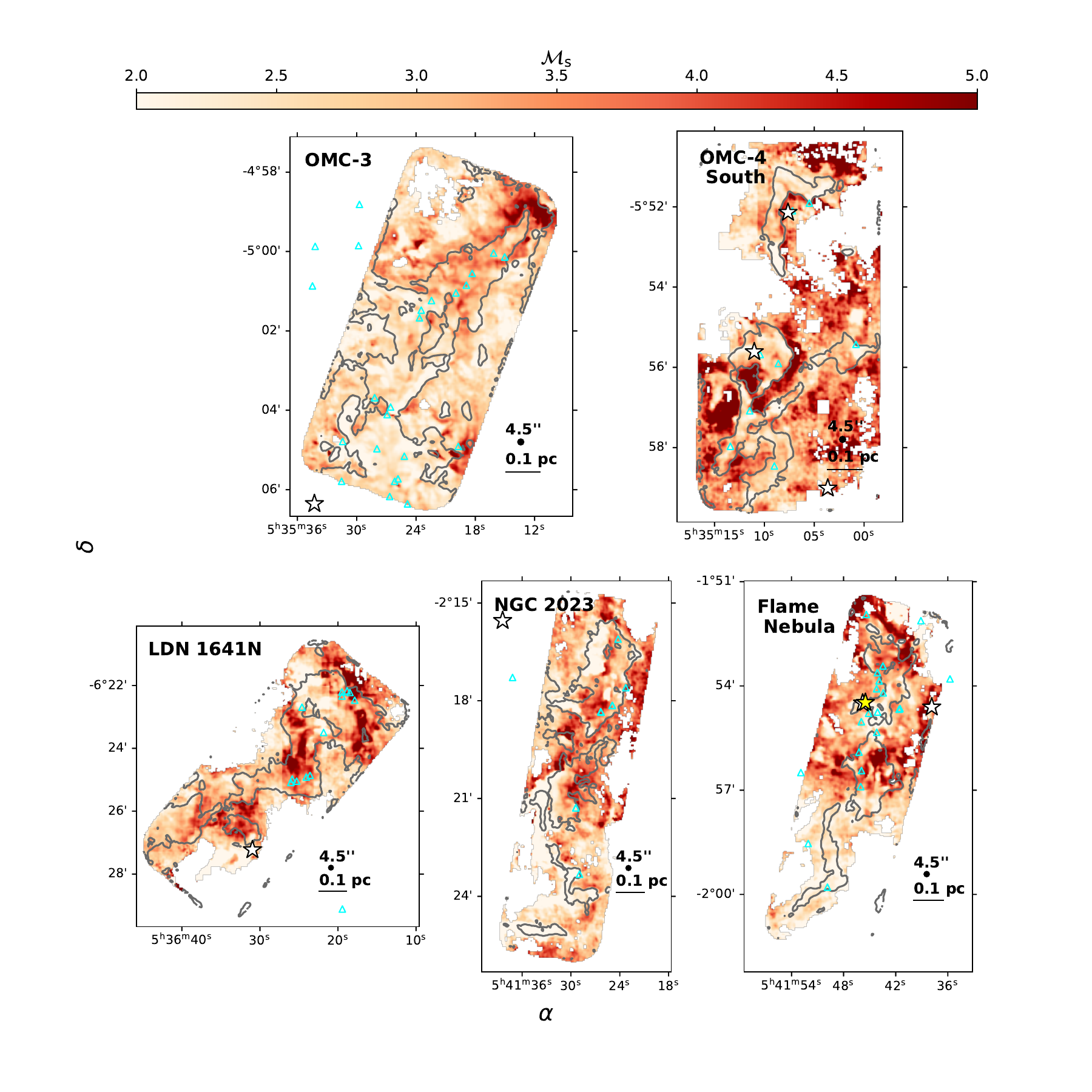}
	\caption{Mach number ($M=\sigma_{nt}/c_s$) maps of the five star-forming regions observed in \hnctrans~in the EMERGE Early ALMA Survey at 4.5~arcsec resolution. The gray contours show the \nthptrans~integrated intensity above 3$\sigma$ (the \nthp~maps are visible in Paper III). Symbols are similar to those in Fig.~\ref{fig:OMC3_maps}. Compared to this first figure, we note that some positions in our maps are masked due to the lack of good temperature estimates in these positions. Beam sizes and scale bars are placed in the bottom right corner.}
	\label{fig:mach_number_maps} 
\end{figure*}

We start our analysis of diffuse gas kinematics by studying the \hnc~linewidths, stressing again that these values have to be considered as upper limits. Fig.~\ref{fig:linewidth_statistics} shows the $\Delta$V distribution for the different regions in our survey. All targets show a similar linewidth distribution, with values within 0.5-5~\kms~and $\ge80\%$ of them below 2.5~\kms~(left panel in Fig.~\ref{fig:linewidth_statistics}). The corresponding median value and the inter-quartile range for the sample $\tilde{\Delta V}=1.8_{-0.7}^{+0.5}$~\kms~is reported in Table~\ref{tab:single_fit_properties}, along with the median values reported for the single targets. OMC-3 and LDN 1641N show the narrowest lines with $\tilde{\Delta V}=1.7$~\kms~and $\tilde{\Delta V}=1.8$~\kms, respectively; instead the Flame Nebula and OMC-4 South show the broadest ones with $\tilde{\Delta V}=2.2$~\kms~and $\tilde{\Delta V}=2.0$~\kms, respectively.

We explored the diffuse gas local behaviour by producing linewidth maps (Fig.~\ref{fig:linewidth_maps}). The linewidth measurement inside the gas dense contours (gray) is overall lower than outside. Slightly broader lines are observed near the YSOs and protostars, possibly due to local feedback (e.g., outflows) or its interaction with the gas. This effect was already discussed in \citet{2011Pineda}, where they found a positive correlation between the location of the YSOs and broader spectral lines. While some positions appear to be affected by local feedback around individual protostars, we find no systematic correlation between the increase of the observed HNC linewidths and the location of these objects in our maps. Comparing these local effects with the large coverage of our maps their statistical influence appears to be limited  in agreement with previous statistical estimates in several of these regions \citep{2019Feddersen}.
A potential cause of line broadening to take into account is large-scale feedback from stellar winds generated by O/B stars. This effect is visible in the Flame Nebula, where the lines become broader toward the northern area where the HII region extends (see Sect.~\ref{sec:feedback}).

The linewidth $\Delta V$ is a direct measurement of the total gas velocity dispersion $\sigma_{los}$ along the line-of-sight (LOS), which includes contributions from both thermal and non-thermal gas motion as $\sigma_{los}=\sqrt{\sigma_{th}^2 + \sigma_{nt}^2}=\Delta V / \sqrt{8\ ln 2}$. The turbulent regime is usually evaluated via the non-thermal motions that needs to be extracted as
\begin{equation}
    \sigma_{nt}=\sqrt{\sigma_{los}- \frac{k_B T_K}{\mu_{HNC}\ m_p}}
\end{equation}
with $\sigma_{th}=\sqrt{k_B T_K/\mu_{HNC}\ m_p}$ the thermal broadening of the line, $\mu_{HNC}=27$~amu the molecular weight of \hnc, $k_B$ the Boltzmann constant, and $m_p$ the proton mass.
The ratio between the non-thermal velocity dispersion and the gas sound speed ($c_s(T_K)=k_B T_K/\mu m_p$) with $\mu=2.33$~amu is the Mach number ($\mathcal{M}_\text{s}=\sigma_{nt}/c_s(T_K)$), which is used to describe the turbulence regime of the gas. The latter can be classified as subsonic ($0\le\mathcal{M}_\text{s}\le1$), transonic ($1\le\mathcal{M}_\text{s}\le2$), and supersonic ($\mathcal{M}_\text{s}\ge2$). 

To derive the Mach number we extracted the temperature value from our IRAM-30m $T_k$ maps at 30~arcsec resolution (panel c in Fig.~\ref{fig:OMC3_maps} and panel b in Fig.~\ref{fig:survey_maps}). Since the IRAM-30m beam is $\sim6$ times larger than in our ALMA+IRAM-30m maps, we assume $T_k$ to be constant within each IRAM-30m pixel and associated this value pixel-by-pixel to our ALMA+IRAM-30m \hnc~maps. We note  that variations on the kinetic temperature will only marginally impact on the Mach number as $\mathcal{M}_\text{s}\propto 1/\sqrt{T_K}$.

Figure~\ref{fig:mach_number_statistics} shows the Mach number distribution in our sample (see Figs.~\ref{fig:mach_hist} and \ref{fig:mach_cdis} for to better compare the behaviour of individual targets with respect to the sample). The values range between 0.5-8, with a clearly supersonic median value $\tilde{ \mathcal{M}_\text{s}}=2.9_{-0.8}^{+1.0}$ that is supersonic (see Table~\ref{tab:single_fit_properties})\footnote{All our Mach number measurements display the median value $\tilde{\mathcal{M}}$ and the 16\%-84\% inter-quartiles ranges.}. OMC-3 has the narrowest distribution with the lowest median value ($\tilde{\mathcal{M}_\text{s}}=2.6_{-0.5}^{+0.7}$), reflecting its narrower linewidths, while OMC-4 South the broadest ($\tilde{\mathcal{M}_\text{s}}=3.3_{-0.8}^{+1.1}$). Comparing Figs.~\ref{fig:linewidth_statistics} and \ref{fig:mach_number_statistics}, we notice that the shape of the $\Delta V$ and $\mathcal{M}_\text{s}$ distributions in our regions changes once corrected by $T_K$. 
The Flame Nebula, in particular, has the largest linewidth median value $\tilde{\Delta V}=2.2$~\kms, but not the highest Mach number median one $\tilde{ \mathcal{M}_\text{s}}=3.0$. This effect is likely tied to the high temperature in this region (see Sect.~\ref{sec:tempmaps}).
Overall, all our targets show similar Mach number distributions peaking around $\mathcal{M}_\text{s}\sim3$, despite the different properties of the sample. In Fig.~\ref{fig:mach_number_statistics} we also display the (total) Mach number distribution of the dense gas obtained from the analysis of \nthp~combining all regions in sample \citep[gray line;][]{2024Soccia}.
As seen by this figure, \nthp~samples a largely quiescent gas, with $>70\%$ of the fields being subsonic and up to $90\%$ within the transonic limit (see cumulative plot in right panel). 
In comparison, the diffuse gas traced by \hnc~shows a factor of $\sim4$ times higher degree of turbulence than the dense gas counterpart traced by \nthp~ ($\tilde{\mathcal{M}_\text{s}}(N_2H^+)=0.74$ vs $\tilde{\mathcal{M}_\text{s}}(HNC)=2.9$; see Table~\ref{tab:sample_properties}).

We now explore the local turbulence of the diffuse gas in the sample by producing Mach number maps (Fig.~\ref{fig:mach_number_maps}). The \hnc~gas is more turbulent than \nthp~everywhere; however, it does not exhibit regular behaviour in any of our targets.
In OMC-3 and 4 South, the \hnc~Mach number is roughly uniform and constant around $\mathcal{M}_\text{s}\sim2$ where the dense fibers are identified, while it grows up to $\mathcal{M}_\text{s}\sim5$ outside the dense gas contours. The $\mathcal{M}_\text{s}$ spatial distribution appears to be highly irregular around the fibers, with either abrupt changes or no visible differences within and outside the dense gas (e.g., the southern region of OMC-3).
In LDN 1641N, NGC 2023 and the Flame Nebula the distribution of the Mach number appears to be even more complex. In LDN 1641N the values grow inside the dense gas contours, showing an opposite behaviour with respect to OMC-3 and 4 South. In the Flame Nebula, instead, the southern part appears as more quiescent than the northern one, as already observed in the linewidth map, probably due to feedback (we will discuss about feedback in Sect.~\ref{sec:feedback}). Finally, NGC 2023 shows a distribution similar to the Flame Nebula, in this case in the East-West direction.

\subsection{The influence of feedback on the diffuse gas kinematics}\label{sec:feedback}

\begin{figure*}[htbp]
	\centering
        \includegraphics[width=\textwidth]{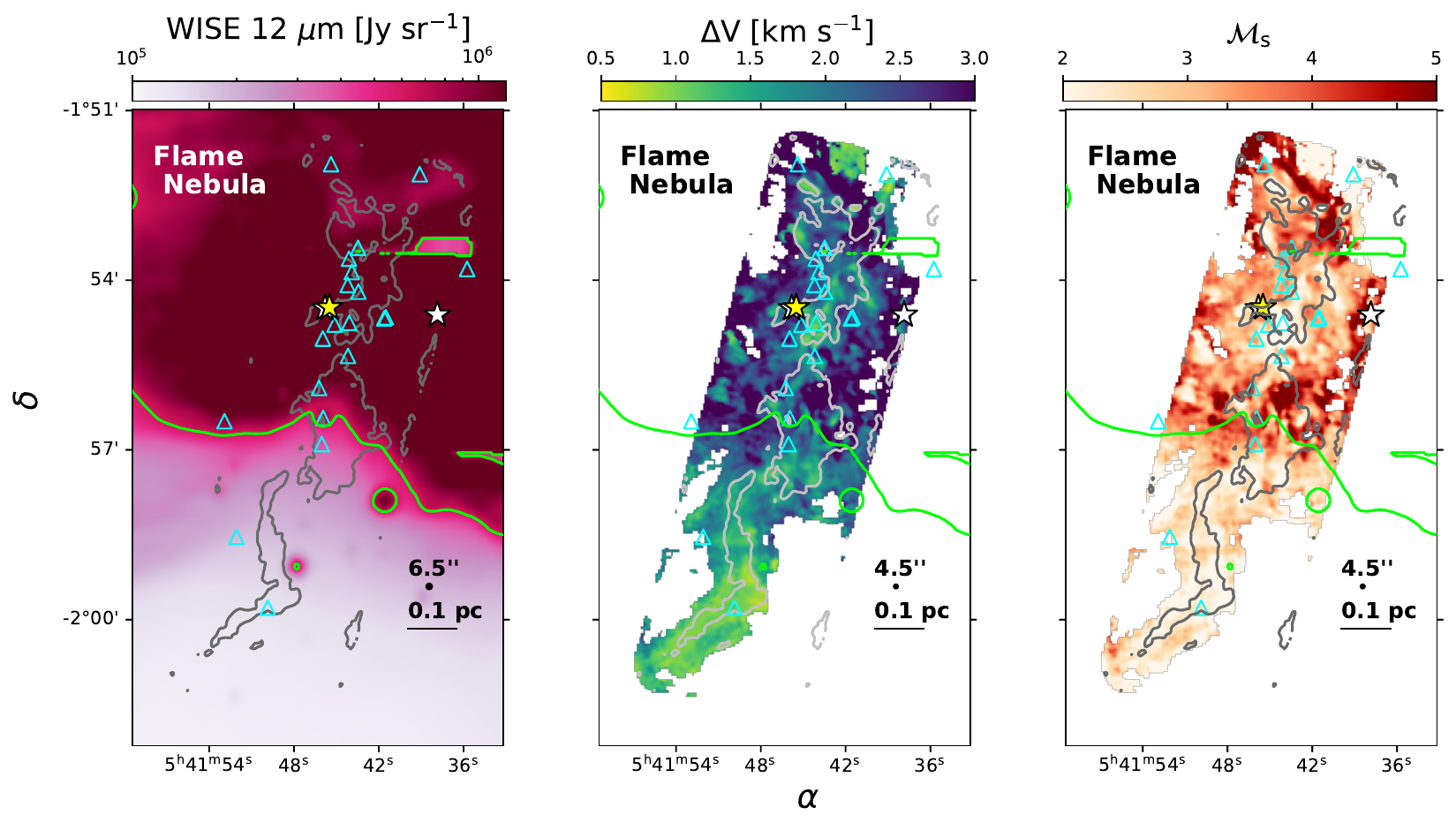}
	\caption{Maps of the Flame Nebula. From left to right: WISE 12$\mu$m map (6.5~arcsec), linewidth and Mach number map. The lime and grey contours define the area affected by feedback and the \nthptrans~integrated intensity above 3$\sigma$, respectively. % show the WISE 12$\mu$m emission at $3.5\times10^3$.  
    Symbols are similar to those in Fig.~\ref{fig:OMC3_maps}. Beam sizes and scale bars are placed in the bottom right corner.}
	\label{fig:feedback} 
\end{figure*}

%While the dense gas acts as "a rock in a river", 
The kinematics of the diffuse material and its turbulence regime appear to be influenced by feedback.
Given the various behaviours of the Mach number in our targets, we try to sample the regions in our targets most affected by feedback and investigate its influence on the gas kinematics. 

We test this approach on the Flame Nebula, since it is exposed to intense extreme-UV ionizing radiation and to strong far-UV dissociating radiation produced by the NGC 2024 IRS-2b (O8V) O-type star \citep[see][]{2008Bally}.
Previous velocity-resolved observations of the far-IR [CII] 158$\mu$m fine-structure line \citep[see][]{2017Pabst,2020Pabst} demonstrated that, despite being less severe, feedback affects other regions in our sample as well, namely OMC-3 \citep[NGC 1977 and 42 Orionis;][]{2008Peterson}, OMC-4 South \citep[NGC 1980 and $\iota$ Ori;][]{2012Alves}, and NGC 2023 \citep[$\sigma$ Ori;][]{1994Brown}.

To identify the regions affected by feedback we use the WISE 12 $\mu$m map at 6.5~arcsec of resolution. We selected the 12 $\mu$m band (W3) of WISE (Fig.~\ref{fig:feedback}, left panel) which, ranging between 7.5-17 $\mu$m, includes the UV-pumped polycyclic aromatic hydrocarbons (PAHs) emission at 8 $\mu$m associated to photo-dissociation \citep{2017Pabst}. 
The bright emission seen in WISE shows how the northern part of the Flame Nebula is largely affected by feedback while the southern is more pristine. We separate these two regions drawing a flux contour at $0.6$~MJy~sr$^{-1}$.
Fig.~\ref{fig:feedback} displays both the linewidth (middle panel) and the $\sigma_{nt}/c_s$ (right panel) maps with the same lime contour separating the two regions: both $\Delta$V and the Mach number values are much smaller and uniform in the South, while in the northern area, affected by the feedback, the distributions are more chaotic and irregular. Once excluded this region affected by feedback, the new median values get smaller $\tilde{\Delta V}=1.5_{-0.4}^{+0.5}$~\kms~and $\tilde{\mathcal{M}_\text{s}}=2.4_{-0.4}^{+0.5}$ ($\tilde{\Delta V}\sim2.4_{-0.6}^{+0.8}$~\kms~and $\tilde{\mathcal{M}_\text{s}}\sim3.4_{-0.9}^{+1.0}$ in the northern region).

Using this simple method, we confirmed that feedback affects the kinematics of the diffuse gas, broadening the linewidth and increasing the turbulent motions. This effect is prominent in the Flame Nebula, being heavily affected by large-scale feedback. The same effect, although less prominently, feedback is seen in other regions in our sample as well. 
An example is NGC 2023  affected by $\zeta$~Ori, where  $\Delta V$ and $\mathcal{M}_\text{s}$  are systematically higher on the West side of the cloud with respect to the East one. 
On the other hand, we did not manage to clearly identify the areas affected by feedback in our other regions, thus we could not properly test the variation on the linewidth median values. 
We attempted similar WISE flux cuts to those in the Flame Nebula in other regions although with limited success due to the different IR background levels in each of these clouds. Because of that, all the values reported in Table~\ref{tab:single_fit_properties} needs to be considered as upper limits. 

\section{Turbulence dissipation and the origin of fibers}

\subsection{Sonic regime and coherent scale}\label{sec:coherence}

Characterizing the turbulent properties of the gas inside molecular clouds is crucial to understand the process of star-formation in the ISM. While molecular clouds at large scales are supported by supersonic motions following a characteristic velocity dispersion-size relation, turbulence needs to be dissipated at smaller scale in order to form stars \citep[see][]{1981Larson}. 
A critical step in this process is the formation of coherent gas, that is the gas which non-thermal gas motions are subsonic and roughly uniform.
Dense cores have been classically interpreted as the first coherent structures formed within clouds as seen via selective tracers such as NH$_3$ \citep{1998Barraco,1998Goodman,2010Pineda} and \nthp~ \citep{2002Caselli}.
In contrast, the lower density gas surrounding cores, traced by diffuse tracers like OH and C$^{18}$O, shows supersonic velocity dispersion and it is assumed to be dominated by turbulence. The transition between these turbulent (diffuse) and more quiescent (dense) gas regimes is proposed to occur at scales comparable to the core radius of $\sim0.1$~pc \citep{1998Goodman} suggesting that the origin of individual cores is linked to the dissipation of the turbulence \citep{2005Klessen}.
The transition to coherence in a single tracer was first directly observed by \citet{2010Pineda} in the B5 region in Perseus using high sensitivity NH$_3$ observations tracing both dense and diffuse gas components. These observations suggest this transition to occur in short lags (sharp transition). Further studies have suggested that this sharp transition might be connected to highly dissipative effects (intermittency) characterized by high velocity gradients of $\nabla V_{lsr}>3$~km~s$^{-1}$~pc$^{-1}$ at small lags \citep[][see also Sect.~\ref{sec:velocity_gradients}]{1990Falgarone,1992Falgarone}. 

To further explore the connection between dense cores and the surrounding diffuse material, \citet{2011Hacar} investigated the L1517 region where several cores embedded in filamentary components have been detected in \nthp~ and C$^{18}$O, respectively. They found that these subsonic and velocity-coherent filamentary regions, now known as fibers, extend over scales significantly larger than cores, and as long as their entire length of $\sim0.5$~pc. Compared to previous results, these findings suggest that the origin of cores may not be directly related to turbulence dissipation. Instead, dense cores appear to inherit their subsonic properties from larger spatial scales, and in particular from their parental fibers. 

Similar subsonic fiber-like regions have been found in other clouds such as L1495/B213 in Taurus \citep{2013Hacar}, and in some cases showing coherent scales up to several parsecs in size \citep[e.g., in Musca][]{2016Hacar}. Interestingly, the gas inside B5 resolves into similar filamentary structures, here embedded in a dense region characterised by subsonic turbulence, when observed at high spatial resolution \citep{2011Pineda}.
These results are also in agreement with previous studies showing a similar gas organization traced by \nthp~at different scales \citep[see][]{2014Lee, 2014Henshaw,2017Hacar,2018Hacar,2019Sokolov} which highlights how the dense gas remains mostly sub-sonic despite the star-formation regime or environmental effects.

The scale in which the transition to coherence appears might also vary between regions.
The recent Green Bank Ammonia Survey (GAS) observed L1688 (Ophiuchus), NGC 1333 (Perseus), B18 (Taurus), and the ISF (Orion A) at a resolution of 30~arcsec. \citet{2017Friesen} found that Orion showed the broadest distribution of velocity dispersions within the sample, and attributed this effect to its intense stellar activity. Focusing on B18 and L1688, \citep{2019Chen_GAS} identified 12 coherent structures, known as droplets, characterized by a sharp transition in velocity dispersion measured by NH$_3$, smaller ($\sim0.04$~pc) than the coherent cores previously identified. A following study based on GAS data smoothed to 1~arcmin resolution, identified 12 coherent cores in L1688 measuring a gradual transition to coherence extending beyond typical core scales up to $\sim0.2$~pc \citep{2021Choudhury}. A recent GAS work in Perseus \citep{2024Chen_GAS} revealed that most filaments observed in NH$_3$ are subsonic, up to $\sim0.3-0.5$~pc scales.
While the transition to coherence appears to happen at the scales of the parent filament embedding the cores, more investigation is required to connect the former to its surrounding material.

Our previous Paper III investigates the distribution and kinematic properties of the dense gas traced by our ALMA+IRAM-30m \nthp~maps \citep{2024Soccia}. This analysis reveals that most of the dense gas is heavily organized in complex networks of elongated, fiber-like structures of $\sim$0.1~pc in size harbouring most cores in these regions. Paper III identified a total of 76 fibers within our maps (152 including OMC-1 and OMC-2), where the majority of these structures shows non-thermal motions within the sonic regime ($\tilde{\mathcal{M}_\text{s}}(N_2H^+)\sim0.74$, see Table~\ref{tab:single_fit_properties} and Fig.~\ref{fig:mach_number_statistics}). This statistical behaviour would agree with fibers being the first structures formed out of the turbulent cascade in the regions sampled by our survey.

The aim of this work is to investigate the turbulence regime of the diffuse gas traced by \hnc~and its interaction with the above quiescent fibers. As seen in Fig.~\ref{fig:mach_number_maps}, those regions with dense gas inside clouds such as OMC-3 and OMC-4 South (enclosed by gray contours in our figures), and where fibers have been identified,  show a roughly uniform and constant Mach number $\mathcal{M}_\text{s}\sim2$.
In contrast, the diffuse gas outside these fibers is overall more turbulent and exhibits supersonic motions with Mach numbers up to $\mathcal{M}_\text{s}\sim5$.
We remark that our analysis does not identify the transition to coherence as the change in Mach number in a single tracer as reported by \citet{2010Pineda} \citep[Type 3 correlation according to][]{1998Goodman}. Instead, our analysis explores the different Mach numbers seen in the different diffuse (\hnc) and dense (\nthp) gas components, respectively \citep[similar to a Type 4 correlation in][]{1998Goodman}. Nonetheless, these results suggest that the transition to coherence may occur at fiber scales. This transition requires, however, turbulence dissipation, which sets the initial conditions for fibers formation and evolution.

 \subsection{Signatures of turbulence dissipation and intermittency}\label{sec:velocity_increments}

\begin{figure*}[htbp]%
	\centering
        \includegraphics[trim={2cm 2cm 2cm 1cm},clip,width=1\textwidth]{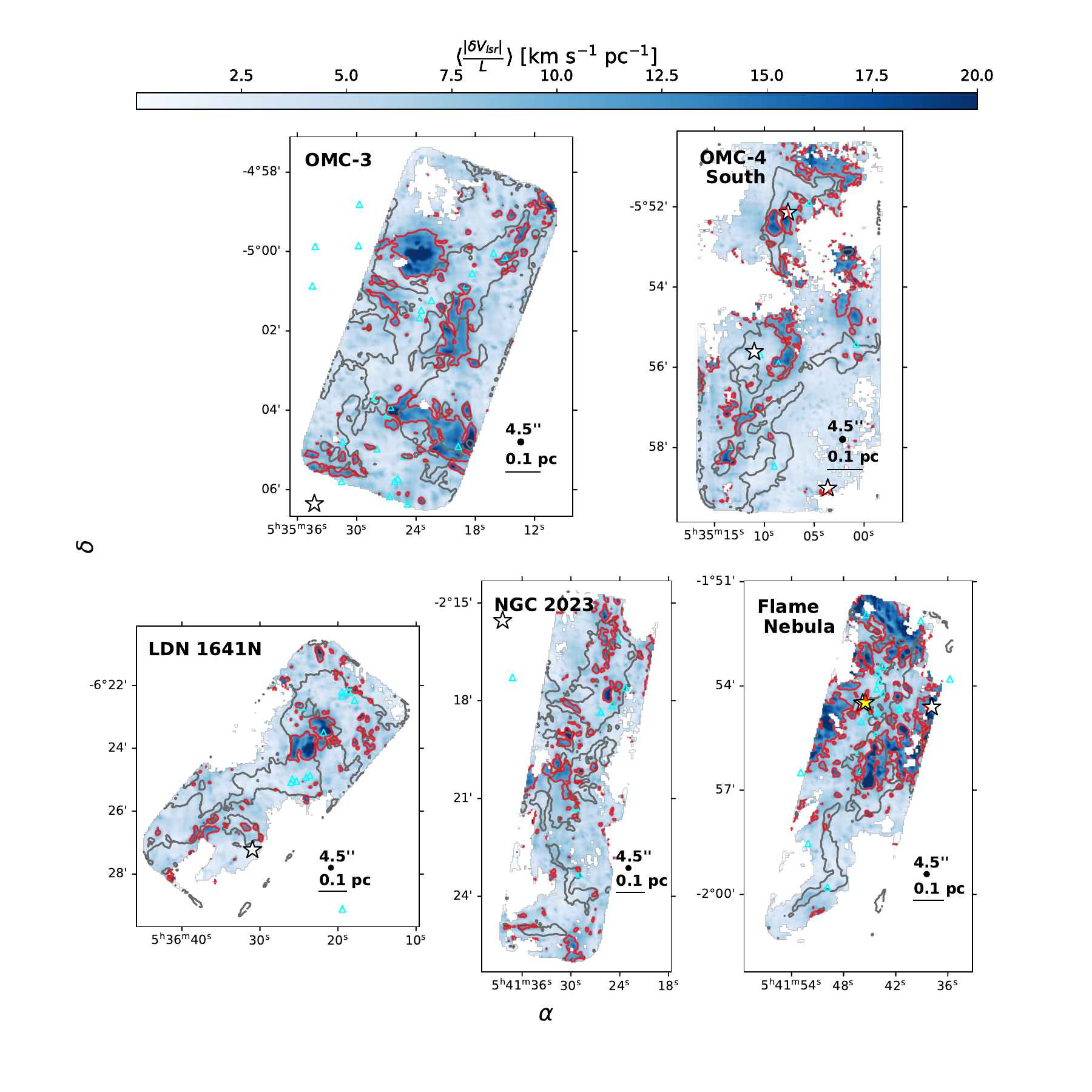}%
	\caption{\hnctrans~$V_\text{lsr}$ increment maps of the five star-forming regions in the EMERGE Early ALMA Survey at $4.5$~arcsec resolution with ALMA+IRAM-30m. The gray contours show the \nthptrans~integrated intensity above $3\sigma$ (the \nthp~maps are visible in Paper III). The red contours display the high velocity gradient regions identified in Fig.~\ref{fig:gradient_maps_increment}. Symbols are similar to those in Fig.~\ref{fig:OMC3_maps}. Beam sizes and scale bars are placed in the bottom right corner.}
	\label{fig:gradient_maps_increment} 
\end{figure*}

\begin{figure*}[htbp]
	\centering
        \includegraphics[width=0.65\textwidth]{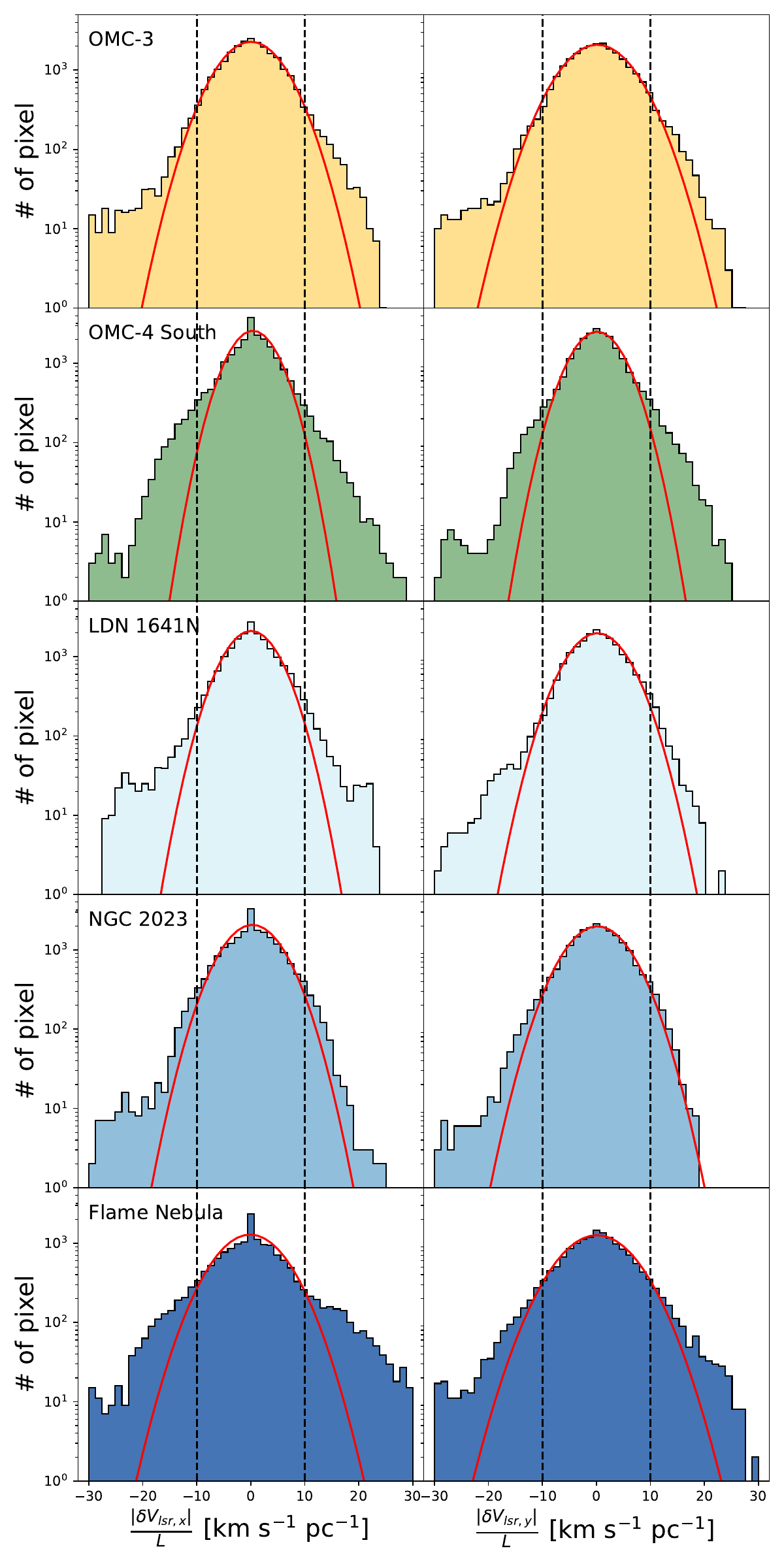}
	\caption{Histogram of the velocity increments evaluated along the $\alpha$ (left) and $\delta$ axes (right) for the targets in our survey. The red lines show the Gaussian fit applied to the distributions.}
	\label{fig:incr_sel} 
\end{figure*}

\begin{figure*}[htbp]
	\centering
        \includegraphics[width=0.8\textwidth]{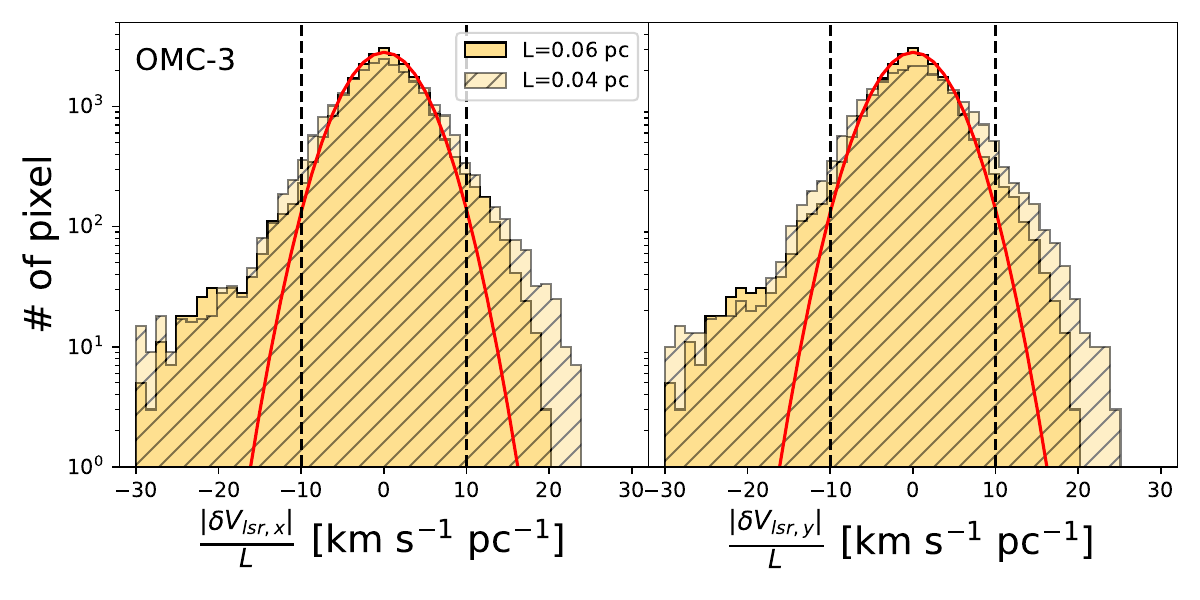}
	\caption{Histogram of the projection of the increments along the $\alpha$ (left) and $\delta$ axes (right) for a maximum offset L=20.25~arcsec (hatched area) and 29.25~arcsec in OMC-3, our showcase. The red lines show the Gaussian fit applied to the distributions.}
	\label{fig:lag_OMC3} 
\end{figure*}

Following the \citet{1941Kolmogorov} model of turbulence, the energy injected at large scales is progressively transferred to smaller scales, until it dissipates by viscosity at much smaller scales \citep[see][for a comprehensive review on turbulence]{2004Elmegreen,2004Scalo}. Investigating how this dissipative process occurs is thus crucial to understand fiber formation. The combination of high spatial resolution of our ALMA+IRAM-30m observations with the high dynamic range in column density of HNC, as a tracer of the diffuse gas around fibers, makes our survey ideal to investigate these processes. 

A pivotal property of turbulence dissipation is its intermittency at small scales \citep{1990Falgarone}, that is the spatio-temporal fluctuations of the density and velocity fields relative to their average values, which increase as the scale decreases \citep{1959Landau,1962Kolmogorov}. The energy cascade is not space-filling, thus this energy transfer to smaller scales occurs in a localized time and space. Intermittency governs the distribution of the subset of space in which turbulence dissipation occurs \citep{1994She,2001Anselmet}.
Although the intermittency nature of turbulence was recognized half a century ago, identifying these intermittent regions is still challenging due to their scarcity in time and space \citep{1992Falgarone,2003Pety,2007HilyBlant_paperI}. 
Our EMERGE Early ALMA Survey provides a large and homogeneous statistical sample of velocity field at high spectral and spatial resolution to probe the turbulent intermittency of the diffuse gas.

A statistical approach is the best way to investigate turbulence, in particular the analysis of the gas velocity field. 
Considering incompressible turbulence, the probability distribution function (PDF) of the velocity field is nearly Gaussian, at least at large scales \citep[see][]{1953Batchelor,1995Miesch}. 
Numerical simulations and laboratory experiments found that intermittency manifests itself through non-Gaussian wings in the PDF of velocity increments or derivatives \citep[see e.g.,][]{1984Anselmet,1990Falgarone,1991Vincent}. 
Simulations show the non-trivial connection between these structures found in the observed position-position-velocity (PPV) and the true position-position-position (PPP) space, respectively \citep{2013Beaumont,2018Clarke}. Nonetheless, additional simulations indicate that optically thin traces can be used to statistically investigate the turbulent properties of the gas \citep{2003Miville-Deschenes}. 
In particular, and although complicated by the above line-of-sight (los) effects, \citet{1996Lis,1998Lis} showed that intermittency can be better found in centroid velocity increment PDFs both in simulations and in CO (2-1) observations in clouds such as $\rho$ Ophiuchi. Similarly, intermittency manifests itself also in other high-order moments in these PDF distributions \citep[e.g., kurtosis;][]{2002Ossenkopf}.

\subsubsection{Statistical analysis using velocity increments}\label{sec:dissipation_statistics}

Following classical works in the literature, we evaluated velocity increments over a lag $\delta x=$L as $|\delta V_{lsr}|=|(V_{lsr}(x_0)-V_{lsr}(x_0+L)|$ for all %points located at a distance L from the centre 
\citep[e.g.,][]{2003Pety}.
For each position on our maps we consider all the adjacent points within a minimum and maximum offsets of L$-0.5$~pix and  L$+0.5$~pix \citep[see also][]{1998Lis}, respectively. We set L=0.04 pc (= 20.25~arcsec or $\sim 9$~pixels or $\sim4.5$~beams) and calculate the velocity increments over a minimum number of 20 independent, Nyquist-sampled pixels ($\sim1/3$ of the 72 independent pixels to ensure the calculation to be statistically significant).
To facilitate the comparison with other methods (see below), we normalize and evaluate the single velocity increment per position as $\frac{| \delta V_{lsr} |}{L}$ expressed \kmspc.

The spatial distribution of the velocity increments is
shown as azimuthal averages over the possible directions $\left(\frac{\langle | \delta V_{lsr} |\rangle}{L}\right)$.
We estimated the error on the velocity increment between $\sim0.03$ and $\sim0.07$~\kmspc.
We display the results in Fig.~\ref{fig:gradient_maps_increment} to investigate the spatial distribution of velocity increments in all our ALMA fields.

%%A small and roughly constant $\frac{|\delta V_{lsr}|}{L}$ characterizes 
Low $\frac{|\delta V_{lsr}|}{L}$ values characterize those regions where the velocity field is uniform, while large increments indicate sharp changes in velocity, probably associated to enhanced dissipation.
We identify typical velocity gradients below $\frac{|\delta V_{lsr}|}{L}\lesssim 5$~\kmspc~ in most parts of our maps, especially within the dense gas boundaries (gray contours) where the values are roughly constant. In contrast, several regions show higher gradients with $\frac{|\delta V_{lsr}|}{L}\gtrsim10$~\kmspc sometimes referred to as high-shear regions \citep[e.g.,][]{2009Hilyblant}.

To better identify and statistically constrain those gas parcels showing high velocity increments and associate them with intermittency, we follow a similar approach to \citet{1996Lis,1998Lis,2003Pety}, using the increments evaluated along the $\alpha$ and $\delta$ axes instead of simple velocity increments. Fig.~\ref{fig:incr_sel} shows the histograms of $\frac{|\delta V_{lsr,x}|}{L}$ and $\frac{|\delta V_{lsr,y}|}{L}$ increments for each of our five targets (see corresponding panels). The PDF of the centroid velocity increments is expected to follow a Gaussian distribution \citet{1996Lis}. Similar to it, most observed increments ($\sim70-90\%$ of the total number of points) can be reproduced by a Gaussian distribution (red curve in all panels) with a typical dispersion of $\sigma(\frac{|\delta V_{lsr,x}|}{L})$ or $\sigma(\frac{|\delta V_{lsr,y}|}{L})$ of $\sim 3-6$~\kmspc, obtained from a least-squares fit performed using \texttt{optimize.curve\_fit} from the \texttt{scipy} package.
However, all regions also show gradient distributions with significant non-Gaussian wings (up to $\sim10-30\%$ of the total number of points). These deviations are analogous to those wings reported in previous works using velocity increments \citep{1996Lis,1998Lis,2003Pety,2007Hily-Blant,2008Hily-Blant,2009Hilyblant}. We notice this departure from the expected normal behaviour to occur roughly around $\pm10$~\kmspc~(black dashed lines) in almost all cases, thus we use this value as a threshold to identify the intermittent regions in our maps. 

\citet{1996Lis,1998Lis,2003Pety,2007Hily-Blant,2008Hily-Blant,2009Hilyblant} claimed that the magnitude of the wings in the centroid velocity PDFs decreases monotonically with increasing separation (lag) in evaluating the increments. We tested this effect on our data by calculating the increments considering a maximum offset of L=0.06~pc (i.e.,29.25~arcsec or 13~pixels). Fig.~\ref{fig:lag_OMC3} shows the histograms of $\frac{|\delta V_{lsr,x}|}{L}$ and $\frac{|\delta V_{lsr,y}|}{L}$ in OMC-3 that compare the results for L=0.04 (hatched area) and 0.06~pc (solid area). We observe the wings decreasing for L=0.06~pc, with a distribution closer to a Gaussian, in agreement with previous works. This result proves that intermittency occurs only at small scales, and thus high spatial resolution observations are required to investigate this phenomenon \citep{1992Falgarone}.

Focusing on the analysis of scales L=0.04~pc we identify all positions in our maps where $\frac{|\delta V_{lsr,x}|}{L}$ and/or $\frac{|\delta V_{lsr,y}|}{L}$ are $\ge10$~\kmspc~and display them within the red contours in Fig.~\ref{fig:gradient_maps_increment}. These points extracted from the non-Gaussian wings (with $\frac{|\delta V_{lsr,x|y}|}{L}\gtrsim 3\times $ $\sigma(\frac{|\delta V_{lsr,x|y}|}{L}$) correspond to high-velocity gradient areas and are potentially associated with intermittency.

\subsubsection{Velocity gradient statistics}\label{sec:velocity_gradients}

\begin{figure*}[htbp]
	\centering
        \includegraphics[trim={2cm 2cm 2cm 1cm},clip,width=1\textwidth]{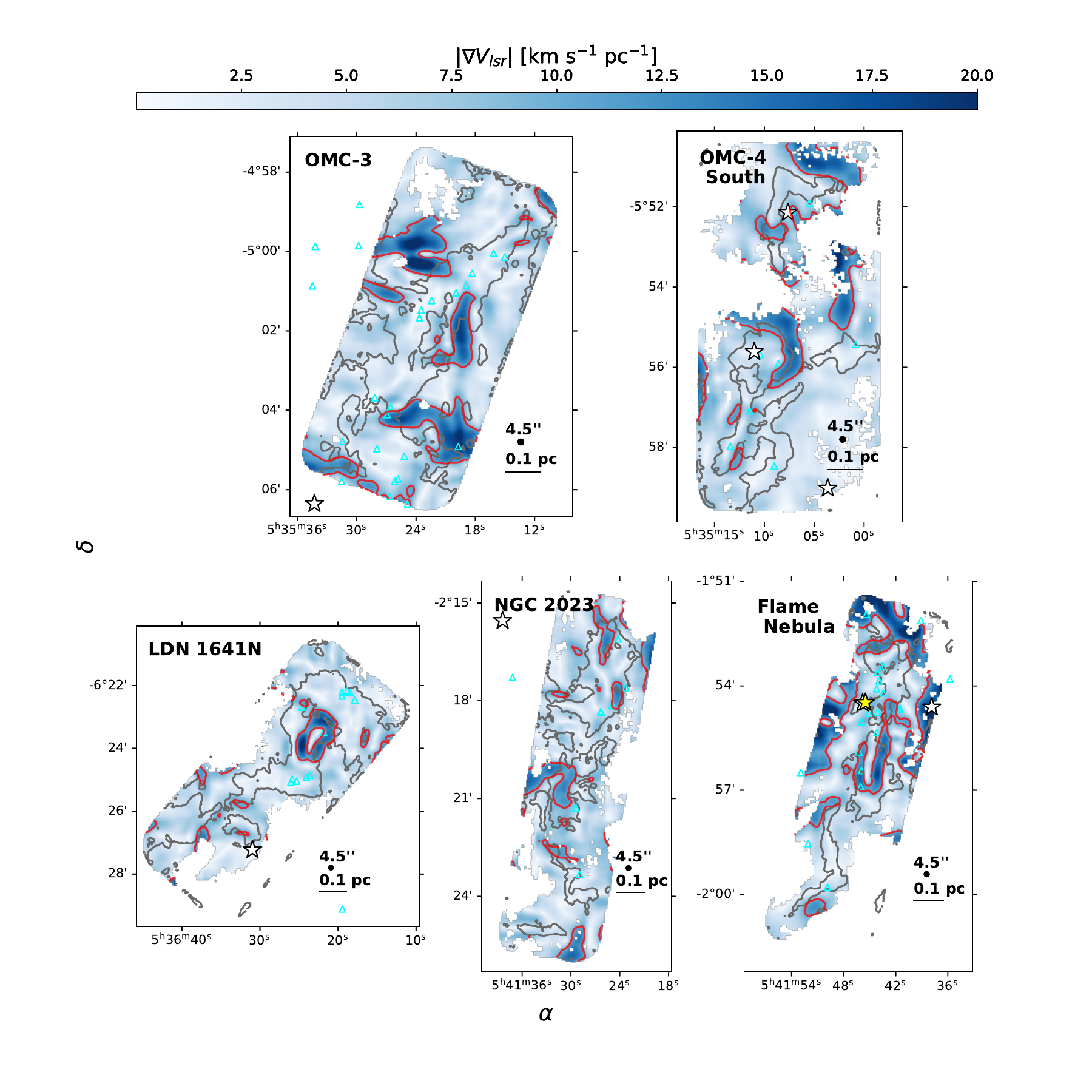}
	\caption{\hnctrans~ $V_\text{lsr}$ gradient maps of the five star-forming regions in the EMERGE Early ALMA Survey at 4.5~arcsec resolution with ALMA+IRAM-30m. The gray contours show the \nthptrans~integrated intensity above 3$\sigma$ (the \nthp~maps are visible in Paper III). The red contours display the high velocity gradient regions identified in Fig.~\ref{fig:grad_sel}. Symbols are similar to those in Fig.~\ref{fig:OMC3_maps}. Beam sizes and scale bars are placed in the bottom right corner.}
	\label{fig:grad_maps} 
\end{figure*}

\begin{figure*}[htbp]
	\centering
        \includegraphics[width=0.65\textwidth]{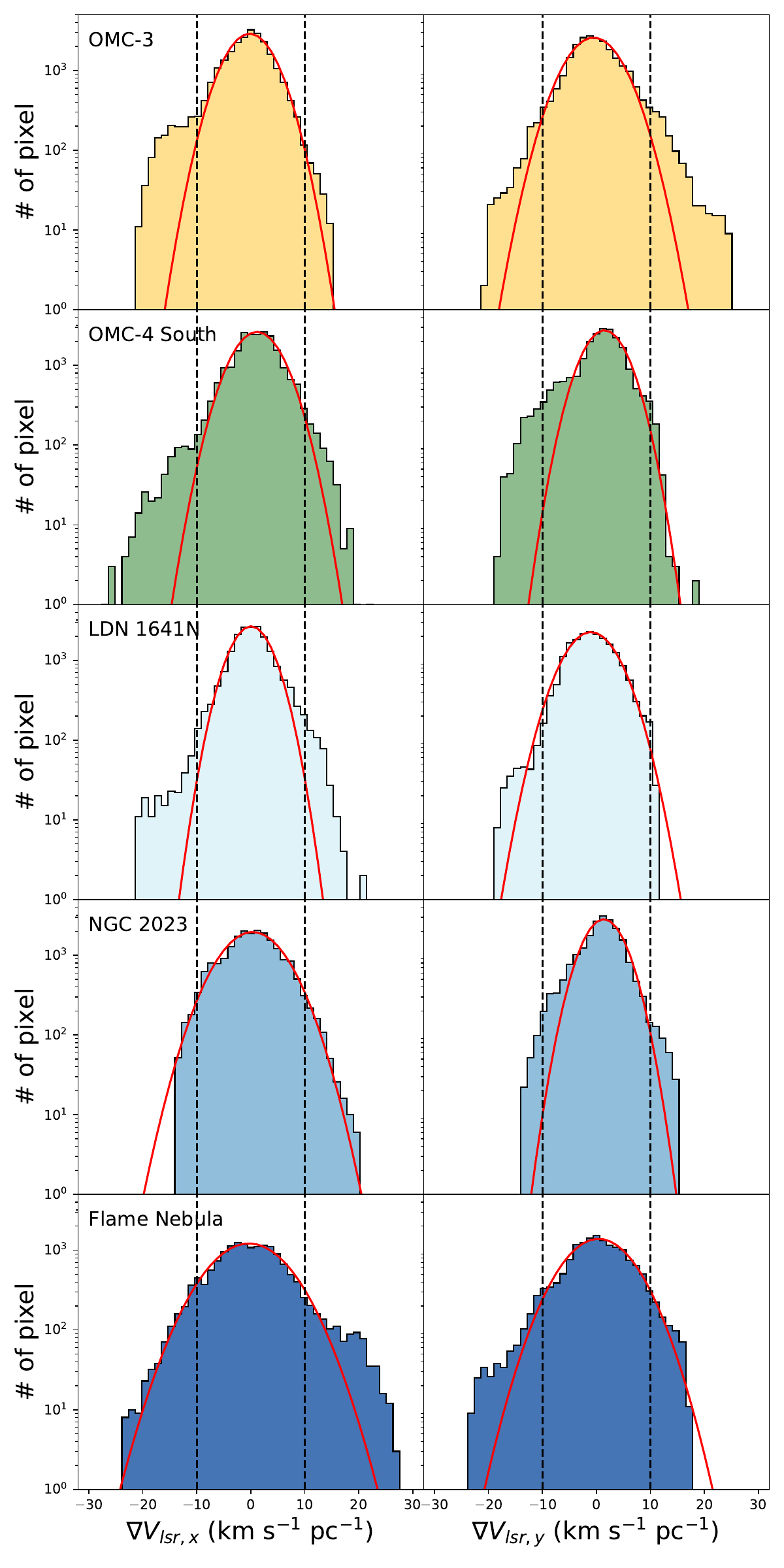}
	\caption{Histogram of the projection of the gradient along the $\alpha$ (left) and $\delta$ axes (right) form Eq.~\ref{eq:vlsr} for the targets in our survey. The red lines show the Gaussian fit applied to the distributions.}
	\label{fig:grad_sel} 
\end{figure*}

\begin{figure*}[htbp]
	\centering
        \includegraphics[width=0.8\textwidth]{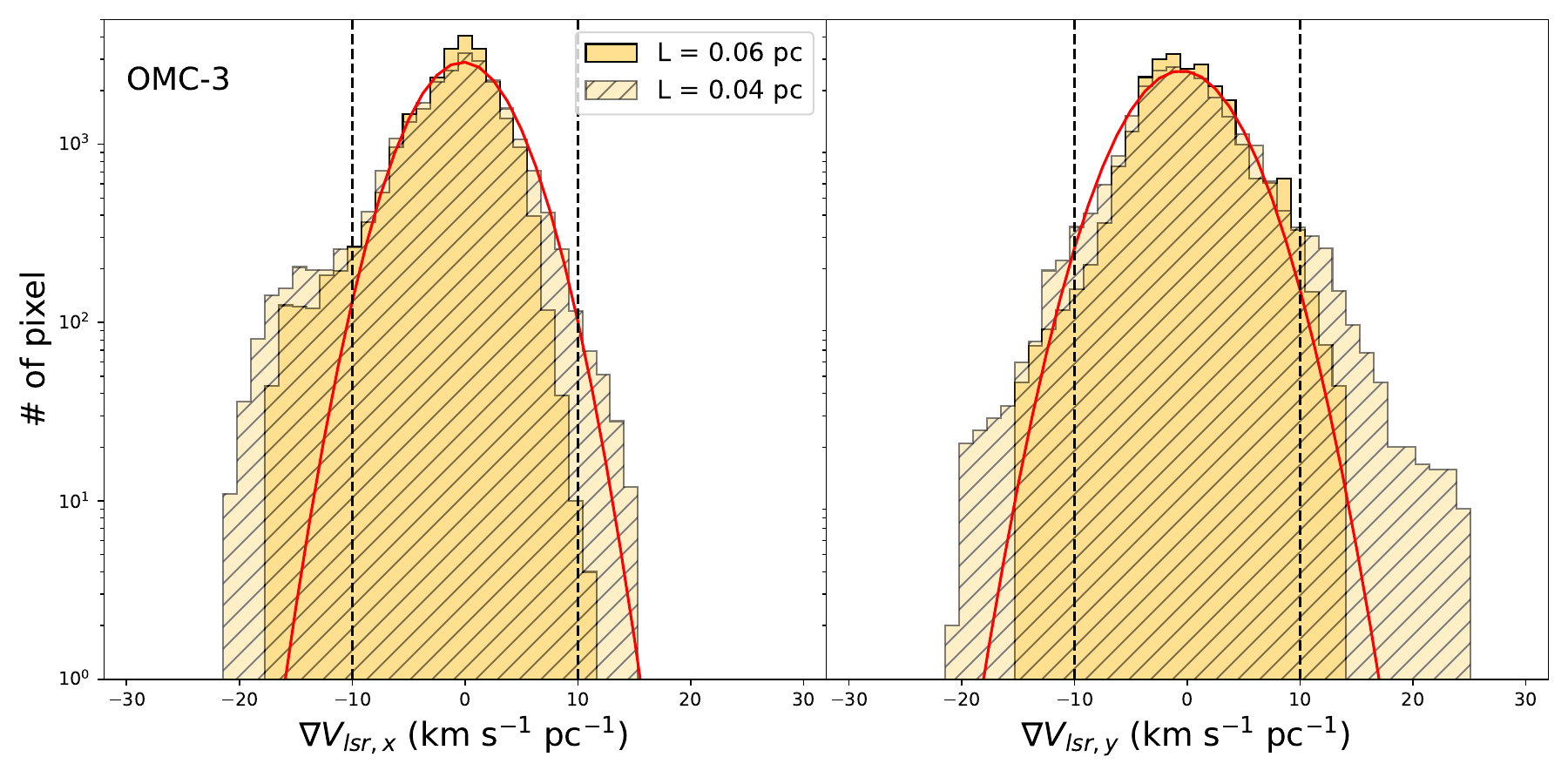}
	\caption{Histogram of the projection of the gradient along the $\alpha$ (left) and $\delta$ axes (right) for a maximum offset L=20.25~arcsec (hatched area) and 29.25~arcsec in OMC-3, our showcase. The red lines show the Gaussian fit applied to the distributions.}
	\label{fig:grad_OMC3} 
\end{figure*}

In this work, we introduce the use of centroid velocity gradients ($\nabla V_{lsr}$) as alternative method to identify regions of high shear and turbulence dissipation.
Following the formulation in \citet{1993Goodman}, the centroid velocity $V_\text{lsr}$ of a given position can be described as 
\begin{equation}
    V_{lsr}=V_0+a\Delta\alpha+b\Delta\delta,
    \label{eq:vlsr}
\end{equation}
with respect to the systemic velocity of the cloud $V_0$, where $\Delta\alpha$ and $\Delta\delta$ are the offset in right ascension and declination expressed in radians to the reference center, and $a$ and $b$ the projection of the gradient per radians along the $\alpha$ and $\delta$ axes, respectively. Following \citet{1993Goodman} we perform a least-squares fit of Eq.~\ref{eq:vlsr} to our centroid velocity maps using the function \texttt{optimize.curve\_fit} from the \texttt{scipy} package \citep{2020SciPy-NMeth} which returns the magnitude of the gradient as
\begin{equation}
    |\nabla V_{lsr}|=(\sqrt{a^2+b^2})/D,
    \label{eq:grad}
\end{equation}
with D as distance to each individual cloud (see Table~\ref{tab:sample_properties}). The estimated error on the magnitude of the gradient estimated form the fit is $\sim0.3$~\kmspc.
For each position on our maps we select all the adjacent positions within a maximum offset of L=0.04 pc (= 20.25~arcsec or $\sim 9$~pixels or $\sim4.5$~beams) both in $\alpha$ and $\delta$ and, and calculate the velocity gradient (i.e., minimize Eq.~\ref{eq:grad}) in those positions with a minimum number of 100 Nyquist-sampled pixels ($\sim1/3$ of the 324 independent pixels to ensure the calculation to be statistically significant).

We note that the above gradient estimates are equivalent to the velocity increments $\delta V_{lsr}$ at a given lag $\delta x$(=L) presented in the previous Sect.~\ref{sec:velocity_increments}.
The difference between these two methods is that the gradient estimate considers all the points within $\leq L$ into account in the calculation, while the velocity increment only the one at a specific scale $=L$. To ensure that these two methods produce similar results we also computed the equivalent velocity gradient in annuli of size L, i.e. in an annulus at a distance L and of extent 1 pix (considering all the points between $L-0.5$~pix and $L+ 0.5$~pix). We present a detailed comparison of the above classical increments, the annulus and our new gradient methods, including the analysis of their relative performance and systematic biases, in Appendix~\ref{sec:appendix_gradients}.

The analysis of gas motions and gradients is potentially biased by line opacity issues. Heavily saturated lines ($\tau>50$) are expected to be less sensitive to the small Doppler shifts of their line centroids as their central velocities are the most affected by opacity. This effect is however less relevant at intermediate opacities (e.g. $\tau\sim 5-10$) as the optically thin line wings significantly contribute to the line fit and compensate the few saturated velocities at the line centroids \citep[see Fig.1 in ][]{2016Hacarb}.
We expect \hnc~ to present low to moderate opacities of $\tau\sim 0.5-5$ in most parts of our maps (see Sect.~\ref{sec:survey}) and thus our gradient measurements to be not significantly biased by opacity.

We display the results obtained by the gradient method in Fig.~\ref{fig:grad_maps} to investigate the spatial distribution of velocity gradients in all our ALMA fields.
As already observed in the increments maps, the magnitude of the observed velocity gradients appears to be consistent with those inside filaments \citep{2023Hacar} and dense cores \citep{1993Goodman, 2002Caselli}. The gradients we measured are below $|\nabla V_{lsr}|\lesssim 5$~\kmspc~ in most parts of our maps, in particular within the dense gas boundaries (gray contours) where the values are roughly constant. In contrast, several regions show higher gradients with $|\nabla V_{lsr}|\gtrsim10$~\kmspc, with a maximum value of $|\nabla V_{lsr,max}|\gtrsim47$~\kmspc, already identified as high-shear regions \citep[e.g.,][]{2009Hilyblant}. Some of these measurements may be locally affected by the presence of protostellar outflows (e.g., MMS7 in OMC-3). Nonetheless, the systematic detection of these features in regions with almost no stellar activity such as NGC~2023 rule out these contamination at large scale. Howsoever, we tested this effect removing the pixels around the protostars (up to 3~beams) and proved that the presence of these objects does not affect the high shear regions.

More important, these high-velocity gradient positions are always concentrated in small regions located next or surrounding the dense gas traced by \nthp~(gray contours in Fig.\ref{fig:grad_maps}). Clear examples can be found in OMC-3, OMC-4 South, and NGC~2023 where we find peak gradients $|\nabla V_{lsr,max}|>15$~\kmspc~close to the dense gas. These gradients are a factor of $>3$ times higher than inside fibers ($|\nabla V_{lsr}|\sim3-5$~\kmspc).
This morphology reinforces the idea of a transition to coherence happening at fibers scales and turbulence dissipation being connected to fiber formation (see Sect.~\ref{sec:fiberorigin}). 

Interestingly, we observe that these high velocity gradient regions are only associated with some fibers, not all of them. This result is in agreement with the intermittent nature of turbulence where dissipation likely operates as irregular phenomenon in time and space \citep{1995Miesch,1999Miesch,1996Lis}. Our results also agree with a similar analysis performed in the Serpens filament using C$^{18}$O  observations \citep{2021Gong}. Following a similar approach, these authors find that $|\nabla V_{lsr}|$ is generally higher at the filament boundaries than at its crest. Our results appear to be more systematic and cover a wider range of physical cloud conditions.

\begin{table}[]
\caption{Turbulence dissipation estimates}
    \centering
    \begin{tabular}{lcccl}
        \toprule
        Target &  $\sigma(\nabla V_{lsr})$ & $(\nabla V_{lsr})_{max}$   &$f_s$& $f_\epsilon$ \\
        & (\kmspc) & (\kmspc) & & \\
        \midrule
         OMC-3  & 5.9 & 26 & 0.13&0.51\\
         OMC-4 South    & 5.3 & 31 &0.11&0.44\\
         LDN 1641N  & 5.4 & 47 & 0.05&0.30\\
         NGC 2023   & 6.2 & 27  & 0.09&0.31\\
         Flame Nebula   & 8.4 & 34 & 0.23&0.64\\
         \bottomrule
         \\
        \multicolumn{5}{l}{(Col.4) \small{Surface filling fraction $f_s=\frac{N_{pix}(high shear)}{N_{pix}(all)}$.}}\\
        \multicolumn{5}{l}{(Col.5) \small{Fraction of turbulent dissipation according to Eq.~\ref{eq:turbdisp}.}}
    \end{tabular}
    \label{tab:turb_dissipation}
\end{table}

To better identify and statistically constrain those gas parcels showing high velocity gradients and associate them with intermittency, we repeat the analysis presented in Sect.~\ref{sec:dissipation_statistics} for velocity gradients. Fig.~\ref{fig:grad_sel} shows the histograms of $\nabla V_{lsr,x}$ ($=a/D$) and $\nabla V_{lsr,y}$ ($=b/D$) gradient projections obtained from Eq.~\ref{eq:vlsr} for each of our five targets (see corresponding panels). 
As already shown in Fig.~\ref{fig:incr_sel}, all regions also show gradient distributions with significant non-Gaussian wings, appearing roughly around the same value identified for velocity increments ($\pm10$~\kmspc~, black dashed lines). Thus we keep this value as a threshold to identify the intermittent regions in our maps.

We tested the effect of considering different lag values in our analysis, running a fitting routine considering a maximum offset of L=0.06~pc (i.e.,29.25~arcsec or 13~pixels).
Fig.~\ref{fig:grad_OMC3} shows the histograms of $\nabla V_{lsr,x}$  and $\nabla V_{lsr,y}$ ($=b/D$) in OMC-3, 
comparing the results for L=0.04 (hatched area) and 0.06 pc
(solid area). In agreement with Fig.~\ref{fig:lag_OMC3}, the wings decrease for L=0.06 pc, with a distribution closer to a Gaussian.

Focusing on the analysis of scales L=0.04~pc we identify all positions in our maps where $|\nabla V_{lsr,x}|$ and/or $|\nabla V_{lsr,y}|$ are $\ge10$~\kmspc~and display them within the red contours in Fig.~\ref{fig:grad_maps}. These points extracted from the non-Gaussian wings (with $|\nabla V_{lsr,x|y}|\gtrsim 3\times $ $\sigma(\nabla V_{lsr,x|y}$) correspond to high-velocity gradient areas and are potentially associated with intermittency. These locations %%are concentrated in specific sub-regions within our maps
coincide with the areas previously identified from their elevated velocity increments (see Fig.~\ref{fig:gradient_maps_increment})
and form elongated structures with typical sizes of $\sim$~0.1-0.3~pc. These structures are only found in few areas of our maps, and usually occupy a small surface filling factor $f_s\lesssim$20\% estimated from the ratio of positions with high-shear with respect to the entire map. In most cases these high velocity gradient areas do not coincide with dense gas (within the gray contours), instead they tend to be located near their boundaries, as already observed by \citet{2003Pety}.

\subsection{Enhanced turbulent dissipation}\label{sec:fiberorigin}

The analysis of our ALMA \hnc~observations (see above) agrees with previous results obtained in SD observations of nearby molecular clouds using different CO isotopologues \citep{2003Pety,2007HilyBlant_paperI,2007Hily-Blant,2008Hily-Blant,2009Hilyblant}. 
The gas showing high velocity gradients (or high shear) is found in small locations (low filling factors) forming narrow ($\sim$few thousand au) and elongated ($\sim0.1$~pc) features in regions of intermediate column densities ($N(\text{H}_2)\sim5\times 10^{21}-5\times 10^{22}$ cm$^{-2}$) in the surrounding of some dense gas structures.
Connected to the intermittent nature of both density and velocity fields in turbulent gas (Sect.~\ref{sec:velocity_increments}), these high shear features are interpreted as regions of high vorticity associated to the dissipation of turbulence in molecular clouds.

Despite their small size, the previously identified high-shear regions may play an important role on the dissipation of turbulence in our clouds.
Following a similar approach to \citet{2003Pety}, we have calculated the enhanced local contribution of turbulence dissipation as $\nabla V_{lsr,max}^2/\sigma(\nabla V_{lsr})^2$ per cloud in Table~\ref{tab:turb_dissipation}, where in this case we use $\nabla V_{lsr}^2=\nabla V_{lsr,x}^2+\nabla V_{lsr,y}^2$ as the total velocity gradient in 2D. We find a maximum factor of 16-75x in turbulence dissipation at the position of these high shear locations compared to an average cloud position. Likewise, we also estimated the fraction of turbulent dissipation occurring in these high shear regions with respect the rest of the cloud such as 
\begin{equation}\label{eq:turbdisp}
    f_\epsilon=\frac{\sum_{shear}(\nabla V_{lsr}^2)}{\sum_{all}(\nabla V_{lsr}^2)}, 
\end{equation}
where $\sum_{all}$ and $\sum_{shear}$ refer to the sum of all contributions or only positions of high shear, respectively. As shown in Table~\ref{tab:turb_dissipation}, those areas identified as high shear regions appear as major contributors to the total turbulent dissipation in our fields with $f_\epsilon>$30\%. In some extreme cases such as OMC-3 and the Flame Nebula, their contribution is even dominant over the rest of the cloud with $f_\epsilon>$50\%. This enhanced contribution is even more surprising considering the limited filling factor of these high shear regions of $f_s\lesssim$~20\% \citep[see][for a similar discussion]{2003Pety}. However, such high values may be partially due to feedback effects.
Previous studies have associated the location of turbulent dissipative structures with the position of individual dense cores \citep{2003Pety,2007Hily-Blant,2008Hily-Blant,2009Hilyblant} in agreement to the previously proposed transition to coherence during the formation of these objects \citep{1998Goodman,2010Pineda}.
Instead, the homogeneous coverage of our maps shows how these high-shear features detected in HNC are systematically associated to the location of many of the dense fibers identified in \nthp~ rather than to individual cores (see Fig.~\ref{fig:grad_maps}). 
Unlike previous findings, our results (Sect.~\ref{sec:coherence}-\ref{sec:velocity_gradients}) thus favour the interpretation of turbulence to be dissipated before the formation of cores and during the formation of these first dense gas structures in molecular clouds.

From the analysis of these velocity gradient maps, we do not observe any preferential direction of the velocity field with respect to the dense structures. This result does not allow us to favour and/or rule out any of the filament formation mechanisms in shock compressed layers \citep[see][for a review]{2023Pineda}. We only observe the intermittent behaviour of shocks and dissipative patterns as they appear near fibers.
A larger statistical sample together with simulations may allow to better constrain the properties of turbulence dissipation and its link to the formation processes of filaments in molecular clouds.

\section{Conclusions}\label{sec:summary}
Our aim in this work is to investigate the kinematics of the diffuse gas and the link between turbulence dissipation and the formation of fibers. 
We characterized the physical and kinematic properties of a reduced sample of the EMERGE Early ALMA Survey, a sample of five star-forming regions in Orion, all homogeneously surveyed in \nthptrans~and \hnctrans~at 4.5~arcsec (or 2000~au) using high-resolution large-scale ALMA+IRAM-30m mosaics (see Paper I).
Our survey includes OMC-3, 4 South and LDN 1641N in Orion A, NGC 2023 and the Flame Nebula in Orion B covering a wide range of star-formation regimes, cloud morphology, and evolutionary stages.
We summarize our main results as follows:
\begin{enumerate}

    \item Our data sample the \hnctrans~gas properties across our survey. The high resolution (2000~au) allows us to investigate the complex organization of the gas in details, revealing that \hnc~trace diffuse ($N(\text{H}_2)\gtrsim10^{22}$) material showing an irregular and chaotic distribution that extends outside the dense material. The organization of the diffuse gas appears similar across our sample, which covers a wide range of different environments (Sect.~\ref{sec:diffuse_gas});
    
    \item We investigated the kinematics of the diffuse gas through the analysis of the linewidth and of the non-thermal velocity dispersion to evaluate its turbulent regime (Sect.~\ref{sec:kinematics}). Our results demonstrate that the diffuse gas is overall more turbulent than the dense material. While fibers are quiescent structures characterized by subsonic motions, $\sigma_{nt}/c_s$ increases at the edges of the \nthp~contours, although not following a regular spatial distribution. This result is in agreement with previous works suggesting that the transition to coherence occurs at fibers and not core level, with fibers being the first structures formed out of the turbulent cascade (Sect.~\ref{sec:coherence}).
    
    \item The kinematics of the diffuse gas is highly influenced by feedback, which results in the broadening of the linewidth and the increase in turbulence motions (Sect.~\ref{sec:feedback}). We tested this effect only on the Flame Nebula, however, we need to consider that, despite being less severe, feedback affects all our targets. Moreover, the distribution of the turbulent motions appears more chaotic and irregular in the region influenced by feedback.

    \item We investigated the turbulent state of the gas within our sample through the statistical analysis of the diffuse gas centroid velocity increments observed in HNC (Sect.~\ref{sec:velocity_increments}). We also introduced the use of velocity gradient statistics for these purposes (Sect.~\ref{sec:velocity_gradients}) which takes advantage of the high spatial sampling provided by our ALMA observations.

    \item Overall, the gas velocity gradients describe a normal distribution with typically $\nabla V_{lsr}\le5$~\kmspc~as expected for a gas dominated by random motions. In contrast, we identified high-shear regions in all our targets showing extreme gradients $|\nabla V_{lsr}|\ge10$~\kmspc~(Sect.~\ref{sec:velocity_gradients}). 
    These points are concentrated forming elongated features at intermediate column densities ($N(\text{H}_2)\sim10^{22}$~cm$^{-2}$) with typical sizes of $\sim$0.1-0.3 pc, occupying a small surface filling factor $f_s \le20$\% of our maps. These high velocity gradient regions are associated with some fibers (not all), in agreement with the intermittent nature of turbulence where dissipation likely operates as irregular phenomenon in time and space.

    \item Despite their small filling factor, these high shear regions are major contributors ($f_\epsilon > 30-60$\%) to the turbulence dissipation in our targets (Sect.~\ref{sec:fiberorigin}).
    We interpreted these high shear features as regions of high vorticity associated to the location of many of the dense fibers, favouring the interpretation of turbulence to be dissipated before the formation of cores and during the formation of these first dense gas structures.

\end{enumerate}

\section*{Data availability}

The ALMA+IRAM-30m \hnctrans~cubes are available as Data Release 3 (DR3) on the following website: \texttt{https://emerge.univie.ac.at/results/data/}.

\begin{acknowledgements}
      This project has received funding from the European Research Council (ERC) under the European Union’s Horizon 2020 research and innovation programme (Grant agreement No. 851435).
      This paper makes use of the following ALMA data: ADS/JAO.ALMA\#2019.1.00641.S., ADS/JAO.ALMA\#2013.1.00662.S.  ALMA is a partnership of ESO (representing its member states), NSF (USA) and NINS (Japan), together with NRC (Canada), MOST and ASIAA (Taiwan), and KASI (Republic of Korea), in cooperation with the Republic of Chile. The Joint ALMA Observatory is operated by ESO, AUI/NRAO and NAOJ.
      This work is based on IRAM-30m telescope observations carried out under project numbers 032-13, 034-16, 120-20, 060-22, and 133-22. IRAM is supported by INSU/CNRS (France), MPG (Germany), and IGN (Spain). This research has made use of the SIMBAD database, operated at CDS, Strasbourg, France. This research has made use of NASA’s Astrophysics Data System.
      This publication makes use of data products from the Wide-field Infrared Survey Explorer, which is a joint project of the University of California, Los Angeles, and the Jet Propulsion Laboratory/California Institute of Technology, funded by the National Aeronautics and Space Administration.        
\end{acknowledgements}
\bibliographystyle{aa}
\bibliography{Mybibliography} 

@ARTICLE{2012Alves,
       author = {{Alves}, J. and {Bouy}, H.},
        title = "{Orion revisited. I. The massive cluster in front of the Orion nebula cluster}",
      journal = {\aap},
         year = 2012,
        month = nov,
       volume = {547},
          eid = {A97},
        pages = {A97},
          doi = {10.1051/0004-6361/201220119},
archivePrefix = {arXiv},
       eprint = {1209.3787},
 primaryClass = {astro-ph.GA},
       adsurl = {https://ui.adsabs.harvard.edu/abs/2012A&A...547A..97A}
}

@ARTICLE{2010Andre,
       author = {{Andr{\'e}}, Ph. and {Men'shchikov}, A. and {Bontemps}, S. and {K{\"o}nyves}, V. and {Motte}, F. and {Schneider}, N. and {Didelon}, P. and {Minier}, V. and {Saraceno}, P. and {Ward-Thompson}, D. and {di Francesco}, J. and {White}, G. and {Molinari}, S. and {Testi}, L. and {Abergel}, A. and {Griffin}, M. and {Henning}, Th. and {Royer}, P. and {Mer{\'\i}n}, B. and {Vavrek}, R. and {Attard}, M. and {Arzoumanian}, D. and {Wilson}, C.~D. and {Ade}, P. and {Aussel}, H. and {Baluteau}, J. -P. and {Benedettini}, M. and {Bernard}, J. -Ph. and {Blommaert}, J.~A.~D.~L. and {Cambr{\'e}sy}, L. and {Cox}, P. and {di Giorgio}, A. and {Hargrave}, P. and {Hennemann}, M. and {Huang}, M. and {Kirk}, J. and {Krause}, O. and {Launhardt}, R. and {Leeks}, S. and {Le Pennec}, J. and {Li}, J.~Z. and {Martin}, P.~G. and {Maury}, A. and {Olofsson}, G. and {Omont}, A. and {Peretto}, N. and {Pezzuto}, S. and {Prusti}, T. and {Roussel}, H. and {Russeil}, D. and {Sauvage}, M. and {Sibthorpe}, B. and {Sicilia-Aguilar}, A. and {Spinoglio}, L. and {Waelkens}, C. and {Woodcraft}, A. and {Zavagno}, A.},
        title = "{From filamentary clouds to prestellar cores to the stellar IMF: Initial highlights from the Herschel Gould Belt Survey}",
      journal = {\aap},
         year = 2010,
        month = jul,
       volume = {518},
          eid = {L102},
        pages = {L102},
          doi = {10.1051/0004-6361/201014666},
archivePrefix = {arXiv},
       eprint = {1005.2618},
 primaryClass = {astro-ph.GA},
       adsurl = {https://ui.adsabs.harvard.edu/abs/2010A&A...518L.102A}
}

@INPROCEEDINGS{2014Andre,
       author = {{Andr{\'e}}, P. and {Di Francesco}, J. and {Ward-Thompson}, D. and {Inutsuka}, S. -I. and {Pudritz}, R.~E. and {Pineda}, J.~E.},
        title = "{From Filamentary Networks to Dense Cores in Molecular Clouds: Toward a New Paradigm for Star Formation}",
    booktitle = {Protostars and Planets VI},
         year = 2014,
       editor = {{Beuther}, Henrik and {Klessen}, Ralf S. and {Dullemond}, Cornelis P. and {Henning}, Thomas},
        month = jan,
        pages = {27-51},
          doi = {10.2458/azu_uapress_9780816531240-ch002},
archivePrefix = {arXiv},
       eprint = {1312.6232},
 primaryClass = {astro-ph.GA},
       adsurl = {https://ui.adsabs.harvard.edu/abs/2014prpl.conf...27A}
}

@ARTICLE{1984Anselmet,
       author = {{Anselmet}, F. and {Gagne}, Y. and {Hopfinger}, E.~J. and {Antonia}, R.~A.},
        title = "{High-order velocity structure functions in turbulent shear flows}",
      journal = {Journal of Fluid Mechanics},
         year = 1984,
        month = mar,
       volume = {140},
        pages = {63-89},
          doi = {10.1017/S0022112084000513},
       adsurl = {https://ui.adsabs.harvard.edu/abs/1984JFM...140...63A}
}

@ARTICLE{2001Anselmet,
       author = {{Anselmet}, F. and {Antonia}, R.~A. and {Danaila}, L.},
        title = "{Turbulent flows and intermittency in laboratory experiments}",
      journal = {\planss},
         year = 2001,
        month = oct,
       volume = {49},
       number = {12},
        pages = {1177-1191},
          doi = {10.1016/S0032-0633(01)00059-9},
       adsurl = {https://ui.adsabs.harvard.edu/abs/2001P&SS...49.1177A}
}

@INCOLLECTION{2008Bally,
       author = {{Bally}, J.},
        title = "{Overview of the Orion Complex}",
    booktitle = {Handbook of Star Forming Regions, Volume I},
         year = 2008,
       editor = {{Reipurth}, B.},
       volume = {4},
        pages = {459},
          doi = {10.48550/arXiv.0812.0046},
       adsurl = {https://ui.adsabs.harvard.edu/abs/2008hsf1.book..459B}
}

@ARTICLE{1907Barnard,
       author = {{Barnard}, E.~E.},
        title = "{On a nebulous groundwork in the constellation Taurus.}",
      journal = {\apj},
         year = 1907,
        month = apr,
       volume = {25},
        pages = {218-225},
          doi = {10.1086/141434},
       adsurl = {https://ui.adsabs.harvard.edu/abs/1907ApJ....25..218B}
}

@ARTICLE{1998Barraco,
       author = {{Barranco}, Joseph A. and {Goodman}, Alyssa A.},
        title = "{Coherent Dense Cores. I. NH$_{3}$ Observations}",
      journal = {\apj},
         year = 1998,
        month = sep,
       volume = {504},
       number = {1},
        pages = {207-222},
          doi = {10.1086/306044},
       adsurl = {https://ui.adsabs.harvard.edu/abs/1998ApJ...504..207B}
}

@BOOK{1953Batchelor,
       author = {{Batchelor}, G.~K.},
        title = "{The Theory of Homogeneous Turbulence}",
         year = 1953,
       adsurl = {https://ui.adsabs.harvard.edu/abs/1953tht..book.....B}
}

@ARTICLE{2013Beaumont,
       author = {{Beaumont}, Christopher N. and {Offner}, Stella S.~R. and {Shetty}, Rahul and {Glover}, Simon C.~O. and {Goodman}, Alyssa A.},
        title = "{Quantifying Observational Projection Effects Using Molecular Cloud Simulations}",
      journal = {\apj},
         year = 2013,
        month = nov,
       volume = {777},
       number = {2},
          eid = {173},
        pages = {173},
          doi = {10.1088/0004-637X/777/2/173},
archivePrefix = {arXiv},
       eprint = {1310.1929},
 primaryClass = {astro-ph.GA},
       adsurl = {https://ui.adsabs.harvard.edu/abs/2013ApJ...777..173B}
}

@ARTICLE{2007Bergin,
       author = {{Bergin}, Edwin A. and {Tafalla}, Mario},
        title = "{Cold Dark Clouds: The Initial Conditions for Star Formation}",
      journal = {\araa},
         year = 2007,
        month = sep,
       volume = {45},
       number = {1},
        pages = {339-396},
          doi = {10.1146/annurev.astro.45.071206.100404},
archivePrefix = {arXiv},
       eprint = {0705.3765},
 primaryClass = {astro-ph},
       adsurl = {https://ui.adsabs.harvard.edu/abs/2007ARA&A..45..339B}
}

@ARTICLE{2024Bonanomi,
       author = {{Bonanomi}, Francesca and {Hacar}, Alvaro and {Socci}, Andrea and {Petry}, Dirk and {Suri}, S{\"u}meyye},
        title = "{Emergence of high-mass stars in complex fiber networks (EMERGE). II. The need for data combination in ALMA observations}",
      journal = {\aap},
         year = 2024,
        month = aug,
       volume = {688},
          eid = {A30},
        pages = {A30},
          doi = {10.1051/0004-6361/202348920},
archivePrefix = {arXiv},
       eprint = {2405.09290},
 primaryClass = {astro-ph.GA},
       adsurl = {https://ui.adsabs.harvard.edu/abs/2024A&A...688A..30B}
}

@ARTICLE{1994Brown,
       author = {{Brown}, A.~G.~A. and {de Geus}, E.~J. and {de Zeeuw}, P.~T.},
        title = "{The Orion OB1 association. I. Stellar content.}",
      journal = {\aap},
         year = 1994,
        month = sep,
       volume = {289},
        pages = {101-120},
          doi = {10.48550/arXiv.astro-ph/9403051},
archivePrefix = {arXiv},
       eprint = {astro-ph/9403051},
 primaryClass = {astro-ph},
       adsurl = {https://ui.adsabs.harvard.edu/abs/1994A&A...289..101B}
}

@ARTICLE{2002Caselli,
       author = {{Caselli}, Paola and {Benson}, Priscilla J. and {Myers}, Philip C. and {Tafalla}, Mario},
        title = "{Dense Cores in Dark Clouds. XIV. N$_{2}$H$^{+}$ (1-0) Maps of Dense Cloud Cores}",
      journal = {\apj},
         year = 2002,
        month = jun,
       volume = {572},
       number = {1},
        pages = {238-263},
          doi = {10.1086/340195},
archivePrefix = {arXiv},
       eprint = {astro-ph/0202173},
 primaryClass = {astro-ph},
       adsurl = {https://ui.adsabs.harvard.edu/abs/2002ApJ...572..238C}
}

@ARTICLE{Chen2019,
       author = {{Chen}, Huei-Ru Vivien and {Zhang}, Qizhou and {Wright}, M.~C.~H. and {Busquet}, Gemma and {Lin}, Yuxin and {Liu}, Hauyu Baobab and {Olguin}, F.~A. and {Sanhueza}, Patricio and {Nakamura}, Fumitaka and {Palau}, Aina and {Ohashi}, Satoshi and {Tatematsu}, Ken'ichi and {Liao}, Li-Wen},
        title = "{Filamentary Accretion Flows in the Infrared Dark Cloud G14.225-0.506 Revealed by ALMA}",
      journal = {\apj},
         year = 2019,
        month = apr,
       volume = {875},
       number = {1},
          eid = {24},
        pages = {24},
          doi = {10.3847/1538-4357/ab0f3e},
archivePrefix = {arXiv},
       eprint = {1903.04376},
 primaryClass = {astro-ph.GA},
       adsurl = {https://ui.adsabs.harvard.edu/abs/2019ApJ...875...24C}
}

@ARTICLE{2019Chen_GAS,
       author = {{Chen}, Hope How-Huan and {Pineda}, Jaime E. and {Goodman}, Alyssa A. and {Burkert}, Andreas and {Offner}, Stella S.~R. and {Friesen}, Rachel K. and {Myers}, Philip C. and {Alves}, Felipe and {Arce}, H{\'e}ctor G. and {Caselli}, Paola and {Chac{\'o}n-Tanarro}, Ana and {Chen}, Michael Chun-Yuan and {Di Francesco}, James and {Ginsburg}, Adam and {Keown}, Jared and {Kirk}, Helen and {Martin}, Peter G. and {Matzner}, Christopher and {Punanova}, Anna and {Redaelli}, Elena and {Rosolowsky}, Erik and {Scibelli}, Samantha and {Seo}, Youngmin and {Shirley}, Yancy and {Singh}, Ayushi and {GAS Collaboration}},
        title = "{Droplets. I. Pressure-dominated Coherent Structures in L1688 and B18}",
      journal = {\apj},
         year = 2019,
        month = jun,
       volume = {877},
       number = {2},
          eid = {93},
        pages = {93},
          doi = {10.3847/1538-4357/ab1a40},
archivePrefix = {arXiv},
       eprint = {1809.10223},
 primaryClass = {astro-ph.GA},
       adsurl = {https://ui.adsabs.harvard.edu/abs/2019ApJ...877...93C}
}

@ARTICLE{2024Chen_GAS,
       author = {{Chen}, Michael Chun-Yuan and {Di Francesco}, James and {Friesen}, Rachel K. and {Pineda}, Jaime E. and {Caselli}, Paola and {Ginsburg}, Adam and {Kirk}, Helen and {Punanova}, Anna and {The GAS Collaboration}},
        title = "{Filament Accretion and Fragmentation in the Perseus Molecular Cloud}",
      journal = {\apj},
         year = 2024,
        month = dec,
       volume = {977},
       number = {1},
          eid = {135},
        pages = {135},
          doi = {10.3847/1538-4357/ad88e8},
archivePrefix = {arXiv},
       eprint = {2410.16372},
 primaryClass = {astro-ph.GA},
       adsurl = {https://ui.adsabs.harvard.edu/abs/2024ApJ...977..135C}
}

@ARTICLE{2021Choudhury,
       author = {{Choudhury}, Spandan and {Pineda}, Jaime E. and {Caselli}, Paola and {Offner}, Stella S.~R. and {Rosolowsky}, Erik and {Friesen}, Rachel K. and {Redaelli}, Elena and {Chac{\'o}n-Tanarro}, Ana and {Shirley}, Yancy and {Punanova}, Anna and {Kirk}, Helen},
        title = "{Transition from coherent cores to surrounding cloud in L1688}",
      journal = {\aap},
         year = 2021,
        month = apr,
       volume = {648},
          eid = {A114},
        pages = {A114},
          doi = {10.1051/0004-6361/202039897},
archivePrefix = {arXiv},
       eprint = {2102.06459},
 primaryClass = {astro-ph.GA},
       adsurl = {https://ui.adsabs.harvard.edu/abs/2021A&A...648A.114C}
}

@ARTICLE{2017Clarke,
       author = {{Clarke}, S.~D. and {Whitworth}, A.~P. and {Duarte-Cabral}, A. and {Hubber}, D.~A.},
        title = "{Filamentary fragmentation in a turbulent medium}",
      journal = {\mnras},
         year = 2017,
        month = jun,
       volume = {468},
       number = {2},
        pages = {2489-2505},
          doi = {10.1093/mnras/stx637},
archivePrefix = {arXiv},
       eprint = {1703.04473},
 primaryClass = {astro-ph.GA},
       adsurl = {https://ui.adsabs.harvard.edu/abs/2017MNRAS.468.2489C}
}

@ARTICLE{2018Clarke,
       author = {{Clarke}, S.~D. and {Whitworth}, A.~P. and {Spowage}, R.~L. and {Duarte-Cabral}, A. and {Suri}, S.~T. and {Jaffa}, S.~E. and {Walch}, S. and {Clark}, P.~C.},
        title = "{Synthetic C$^{18}$O observations of fibrous filaments: the problems of mapping from PPV to PPP}",
      journal = {\mnras},
         year = 2018,
        month = sep,
       volume = {479},
       number = {2},
        pages = {1722-1746},
          doi = {10.1093/mnras/sty1675},
archivePrefix = {arXiv},
       eprint = {1806.08564},
 primaryClass = {astro-ph.GA},
       adsurl = {https://ui.adsabs.harvard.edu/abs/2018MNRAS.479.1722C}
}

@ARTICLE{2004Elmegreen,
       author = {{Elmegreen}, Bruce G. and {Scalo}, John},
        title = "{Interstellar Turbulence I: Observations and Processes}",
      journal = {\araa},
         year = 2004,
        month = sep,
       volume = {42},
       number = {1},
        pages = {211-273},
          doi = {10.1146/annurev.astro.41.011802.094859},
archivePrefix = {arXiv},
       eprint = {astro-ph/0404451},
 primaryClass = {astro-ph},
       adsurl = {https://ui.adsabs.harvard.edu/abs/2004ARA&A..42..211E}
}

@ARTICLE{2009Evans,
       author = {{Evans}, II, Neal J. and {Dunham}, Michael M. and {J{\o}rgensen}, Jes K. and {Enoch}, Melissa L. and {Mer{\'\i}n}, Bruno and {van Dishoeck}, Ewine F. and {Alcal{\'a}}, Juan M. and {Myers}, Philip C. and {Stapelfeldt}, Karl R. and {Huard}, Tracy L. and {Allen}, Lori E. and {Harvey}, Paul M. and {van Kempen}, Tim and {Blake}, Geoffrey A. and {Koerner}, David W. and {Mundy}, Lee G. and {Padgett}, Deborah L. and {Sargent}, Anneila I.},
        title = "{The Spitzer c2d Legacy Results: Star-Formation Rates and Efficiencies; Evolution and Lifetimes}",
      journal = {\apjs},
         year = 2009,
        month = apr,
       volume = {181},
       number = {2},
        pages = {321-350},
          doi = {10.1088/0067-0049/181/2/321},
archivePrefix = {arXiv},
       eprint = {0811.1059},
 primaryClass = {astro-ph},
       adsurl = {https://ui.adsabs.harvard.edu/abs/2009ApJS..181..321E}
}

@ARTICLE{1990Falgarone,
       author = {{Falgarone}, E. and {Phillips}, T.~G.},
        title = "{A Signature of the Intermittency of Interstellar Turbulence: The Wings of Molecular Line Profiles}",
      journal = {\apj},
         year = 1990,
        month = aug,
       volume = {359},
        pages = {344},
          doi = {10.1086/169068},
       adsurl = {https://ui.adsabs.harvard.edu/abs/1990ApJ...359..344F}
}

@ARTICLE{1992Falgarone,
       author = {{Falgarone}, E. and {Puget}, J. -L. and {Perault}, M.},
        title = "{The small-scale density and velocity structure of quiescent molecular clouds.}",
      journal = {\aap},
         year = 1992,
        month = apr,
       volume = {257},
        pages = {715-730},
       adsurl = {https://ui.adsabs.harvard.edu/abs/1992A&A...257..715F}
}

@ARTICLE{2019Feddersen,
       author = {{Feddersen}, Jesse R. and {Arce}, H{\'e}ctor G. and {Kong}, Shuo and {Ossenkopf-Okada}, Volker and {Carpenter}, John M.},
        title = "{The CARMA-NRO Orion Survey: Statistical Signatures of Feedback in the Orion A Molecular Cloud}",
      journal = {\apj},
         year = 2019,
        month = apr,
       volume = {875},
       number = {2},
          eid = {162},
        pages = {162},
          doi = {10.3847/1538-4357/ab0e7d},
archivePrefix = {arXiv},
       eprint = {1903.05104},
 primaryClass = {astro-ph.GA},
       adsurl = {https://ui.adsabs.harvard.edu/abs/2019ApJ...875..162F}
}

@ARTICLE{2017Friesen,
       author = {{Friesen}, Rachel K. and {Pineda}, Jaime E. and {co-PIs} and {Rosolowsky}, Erik and {Alves}, Felipe and {Chac{\'o}n-Tanarro}, Ana and {How-Huan Chen}, Hope and {Chun-Yuan Chen}, Michael and {Di Francesco}, James and {Keown}, Jared and {Kirk}, Helen and {Punanova}, Anna and {Seo}, Youngmin and {Shirley}, Yancy and {Ginsburg}, Adam and {Hall}, Christine and {Offner}, Stella S.~R. and {Singh}, Ayushi and {Arce}, H{\'e}ctor G. and {Caselli}, Paola and {Goodman}, Alyssa A. and {Martin}, Peter G. and {Matzner}, Christopher and {Myers}, Philip C. and {Redaelli}, Elena and {GAS Collaboration}},
        title = "{The Green Bank Ammonia Survey: First Results of NH$_{3}$ Mapping of the Gould Belt}",
      journal = {\apj},
         year = 2017,
        month = jul,
       volume = {843},
       number = {1},
          eid = {63},
        pages = {63},
          doi = {10.3847/1538-4357/aa6d58},
archivePrefix = {arXiv},
       eprint = {1704.06318},
 primaryClass = {astro-ph.GA},
       adsurl = {https://ui.adsabs.harvard.edu/abs/2017ApJ...843...63F}
}

@ARTICLE{2016Furlan,
       author = {{Furlan}, E. and {Fischer}, W.~J. and {Ali}, B. and {Stutz}, A.~M. and {Stanke}, T. and {Tobin}, J.~J. and {Megeath}, S.~T. and {Osorio}, M. and {Hartmann}, L. and {Calvet}, N. and {Poteet}, C.~A. and {Booker}, J. and {Manoj}, P. and {Watson}, D.~M. and {Allen}, L.},
        title = "{The Herschel Orion Protostar Survey: Spectral Energy Distributions and Fits Using a Grid of Protostellar Models}",
      journal = {\apjs},
         year = 2016,
        month = may,
       volume = {224},
       number = {1},
          eid = {5},
        pages = {5},
          doi = {10.3847/0067-0049/224/1/5},
archivePrefix = {arXiv},
       eprint = {1602.07314},
 primaryClass = {astro-ph.SR},
       adsurl = {https://ui.adsabs.harvard.edu/abs/2016ApJS..224....5F}
}

@ARTICLE{Gaches2025,
       author = {{Gaches}, Brandt A.~L.},
        title = "{An analytic formalism to describe the N$_{eff}$(H)‑n$_{H}$ relationship in molecular clouds}",
      journal = {\aap},
         year = 2025,
        month = aug,
       volume = {700},
          eid = {L16},
        pages = {L16},
          doi = {10.1051/0004-6361/202555937},
archivePrefix = {arXiv},
       eprint = {2507.16931},
 primaryClass = {astro-ph.GA},
       adsurl = {https://ui.adsabs.harvard.edu/abs/2025A&A...700L..16G}
}

@software{2013GildasTeam,
       author = {{Gildas Team}},
        title = "{GILDAS: Grenoble Image and Line Data Analysis Software}",
 howpublished = {Astrophysics Source Code Library, record ascl:1305.010},
         year = 2013,
        month = may,
          eid = {ascl:1305.010},
       adsurl = {https://ui.adsabs.harvard.edu/abs/2013ascl.soft05010G}
}

@ARTICLE{1993Goodman,
       author = {{Goodman}, A.~A. and {Benson}, P.~J. and {Fuller}, G.~A. and {Myers}, P.~C.},
        title = "{Dense Cores in Dark Clouds. VIII. Velocity Gradients}",
      journal = {\apj},
         year = 1993,
        month = apr,
       volume = {406},
        pages = {528},
          doi = {10.1086/172465},
       adsurl = {https://ui.adsabs.harvard.edu/abs/1993ApJ...406..528G}
}

@ARTICLE{1998Goodman,
       author = {{Goodman}, Alyssa A. and {Barranco}, Joseph A. and {Wilner}, David J. and {Heyer}, Mark H.},
        title = "{Coherence in Dense Cores. II. The Transition to Coherence}",
      journal = {\apj},
         year = 1998,
        month = sep,
       volume = {504},
       number = {1},
        pages = {223-246},
          doi = {10.1086/306045},
       adsurl = {https://ui.adsabs.harvard.edu/abs/1998ApJ...504..223G}
}

@ARTICLE{2021Gong,
       author = {{Gong}, Y. and {Belloche}, A. and {Du}, F.~J. and {Menten}, K.~M. and {Henkel}, C. and {Li}, G.~X. and {Wyrowski}, F. and {Mao}, R.~Q.},
        title = "{Physical and chemical structure of the Serpens filament: Fast formation and gravity-driven accretion}",
      journal = {\aap},
         year = 2021,
        month = feb,
       volume = {646},
          eid = {A170},
        pages = {A170},
          doi = {10.1051/0004-6361/202039465},
archivePrefix = {arXiv},
       eprint = {2012.11924},
 primaryClass = {astro-ph.GA},
       adsurl = {https://ui.adsabs.harvard.edu/abs/2021A&A...646A.170G}
}

@ARTICLE{2011Hacar,
       author = {{Hacar}, A. and {Tafalla}, M.},
        title = "{Dense core formation by fragmentation of velocity-coherent filaments in L1517}",
      journal = {\aap},
         year = 2011,
        month = sep,
       volume = {533},
          eid = {A34},
        pages = {A34},
          doi = {10.1051/0004-6361/201117039},
archivePrefix = {arXiv},
       eprint = {1107.0971},
 primaryClass = {astro-ph.GA},
       adsurl = {https://ui.adsabs.harvard.edu/abs/2011A&A...533A..34H}
}

@ARTICLE{2013Hacar,
       author = {{Hacar}, A. and {Tafalla}, M. and {Kauffmann}, J. and {Kov{\'a}cs}, A.},
        title = "{Cores, filaments, and bundles: hierarchical core formation in the L1495/B213 Taurus region}",
      journal = {\aap},
         year = 2013,
        month = jun,
       volume = {554},
          eid = {A55},
        pages = {A55},
          doi = {10.1051/0004-6361/201220090},
archivePrefix = {arXiv},
       eprint = {1303.2118},
 primaryClass = {astro-ph.GA},
       adsurl = {https://ui.adsabs.harvard.edu/abs/2013A&A...554A..55H}
}

@ARTICLE{2016Hacar,
       author = {{Hacar}, A. and {Kainulainen}, J. and {Tafalla}, M. and {Beuther}, H. and {Alves}, J.},
        title = "{The Musca cloud: A 6 pc-long velocity-coherent, sonic filament}",
      journal = {\aap},
         year = 2016,
        month = mar,
       volume = {587},
          eid = {A97},
        pages = {A97},
          doi = {10.1051/0004-6361/201526015},
archivePrefix = {arXiv},
       eprint = {1511.06370},
 primaryClass = {astro-ph.GA},
       adsurl = {https://ui.adsabs.harvard.edu/abs/2016A&A...587A..97H}
}

@ARTICLE{2016Hacarb,
       author = {{Hacar}, A. and {Alves}, J. and {Burkert}, A. and {Goldsmith}, P.},
        title = "{Opacity broadening and interpretation of suprathermal CO linewidths: Macroscopic turbulence and tangled molecular clouds}",
      journal = {\aap},
         year = 2016,
        month = jun,
       volume = {591},
          eid = {A104},
        pages = {A104},
          doi = {10.1051/0004-6361/201527319},
archivePrefix = {arXiv},
       eprint = {1603.08521},
 primaryClass = {astro-ph.GA},
       adsurl = {https://ui.adsabs.harvard.edu/abs/2016A&A...591A.104H}
}

@ARTICLE{2017Hacar,
       author = {{Hacar}, A. and {Tafalla}, M. and {Alves}, J.},
        title = "{Fibers in the NGC 1333 proto-cluster}",
      journal = {\aap},
         year = 2017,
        month = oct,
       volume = {606},
          eid = {A123},
        pages = {A123},
          doi = {10.1051/0004-6361/201630348},
archivePrefix = {arXiv},
       eprint = {1703.07029},
 primaryClass = {astro-ph.GA},
       adsurl = {https://ui.adsabs.harvard.edu/abs/2017A&A...606A.123H}
}

@ARTICLE{2018Hacar,
       author = {{Hacar}, A. and {Tafalla}, M. and {Forbrich}, J. and {Alves}, J. and {Meingast}, S. and {Grossschedl}, J. and {Teixeira}, P.~S.},
        title = "{An ALMA study of the Orion Integral Filament. I. Evidence for narrow fibers in a massive cloud}",
      journal = {\aap},
         year = 2018,
        month = mar,
       volume = {610},
          eid = {A77},
        pages = {A77},
          doi = {10.1051/0004-6361/201731894},
archivePrefix = {arXiv},
       eprint = {1801.01500},
 primaryClass = {astro-ph.GA},
       adsurl = {https://ui.adsabs.harvard.edu/abs/2018A&A...610A..77H}
}

@ARTICLE{2020Hacar,
       author = {{Hacar}, A. and {Bosman}, A.~D. and {van Dishoeck}, E.~F.},
        title = "{HCN-to-HNC intensity ratio: a new chemical thermometer for the molecular ISM}",
      journal = {\aap},
         year = 2020,
        month = mar,
       volume = {635},
          eid = {A4},
        pages = {A4},
          doi = {10.1051/0004-6361/201936516},
archivePrefix = {arXiv},
       eprint = {1910.13754},
 primaryClass = {astro-ph.GA},
       adsurl = {https://ui.adsabs.harvard.edu/abs/2020A&A...635A...4H}
}

@INPROCEEDINGS{2023Hacar,
       author = {{Hacar}, A. and {Clark}, S.~E. and {Heitsch}, F. and {Kainulainen}, J. and {Panopoulou}, G.~V. and {Seifried}, D. and {Smith}, R.},
        title = "{Initial Conditions for Star Formation: a Physical Description of the Filamentary ISM}",
    booktitle = {Protostars and Planets VII},
         year = 2023,
       editor = {{Inutsuka}, S. and {Aikawa}, Y. and {Muto}, T. and {Tomida}, K. and {Tamura}, M.},
       series = {Astronomical Society of the Pacific Conference Series},
       volume = {534},
        month = jul,
        pages = {153},
          doi = {10.48550/arXiv.2203.09562},
archivePrefix = {arXiv},
       eprint = {2203.09562},
 primaryClass = {astro-ph.GA},
       adsurl = {https://ui.adsabs.harvard.edu/abs/2023ASPC..534..153H}
}

@ARTICLE{2024Hacara,
       author = {{Hacar}, A. and {Socci}, A. and {Bonanomi}, F. and {Petry}, D. and {Tafalla}, M. and {Harsono}, D. and {Forbrich}, J. and {Alves}, J. and {Grossschedl}, J. and {Goicoechea}, J.~R. and {Pety}, J. and {Burkert}, A. and {Li}, G.~X.},
        title = "{Emergence of high-mass stars in complex fiber networks (EMERGE). I. Early ALMA Survey: Observations and massive data reduction}",
      journal = {\aap},
         year = 2024,
        month = jul,
       volume = {687},
          eid = {A140},
        pages = {A140},
          doi = {10.1051/0004-6361/202348565},
archivePrefix = {arXiv},
       eprint = {2403.08091},
 primaryClass = {astro-ph.GA},
       adsurl = {https://ui.adsabs.harvard.edu/abs/2024A&A...687A.140H}
}

@ARTICLE{2001Heitsch,
       author = {{Heitsch}, Fabian and {Mac Low}, Mordecai-Mark and {Klessen}, Ralf S.},
        title = "{Gravitational Collapse in Turbulent Molecular Clouds. II. Magnetohydrodynamical Turbulence}",
      journal = {\apj},
         year = 2001,
        month = jan,
       volume = {547},
       number = {1},
        pages = {280-291},
          doi = {10.1086/318335},
archivePrefix = {arXiv},
       eprint = {astro-ph/0009227},
 primaryClass = {astro-ph},
       adsurl = {https://ui.adsabs.harvard.edu/abs/2001ApJ...547..280H}
}

@ARTICLE{2014Henshaw,
       author = {{Henshaw}, J.~D. and {Caselli}, P. and {Fontani}, F. and {Jim{\'e}nez-Serra}, I. and {Tan}, J.~C.},
        title = "{The dynamical properties of dense filaments in the infrared dark cloud G035.39-00.33}",
      journal = {\mnras},
         year = 2014,
        month = may,
       volume = {440},
       number = {3},
        pages = {2860-2881},
          doi = {10.1093/mnras/stu446},
archivePrefix = {arXiv},
       eprint = {1403.1444},
 primaryClass = {astro-ph.SR},
       adsurl = {https://ui.adsabs.harvard.edu/abs/2014MNRAS.440.2860H}
}

@ARTICLE{2013Hennebelle,
       author = {{Hennebelle}, Patrick},
        title = "{On the origin of non-self-gravitating filaments in the ISM}",
      journal = {\aap},
         year = 2013,
        month = aug,
       volume = {556},
          eid = {A153},
        pages = {A153},
          doi = {10.1051/0004-6361/201321292},
archivePrefix = {arXiv},
       eprint = {1306.5452},
 primaryClass = {astro-ph.GA},
       adsurl = {https://ui.adsabs.harvard.edu/abs/2013A&A...556A.153H}
}

@INPROCEEDINGS{2007HilyBlant_paperI,
       author = {{Hily-Blant}, P. and {Pety}, J. and {Falgarone}, E.},
        title = "{Small-Scale Dissipative Structures of Diffuse ISM Turbulence: I -- CO Diagnostics}",
    booktitle = {SINS - Small Ionized and Neutral Structures in the Diffuse Interstellar Medium},
         year = 2007,
       editor = {{Haverkorn}, M. and {Goss}, W.~M.},
       series = {Astronomical Society of the Pacific Conference Series},
       volume = {365},
        month = jul,
        pages = {184},
          doi = {10.48550/arXiv.astro-ph/0701326},
archivePrefix = {arXiv},
       eprint = {astro-ph/0701326},
 primaryClass = {astro-ph},
       adsurl = {https://ui.adsabs.harvard.edu/abs/2007ASPC..365..184H}
}

@ARTICLE{2007Hily-Blant,
       author = {{Hily-Blant}, P. and {Falgarone}, E.},
        title = "{Dissipative structures of diffuse molecular gas. II. The translucent environment of a dense core}",
      journal = {\aap},
         year = 2007,
        month = jul,
       volume = {469},
       number = {1},
        pages = {173-187},
          doi = {10.1051/0004-6361:20054565},
       adsurl = {https://ui.adsabs.harvard.edu/abs/2007A&A...469..173H}
}

@ARTICLE{2008Hily-Blant,
       author = {{Hily-Blant}, P. and {Falgarone}, E. and {Pety}, J.},
        title = "{Dissipative structures of diffuse molecular gas. III. Small-scale intermittency of intense velocity-shears}",
      journal = {\aap},
         year = 2008,
        month = apr,
       volume = {481},
       number = {2},
        pages = {367-380},
          doi = {10.1051/0004-6361:20078423},
archivePrefix = {arXiv},
       eprint = {0802.0758},
 primaryClass = {astro-ph},
       adsurl = {https://ui.adsabs.harvard.edu/abs/2008A&A...481..367H}
}

@ARTICLE{2009Hilyblant,
       author = {{Hily-Blant}, P. and {Falgarone}, E.},
        title = "{Intermittency of interstellar turbulence: parsec-scale coherent structure of intense, velocity shear}",
      journal = {\aap},
         year = 2009,
        month = jun,
       volume = {500},
       number = {2},
        pages = {L29-L32},
          doi = {10.1051/0004-6361/200912296},
archivePrefix = {arXiv},
       eprint = {0905.0368},
 primaryClass = {astro-ph.GA},
       adsurl = {https://ui.adsabs.harvard.edu/abs/2009A&A...500L..29H}
}

@ARTICLE{2017Kauffmann,
       author = {{Kauffmann}, Jens and {Goldsmith}, Paul F. and {Melnick}, Gary and {Tolls}, Volker and {Guzman}, Andres and {Menten}, Karl M.},
        title = "{Molecular Line Emission as a Tool for Galaxy Observations (LEGO). I. HCN as a tracer of moderate gas densities in molecular clouds and galaxies}",
      journal = {\aap},
         year = 2017,
        month = sep,
       volume = {605},
          eid = {L5},
        pages = {L5},
          doi = {10.1051/0004-6361/201731123},
archivePrefix = {arXiv},
       eprint = {1707.05352},
 primaryClass = {astro-ph.GA},
       adsurl = {https://ui.adsabs.harvard.edu/abs/2017A&A...605L...5K}
}

@ARTICLE{2005Klessen,
       author = {{Klessen}, Ralf S. and {Ballesteros-Paredes}, Javier and {V{\'a}zquez-Semadeni}, Enrique and {Dur{\'a}n-Rojas}, Carolina},
        title = "{Quiescent and Coherent Cores from Gravoturbulent Fragmentation}",
      journal = {\apj},
         year = 2005,
        month = feb,
       volume = {620},
       number = {2},
        pages = {786-794},
          doi = {10.1086/427255},
archivePrefix = {arXiv},
       eprint = {astro-ph/0306055},
 primaryClass = {astro-ph},
       adsurl = {https://ui.adsabs.harvard.edu/abs/2005ApJ...620..786K}
}

@ARTICLE{1941Kolmogorov,
       author = {{Kolmogorov}, Andrey Nikolaevich},
        title = "{Dissipation of Energy in Locally Isotropic Turbulence}",
      journal = {Akademiia Nauk SSSR Doklady},
         year = 1941,
        month = apr,
       volume = {32},
        pages = {16},
       adsurl = {https://ui.adsabs.harvard.edu/abs/1941DoSSR..32...16K}
}

@ARTICLE{1962Kolmogorov,
       author = {{Kolmogorov}, A.~N.},
        title = "{A refinement of previous hypotheses concerning the local structure of turbulence in a viscous incompressible fluid at high Reynolds number}",
      journal = {Journal of Fluid Mechanics},
         year = 1962,
        month = jan,
       volume = {13},
        pages = {82-85},
          doi = {10.1017/S0022112062000518},
       adsurl = {https://ui.adsabs.harvard.edu/abs/1962JFM....13...82K}
}

@BOOK{1959Landau,
       author = {{Landau}, Lev Davidovich and {Lifshitz}, E.~M.},
        title = "{Fluid mechanics}",
         year = 1959,
       adsurl = {https://ui.adsabs.harvard.edu/abs/1959flme.book.....L}
}

@ARTICLE{1981Larson,
       author = {{Larson}, R.~B.},
        title = "{Turbulence and star formation in molecular clouds.}",
      journal = {\mnras},
         year = 1981,
        month = mar,
       volume = {194},
        pages = {809-826},
          doi = {10.1093/mnras/194.4.809},
       adsurl = {https://ui.adsabs.harvard.edu/abs/1981MNRAS.194..809L}
}

@ARTICLE{1985Larson,
       author = {{Larson}, R.~B.},
        title = "{Cloud fragmentation and stellar masses.}",
      journal = {\mnras},
         year = 1985,
        month = jun,
       volume = {214},
        pages = {379-398},
          doi = {10.1093/mnras/214.3.379},
       adsurl = {https://ui.adsabs.harvard.edu/abs/1985MNRAS.214..379L}
}

@ARTICLE{2014Lee,
       author = {{Lee}, Katherine I. and {Fern{\'a}ndez-L{\'o}pez}, Manuel and {Storm}, Shaye and {Looney}, Leslie W. and {Mundy}, Lee G. and {Segura-Cox}, Dominique and {Teuben}, Peter and {Rosolowsky}, Erik and {Arce}, H{\'e}ctor G. and {Ostriker}, Eve C. and {Shirley}, Yancy L. and {Kwon}, Woojin and {Kauffmann}, Jens and {Tobin}, John J. and {Plunkett}, Adele L. and {Pound}, Marc W. and {Salter}, Demerese M. and {Volgenau}, N.~H. and {Chen}, Che-Yu and {Tassis}, Konstantinos and {Isella}, Andrea and {Crutcher}, Richard M. and {Gammie}, Charles F. and {Testi}, Leonardo},
        title = "{CARMA Large Area Star Formation Survey: Structure and Kinematics of Dense Gas in Serpens Main}",
      journal = {\apj},
         year = 2014,
        month = dec,
       volume = {797},
       number = {2},
          eid = {76},
        pages = {76},
          doi = {10.1088/0004-637X/797/2/76},
archivePrefix = {arXiv},
       eprint = {1410.3514},
 primaryClass = {astro-ph.GA},
       adsurl = {https://ui.adsabs.harvard.edu/abs/2014ApJ...797...76L}
}

@ARTICLE{1996Lis,
       author = {{Lis}, D.~C. and {Pety}, J. and {Phillips}, T.~G. and {Falgarone}, E.},
        title = "{Statistical Properties of Line Centroid Velocities and Centroid Velocity Increments in Compressible Turbulence}",
      journal = {\apj},
         year = 1996,
        month = jun,
       volume = {463},
        pages = {623},
          doi = {10.1086/177276},
       adsurl = {https://ui.adsabs.harvard.edu/abs/1996ApJ...463..623L}
}

@ARTICLE{1998Lis,
       author = {{Lis}, D.~C. and {Keene}, Jocelyn and {Li}, Y. and {Phillips}, T.~G. and {Pety}, J.},
        title = "{Statistical Properties of Line Centroid Velocity Increments in the {\ensuremath{\rho}} Ophiuchi Cloud}",
      journal = {\apj},
         year = 1998,
        month = sep,
       volume = {504},
       number = {2},
        pages = {889-899},
          doi = {10.1086/306096},
       adsurl = {https://ui.adsabs.harvard.edu/abs/1998ApJ...504..889L}
}

@ARTICLE{2014Lombardi,
       author = {{Lombardi}, Marco and {Bouy}, Herv{\'e} and {Alves}, Jo{\~a}o and {Lada}, Charles J.},
        title = "{Herschel-Planck dust optical-depth and column-density maps. I. Method description and results for Orion}",
      journal = {\aap},
         year = 2014,
        month = jun,
       volume = {566},
          eid = {A45},
        pages = {A45},
          doi = {10.1051/0004-6361/201323293},
archivePrefix = {arXiv},
       eprint = {1404.0032},
 primaryClass = {astro-ph.SR},
       adsurl = {https://ui.adsabs.harvard.edu/abs/2014A&A...566A..45L}
}

@ARTICLE{2023Martinez,
       author = {{Martinez}, N.~C. and {Paron}, S.},
        title = "{An analysis of the isomers HCN and HNC in the evolution of high-mass star-forming regions}",
      journal = {Boletin de la Asociacion Argentina de Astronomia La Plata Argentina},
         year = 2023,
        month = aug,
       volume = {64},
        pages = {112-114},
          doi = {10.48550/arXiv.2305.05073},
archivePrefix = {arXiv},
       eprint = {2305.05073},
 primaryClass = {astro-ph.GA},
       adsurl = {https://ui.adsabs.harvard.edu/abs/2023BAAA...64..112M}
}

@ARTICLE{2012Megeath,
       author = {{Megeath}, S.~T. and {Gutermuth}, R. and {Muzerolle}, J. and {Kryukova}, E. and {Flaherty}, K. and {Hora}, J.~L. and {Allen}, L.~E. and {Hartmann}, L. and {Myers}, P.~C. and {Pipher}, J.~L. and {Stauffer}, J. and {Young}, E.~T. and {Fazio}, G.~G.},
        title = "{The Spitzer Space Telescope Survey of the Orion A and B Molecular Clouds. I. A Census of Dusty Young Stellar Objects and a Study of Their Mid-infrared Variability}",
      journal = {\aj},
         year = 2012,
        month = dec,
       volume = {144},
       number = {6},
          eid = {192},
        pages = {192},
          doi = {10.1088/0004-6256/144/6/192},
archivePrefix = {arXiv},
       eprint = {1209.3826},
 primaryClass = {astro-ph.GA},
       adsurl = {https://ui.adsabs.harvard.edu/abs/2012AJ....144..192M}
}

@ARTICLE{2016Megeath,
       author = {{Megeath}, S.~T. and {Gutermuth}, R. and {Muzerolle}, J. and {Kryukova}, E. and {Hora}, J.~L. and {Allen}, L.~E. and {Flaherty}, K. and {Hartmann}, L. and {Myers}, P.~C. and {Pipher}, J.~L. and {Stauffer}, J. and {Young}, E.~T. and {Fazio}, G.~G.},
        title = "{The Spitzer Space Telescope Survey of the Orion A and B Molecular Clouds. II. The Spatial Distribution and Demographics of Dusty Young Stellar Objects}",
      journal = {\aj},
         year = 2016,
        month = jan,
       volume = {151},
       number = {1},
          eid = {5},
        pages = {5},
          doi = {10.3847/0004-6256/151/1/5},
archivePrefix = {arXiv},
       eprint = {1511.01202},
 primaryClass = {astro-ph.GA},
       adsurl = {https://ui.adsabs.harvard.edu/abs/2016AJ....151....5M}
}

@ARTICLE{2007Menten,
       author = {{Menten}, K.~M. and {Reid}, M.~J. and {Forbrich}, J. and {Brunthaler}, A.},
        title = "{The distance to the Orion Nebula}",
      journal = {\aap},
         year = 2007,
        month = nov,
       volume = {474},
       number = {2},
        pages = {515-520},
          doi = {10.1051/0004-6361:20078247},
archivePrefix = {arXiv},
       eprint = {0709.0485},
 primaryClass = {astro-ph},
       adsurl = {https://ui.adsabs.harvard.edu/abs/2007A&A...474..515M}
}

@ARTICLE{1995Miesch,
       author = {{Miesch}, Mark S. and {Scalo}, John M.},
        title = "{Exponential Tails in the Centroid Velocity Distributions of Star-Forming Regions}",
      journal = {\apjl},
         year = 1995,
        month = sep,
       volume = {450},
        pages = {L27},
          doi = {10.1086/309661},
archivePrefix = {arXiv},
       eprint = {astro-ph/9412042},
 primaryClass = {astro-ph},
       adsurl = {https://ui.adsabs.harvard.edu/abs/1995ApJ...450L..27M}
}

@ARTICLE{1999Miesch,
       author = {{Miesch}, Mark S. and {Scalo}, John and {Bally}, John},
        title = "{Velocity Field Statistics in Star-forming Regions. I. Centroid Velocity Observations}",
      journal = {\apj},
         year = 1999,
        month = oct,
       volume = {524},
       number = {2},
        pages = {895-922},
          doi = {10.1086/307824},
archivePrefix = {arXiv},
       eprint = {astro-ph/9810427},
 primaryClass = {astro-ph},
       adsurl = {https://ui.adsabs.harvard.edu/abs/1999ApJ...524..895M}
}

@ARTICLE{2003Miville-Deschenes,
       author = {{Miville-Desch{\^e}nes}, M. -A. and {Levrier}, F. and {Falgarone}, E.},
        title = "{On the Use of Fractional Brownian Motion Simulations to Determine the Three-dimensional Statistical Properties of Interstellar Gas}",
      journal = {\apj},
         year = 2003,
        month = aug,
       volume = {593},
       number = {2},
        pages = {831-847},
          doi = {10.1086/376603},
archivePrefix = {arXiv},
       eprint = {astro-ph/0304539},
 primaryClass = {astro-ph},
       adsurl = {https://ui.adsabs.harvard.edu/abs/2003ApJ...593..831M}
}

@ARTICLE{1987Miyama,
       author = {{Miyama}, S.~M. and {Narita}, S. and {Hayashi}, C.},
        title = "{Fragmentation of Isothermal Sheet-Like Clouds. I ---Solutions of Linear and Second-Order Perturbation Equations---}",
      journal = {Progress of Theoretical Physics},
         year = 1987,
        month = nov,
       volume = {78},
       number = {5},
        pages = {1051-1064},
          doi = {10.1143/PTP.78.1051},
       adsurl = {https://ui.adsabs.harvard.edu/abs/1987PThPh..78.1051M}
}

@ARTICLE{1987Miyamab,
       author = {{Miyama}, S.~M. and {Narita}, S. and {Hayashi}, C.},
        title = "{Fragmentation of Isothermal Sheet-Like Clouds. II ---Full Nonlinear Numerical Simulations---}",
      journal = {Progress of Theoretical Physics},
         year = 1987,
        month = dec,
       volume = {78},
       number = {6},
        pages = {1273-1287},
          doi = {10.1143/PTP.78.1273},
       adsurl = {https://ui.adsabs.harvard.edu/abs/1987PThPh..78.1273M}
}

@ARTICLE{1998Nagai,
       author = {{Nagai}, Tomoya and {Inutsuka}, Shu-ichiro and {Miyama}, Shoken M.},
        title = "{An Origin of Filamentary Structure in Molecular Clouds}",
      journal = {\apj},
         year = 1998,
        month = oct,
       volume = {506},
       number = {1},
        pages = {306-322},
          doi = {10.1086/306249},
       adsurl = {https://ui.adsabs.harvard.edu/abs/1998ApJ...506..306N}
}

@ARTICLE{2024Nanase,
       author = {{Harada}, Nanase and {Saito}, Toshiki and {Nishimura}, Yuri and {Watanabe}, Yoshimasa and {Sakamoto}, Kazushi},
        title = "{A Temperature or Far-ultraviolet Tracer? The HNC/HCN Ratio in M83 on the Scale of Giant Molecular Clouds}",
      journal = {\apj},
         year = 2024,
        month = jul,
       volume = {969},
       number = {2},
          eid = {82},
        pages = {82},
          doi = {10.3847/1538-4357/ad4639},
archivePrefix = {arXiv},
       eprint = {2405.09029},
 primaryClass = {astro-ph.GA},
       adsurl = {https://ui.adsabs.harvard.edu/abs/2024ApJ...969...82H}
}

@ARTICLE{2002Ossenkopf,
       author = {{Ossenkopf}, V. and {Mac Low}, M. -M.},
        title = "{Turbulent velocity structure in molecular clouds}",
      journal = {\aap},
         year = 2002,
        month = jul,
       volume = {390},
        pages = {307-326},
          doi = {10.1051/0004-6361:20020629},
       adsurl = {https://ui.adsabs.harvard.edu/abs/2002A&A...390..307O}
}

@ARTICLE{2017Pabst,
       author = {{Pabst}, C.~H.~M. and {Goicoechea}, J.~R. and {Teyssier}, D. and {Bern{\'e}}, O. and {Ochsendorf}, B.~B. and {Wolfire}, M.~G. and {Higgins}, R.~D. and {Riquelme}, D. and {Risacher}, C. and {Pety}, J. and {Le Petit}, F. and {Roueff}, E. and {Bron}, E. and {Tielens}, A.~G.~G.~M.},
        title = "{[C II] emission from L1630 in the Orion B molecular cloud}",
      journal = {\aap},
         year = 2017,
        month = oct,
       volume = {606},
          eid = {A29},
        pages = {A29},
          doi = {10.1051/0004-6361/201730881},
archivePrefix = {arXiv},
       eprint = {1707.05976},
 primaryClass = {astro-ph.GA},
       adsurl = {https://ui.adsabs.harvard.edu/abs/2017A&A...606A..29P}
}

@ARTICLE{2020Pabst,
       author = {{Pabst}, C.~H.~M. and {Goicoechea}, J.~R. and {Teyssier}, D. and {Bern{\'e}}, O. and {Higgins}, R.~D. and {Chambers}, E.~T. and {Kabanovic}, S. and {G{\"u}sten}, R. and {Stutzki}, J. and {Tielens}, A.~G.~G.~M.},
        title = "{Expanding bubbles in Orion A: [C II] observations of M 42, M 43, and NGC 1977}",
      journal = {\aap},
         year = 2020,
        month = jul,
       volume = {639},
          eid = {A2},
        pages = {A2},
          doi = {10.1051/0004-6361/202037560},
archivePrefix = {arXiv},
       eprint = {2005.03917},
 primaryClass = {astro-ph.GA},
       adsurl = {https://ui.adsabs.harvard.edu/abs/2020A&A...639A...2P}
}

@ARTICLE{1999Padoan,
       author = {{Padoan}, Paolo and {Nordlund}, {\r{A}}ke},
        title = "{A Super-Alfv{\'e}nic Model of Dark Clouds}",
      journal = {\apj},
         year = 1999,
        month = nov,
       volume = {526},
       number = {1},
        pages = {279-294},
          doi = {10.1086/307956},
archivePrefix = {arXiv},
       eprint = {astro-ph/9901288},
 primaryClass = {astro-ph},
       adsurl = {https://ui.adsabs.harvard.edu/abs/1999ApJ...526..279P}
}

@ARTICLE{2001Padoan,
       author = {{Padoan}, Paolo and {Juvela}, Mika and {Goodman}, Alyssa A. and {Nordlund}, {\r{A}}ke},
        title = "{The Turbulent Shock Origin of Proto-Stellar Cores}",
      journal = {\apj},
         year = 2001,
        month = may,
       volume = {553},
       number = {1},
        pages = {227-234},
          doi = {10.1086/320636},
archivePrefix = {arXiv},
       eprint = {astro-ph/0011122},
 primaryClass = {astro-ph},
       adsurl = {https://ui.adsabs.harvard.edu/abs/2001ApJ...553..227P}
}

@INCOLLECTION{2008Peterson,
       author = {{Peterson}, D.~E. and {Megeath}, S.~T.},
        title = "{The Orion Molecular Cloud 2/3 and NGC 1977 Regions}",
    booktitle = {Handbook of Star Forming Regions, Volume I},
         year = 2008,
       editor = {{Reipurth}, B.},
       volume = {4},
        pages = {590},
          doi = {10.48550/arXiv.0809.4006},
       adsurl = {https://ui.adsabs.harvard.edu/abs/2008hsf1.book..590P}
}

@ARTICLE{2003Pety,
       author = {{Pety}, J. and {Falgarone}, E.},
        title = "{Non-Gaussian velocity shears  in the environment of low mass dense cores}",
      journal = {\aap},
         year = 2003,
        month = dec,
       volume = {412},
        pages = {417-430},
          doi = {10.1051/0004-6361:20031474},
archivePrefix = {arXiv},
       eprint = {astro-ph/0310063},
 primaryClass = {astro-ph},
       adsurl = {https://ui.adsabs.harvard.edu/abs/2003A&A...412..417P}
}

@INPROCEEDINGS{2005Pety,
       author = {{Pety}, J.},
        title = "{Successes of and Challenges to GILDAS, a State-of-the-Art Radioastronomy Toolkit}",
    booktitle = {SF2A-2005: Semaine de l'Astrophysique Francaise},
         year = 2005,
       editor = {{Casoli}, F. and {Contini}, T. and {Hameury}, J.~M. and {Pagani}, L.},
        month = dec,
        pages = {721},
       adsurl = {https://ui.adsabs.harvard.edu/abs/2005sf2a.conf..721P}
}

@ARTICLE{2017Pety,
       author = {{Pety}, J{\'e}r{\^o}me and {Guzm{\'a}n}, Viviana V. and {Orkisz}, Jan H. and {Liszt}, Harvey S. and {Gerin}, Maryvonne and {Bron}, Emeric and {Bardeau}, S{\'e}bastien and {Goicoechea}, Javier R. and {Gratier}, Pierre and {Le Petit}, Franck and {Levrier}, Fran{\c{c}}ois and {{\"O}berg}, Karin I. and {Roueff}, Evelyne and {Sievers}, Albrecht},
        title = "{The anatomy of the Orion B giant molecular cloud: A local template for studies of nearby galaxies}",
      journal = {\aap},
         year = 2017,
        month = jan,
       volume = {599},
          eid = {A98},
        pages = {A98},
          doi = {10.1051/0004-6361/201629862},
archivePrefix = {arXiv},
       eprint = {1611.04037},
 primaryClass = {astro-ph.GA},
       adsurl = {https://ui.adsabs.harvard.edu/abs/2017A&A...599A..98P}
}

@ARTICLE{2010Pineda,
       author = {{Pineda}, Jaime E. and {Goodman}, Alyssa A. and {Arce}, H{\'e}ctor G. and {Caselli}, Paola and {Foster}, Jonathan B. and {Myers}, Philip C. and {Rosolowsky}, Erik W.},
        title = "{Direct Observation of a Sharp Transition to Coherence in Dense Cores}",
      journal = {\apjl},
         year = 2010,
        month = mar,
       volume = {712},
       number = {1},
        pages = {L116-L121},
          doi = {10.1088/2041-8205/712/1/L116},
archivePrefix = {arXiv},
       eprint = {1002.2946},
 primaryClass = {astro-ph.GA},
       adsurl = {https://ui.adsabs.harvard.edu/abs/2010ApJ...712L.116P}
}

@ARTICLE{2011Pineda,
       author = {{Pineda}, Jaime E. and {Goodman}, Alyssa A. and {Arce}, H{\'e}ctor G. and {Caselli}, Paola and {Longmore}, Steven and {Corder}, Stuartt},
        title = "{Expanded Very Large Array Observations of the Barnard 5 Star-forming Core: Embedded Filaments Revealed}",
      journal = {\apjl},
         year = 2011,
        month = sep,
       volume = {739},
       number = {1},
          eid = {L2},
        pages = {L2},
          doi = {10.1088/2041-8205/739/1/L2},
archivePrefix = {arXiv},
       eprint = {1106.5474},
 primaryClass = {astro-ph.GA},
       adsurl = {https://ui.adsabs.harvard.edu/abs/2011ApJ...739L...2P}
}

@INPROCEEDINGS{2023Pineda,
       author = {{Pineda}, J.~E. and {Arzoumanian}, D. and {Andre}, P. and {Friesen}, R.~K. and {Zavagno}, A. and {Clarke}, S.~D. and {Inoue}, T. and {Chen}, C. and {Lee}, Y. and {Soler}, J.~D. and {Kuffmeier}, M.},
        title = "{From Bubbles and Filaments to Cores and Disks: Gas Gathering and Growth of Structure Leading to the Formation of Stellar Systems}",
    booktitle = {Protostars and Planets VII},
         year = 2023,
       editor = {{Inutsuka}, S. and {Aikawa}, Y. and {Muto}, T. and {Tomida}, K. and {Tamura}, M.},
       series = {Astronomical Society of the Pacific Conference Series},
       volume = {534},
        month = jul,
        pages = {233},
          doi = {10.48550/arXiv.2205.03935},
archivePrefix = {arXiv},
       eprint = {2205.03935},
 primaryClass = {astro-ph.GA},
       adsurl = {https://ui.adsabs.harvard.edu/abs/2023ASPC..534..233P}
}

@ARTICLE{1994Porter,
       author = {{Porter}, D.~H. and {Pouquet}, A. and {Woodward}, P.~R.},
        title = "{Kolmogorov-like spectra in decaying three-dimensional supersonic flows}",
      journal = {Physics of Fluids},
         year = 1994,
        month = jun,
       volume = {6},
       number = {6},
        pages = {2133-2142},
          doi = {10.1063/1.868217},
       adsurl = {https://ui.adsabs.harvard.edu/abs/1994PhFl....6.2133P}
}

@ARTICLE{2013Pudritz,
       author = {{Pudritz}, R.~E. and {Kevlahan}, N.~K. -R.},
        title = "{Shock interactions, turbulence and the origin of the stellar mass spectrum}",
      journal = {Philosophical Transactions of the Royal Society of London Series A},
         year = 2013,
        month = oct,
       volume = {371},
       number = {2003},
        pages = {20120248-20120248},
          doi = {10.1098/rsta.2012.0248},
archivePrefix = {arXiv},
       eprint = {1201.2650},
 primaryClass = {astro-ph.GA},
       adsurl = {https://ui.adsabs.harvard.edu/abs/2013RSPTA.37120248P}
}

@ARTICLE{2023Santa-Maria,
       author = {{Santa-Maria}, M.~G. and {Goicoechea}, J.~R. and {Pety}, J. and {Gerin}, M. and {Orkisz}, J.~H. and {Le Petit}, F. and {Einig}, L. and {Palud}, P. and {de Souza Magalhaes}, V. and {Be{\v{s}}li{\'c}}, I. and {Segal}, L. and {Bardeau}, S. and {Bron}, E. and {Chainais}, P. and {Chanussot}, J. and {Gratier}, P. and {Guzm{\'a}n}, V.~V. and {Hughes}, A. and {Languignon}, D. and {Levrier}, F. and {Lis}, D.~C. and {Liszt}, H.~S. and {Le Bourlot}, J. and {Oya}, Y. and {{\"O}berg}, K. and {Peretto}, N. and {Roueff}, E. and {Roueff}, A. and {Sievers}, A. and {Thouvenin}, P. -A. and {Yamamoto}, S.},
        title = "{HCN emission from translucent gas and UV-illuminated cloud edges revealed by wide-field IRAM 30 m maps of the Orion B GMC. Revisiting its role as a tracer of the dense gas reservoir for star formation}",
      journal = {\aap},
         year = 2023,
        month = nov,
       volume = {679},
          eid = {A4},
        pages = {A4},
          doi = {10.1051/0004-6361/202346598},
archivePrefix = {arXiv},
       eprint = {2309.03186},
 primaryClass = {astro-ph.GA},
       adsurl = {https://ui.adsabs.harvard.edu/abs/2023A&A...679A...4S}
}

@ARTICLE{2004Scalo,
       author = {{Scalo}, John and {Elmegreen}, Bruce G.},
        title = "{Interstellar Turbulence II: Implications and Effects}",
      journal = {\araa},
         year = 2004,
        month = sep,
       volume = {42},
       number = {1},
        pages = {275-316},
          doi = {10.1146/annurev.astro.42.120403.143327},
archivePrefix = {arXiv},
       eprint = {astro-ph/0404452},
 primaryClass = {astro-ph},
       adsurl = {https://ui.adsabs.harvard.edu/abs/2004ARA&A..42..275S}
}

@ARTICLE{2015Shirley,
       author = {{Shirley}, Yancy L.},
        title = "{The Critical Density and the Effective Excitation Density of Commonly Observed Molecular Dense Gas Tracers}",
      journal = {\pasp},
         year = 2015,
        month = mar,
       volume = {127},
       number = {949},
        pages = {299},
          doi = {10.1086/680342},
archivePrefix = {arXiv},
       eprint = {1501.01629},
 primaryClass = {astro-ph.IM},
       adsurl = {https://ui.adsabs.harvard.edu/abs/2015PASP..127..299S}
}

@ARTICLE{1994She,
       author = {{She}, Zhen-Su and {Leveque}, Emmanuel},
        title = "{Universal scaling laws in fully developed turbulence}",
      journal = {\prl},
         year = 1994,
        month = jan,
       volume = {72},
       number = {3},
        pages = {336-339},
          doi = {10.1103/PhysRevLett.72.336},
       adsurl = {https://ui.adsabs.harvard.edu/abs/1994PhRvL..72..336S}
}

@ARTICLE{2020SciPy-NMeth,
  author  = {Virtanen, Pauli and Gommers, Ralf and Oliphant, Travis E. and
            Haberland, Matt and Reddy, Tyler and Cournapeau, David and
            Burovski, Evgeni and Peterson, Pearu and Weckesser, Warren and
            Bright, Jonathan and {van der Walt}, St{\'e}fan J. and
            Brett, Matthew and Wilson, Joshua and Millman, K. Jarrod and
            Mayorov, Nikolay and Nelson, Andrew R. J. and Jones, Eric and
            Kern, Robert and Larson, Eric and Carey, C J and
            Polat, {\.I}lhan and Feng, Yu and Moore, Eric W. and
            {VanderPlas}, Jake and Laxalde, Denis and Perktold, Josef and
            Cimrman, Robert and Henriksen, Ian and Quintero, E. A. and
            Harris, Charles R. and Archibald, Anne M. and
            Ribeiro, Ant{\^o}nio H. and Pedregosa, Fabian and
            {van Mulbregt}, Paul and {SciPy 1.0 Contributors}},
  title   = {{{SciPy} 1.0: Fundamental Algorithms for Scientific
            Computing in Python}},
  journal = {Nature Methods},
  year    = {2020},
  volume  = {17},
  pages   = {261--272},
  adsurl  = {https://rdcu.be/b08Wh},
  doi     = {10.1038/s41592-019-0686-2},
}

@ARTICLE{2014Smith,
       author = {{Smith}, Rowan J. and {Glover}, Simon C.~O. and {Klessen}, Ralf. S.},
        title = "{On the nature of star-forming filaments - I. Filament morphologies}",
      journal = {\mnras},
         year = 2014,
        month = dec,
       volume = {445},
       number = {3},
        pages = {2900-2917},
          doi = {10.1093/mnras/stu1915},
archivePrefix = {arXiv},
       eprint = {1407.6716},
 primaryClass = {astro-ph.GA},
       adsurl = {https://ui.adsabs.harvard.edu/abs/2014MNRAS.445.2900S}
}

@ARTICLE{2016Smith,
       author = {{Smith}, Rowan J. and {Glover}, Simon C.~O. and {Klessen}, Ralf S. and {Fuller}, Gary A.},
        title = "{On the nature of star-forming filaments - II. Subfilaments and velocities}",
      journal = {\mnras},
         year = 2016,
        month = feb,
       volume = {455},
       number = {4},
        pages = {3640-3655},
          doi = {10.1093/mnras/stv2559},
archivePrefix = {arXiv},
       eprint = {1509.03321},
 primaryClass = {astro-ph.GA},
       adsurl = {https://ui.adsabs.harvard.edu/abs/2016MNRAS.455.3640S}
}

@ARTICLE{2024Soccia,
       author = {{Socci}, A. and {Hacar}, A. and {Bonanomi}, F. and {Tafalla}, M. and {Suri}, S.},
        title = "{Emergence of high-mass stars in complex fiber networks (EMERGE): III. Fiber networks in Orion}",
      journal = {\aap},
         year = 2024,
        month = oct,
       volume = {690},
          eid = {A375},
        pages = {A375},
          doi = {10.1051/0004-6361/202449316},
archivePrefix = {arXiv},
       eprint = {2409.01321},
 primaryClass = {astro-ph.GA},
       adsurl = {https://ui.adsabs.harvard.edu/abs/2024A&A...690A.375S}
}

@ARTICLE{2024Soccib,
       author = {{Socci}, A. and {Hacar}, A. and {Bonanomi}, F. and {Tafalla}, M. and {Suri}, S.},
        title = "{Emergence of high-mass stars in complex fiber networks (EMERGE): IV. Environmental dependence of the fiber widths}",
      journal = {\aap},
         year = 2024,
        month = oct,
       volume = {690},
          eid = {A376},
        pages = {A376},
          doi = {10.1051/0004-6361/202451177},
archivePrefix = {arXiv},
       eprint = {2408.16427},
 primaryClass = {astro-ph.GA},
       adsurl = {https://ui.adsabs.harvard.edu/abs/2024A&A...690A.376S}
}

@ARTICLE{2019Sokolov,
       author = {{Sokolov}, Vlas and {Wang}, Ke and {Pineda}, Jaime E. and {Caselli}, Paola and {Henshaw}, Jonathan D. and {Barnes}, Ashley T. and {Tan}, Jonathan C. and {Fontani}, Francesco and {Jim{\'e}nez-Serra}, Izaskun},
        title = "{Multicomponent Kinematics in a Massive Filamentary Infrared Dark Cloud}",
      journal = {\apj},
         year = 2019,
        month = feb,
       volume = {872},
       number = {1},
          eid = {30},
        pages = {30},
          doi = {10.3847/1538-4357/aafaff},
archivePrefix = {arXiv},
       eprint = {1812.09581},
 primaryClass = {astro-ph.GA},
       adsurl = {https://ui.adsabs.harvard.edu/abs/2019ApJ...872...30S}
}

@ARTICLE{2013Stutz,
       author = {{Stutz}, Amelia M. and {Tobin}, John J. and {Stanke}, Thomas and {Megeath}, S. Thomas and {Fischer}, William J. and {Robitaille}, Thomas and {Henning}, Thomas and {Ali}, Babar and {di Francesco}, James and {Furlan}, Elise and {Hartmann}, Lee and {Osorio}, Mayra and {Wilson}, Thomas L. and {Allen}, Lori and {Krause}, Oliver and {Manoj}, P.},
        title = "{A Herschel and APEX Census of the Reddest Sources in Orion: Searching for the Youngest Protostars}",
      journal = {\apj},
         year = 2013,
        month = apr,
       volume = {767},
       number = {1},
          eid = {36},
        pages = {36},
          doi = {10.1088/0004-637X/767/1/36},
archivePrefix = {arXiv},
       eprint = {1302.1203},
 primaryClass = {astro-ph.SR},
       adsurl = {https://ui.adsabs.harvard.edu/abs/2013ApJ...767...36S}
}

@ARTICLE{2002Tafalla,
       author = {{Tafalla}, M. and {Myers}, P.~C. and {Caselli}, P. and {Walmsley}, C.~M. and {Comito}, C.},
        title = "{Systematic Molecular Differentiation in Starless Cores}",
      journal = {\apj},
         year = 2002,
        month = apr,
       volume = {569},
       number = {2},
        pages = {815-835},
          doi = {10.1086/339321},
archivePrefix = {arXiv},
       eprint = {astro-ph/0112487},
 primaryClass = {astro-ph},
       adsurl = {https://ui.adsabs.harvard.edu/abs/2002ApJ...569..815T}
}

@ARTICLE{2015Tafalla,
       author = {{Tafalla}, M. and {Hacar}, A.},
        title = "{Chains of dense cores in the Taurus L1495/B213 complex}",
      journal = {\aap},
         year = 2015,
        month = feb,
       volume = {574},
          eid = {A104},
        pages = {A104},
          doi = {10.1051/0004-6361/201424576},
archivePrefix = {arXiv},
       eprint = {1412.1083},
 primaryClass = {astro-ph.GA},
       adsurl = {https://ui.adsabs.harvard.edu/abs/2015A&A...574A.104T}
}

@ARTICLE{2021Tafalla,
       author = {{Tafalla}, M. and {Usero}, A. and {Hacar}, A.},
        title = "{Characterizing the line emission from molecular clouds. Stratified random sampling of the Perseus cloud}",
      journal = {\aap},
         year = 2021,
        month = feb,
       volume = {646},
          eid = {A97},
        pages = {A97},
          doi = {10.1051/0004-6361/202038727},
archivePrefix = {arXiv},
       eprint = {2101.02710},
 primaryClass = {astro-ph.GA},
       adsurl = {https://ui.adsabs.harvard.edu/abs/2021A&A...646A..97T}
}

@ARTICLE{2023Tafalla,
       author = {{Tafalla}, M. and {Usero}, A. and {Hacar}, A.},
        title = "{Characterizing the line emission from molecular clouds. II. A comparative study of California, Perseus, and Orion A}",
      journal = {\aap},
         year = 2023,
        month = nov,
       volume = {679},
          eid = {A112},
        pages = {A112},
          doi = {10.1051/0004-6361/202346136},
archivePrefix = {arXiv},
       eprint = {2309.14414},
 primaryClass = {astro-ph.GA},
       adsurl = {https://ui.adsabs.harvard.edu/abs/2023A&A...679A.112T}
}

@ARTICLE{1983Tomisaka,
       author = {{Tomisaka}, K. and {Ikeuchi}, S.},
        title = "{Gravitational instability of isothermal gas layers - Effect of curvature and magnetic field}",
      journal = {\pasj},
         year = 1983,
        month = jan,
       volume = {35},
       number = {2},
        pages = {187-208},
       adsurl = {https://ui.adsabs.harvard.edu/abs/1983PASJ...35..187T}
}

@ARTICLE{2007_vanderTak_RADEX,
       author = {{van der Tak}, F.~F.~S. and {Black}, J.~H. and {Sch{\"o}ier}, F.~L. and {Jansen}, D.~J. and {van Dishoeck}, E.~F.},
        title = "{A computer program for fast non-LTE analysis of interstellar line spectra. With diagnostic plots to interpret observed line intensity ratios}",
      journal = {\aap},
         year = 2007,
        month = jun,
       volume = {468},
       number = {2},
        pages = {627-635},
          doi = {10.1051/0004-6361:20066820},
archivePrefix = {arXiv},
       eprint = {0704.0155},
 primaryClass = {astro-ph},
       adsurl = {https://ui.adsabs.harvard.edu/abs/2007A&A...468..627V}
}

@ARTICLE{2014VanLoo,
       author = {{Van Loo}, Sven and {Keto}, Eric and {Zhang}, Qizhou},
        title = "{Core and Filament Formation in Magnetized, Self-gravitating Isothermal Layers}",
      journal = {\apj},
         year = 2014,
        month = jul,
       volume = {789},
       number = {1},
          eid = {37},
        pages = {37},
          doi = {10.1088/0004-637X/789/1/37},
archivePrefix = {arXiv},
       eprint = {1405.1013},
 primaryClass = {astro-ph.SR},
       adsurl = {https://ui.adsabs.harvard.edu/abs/2014ApJ...789...37V}
}

@ARTICLE{1991Vincent,
       author = {{Vincent}, A. and {Meneguzzi}, M.},
        title = "{The spatial structure and statistical properties of homogeneous turbulence}",
      journal = {Journal of Fluid Mechanics},
         year = 1991,
        month = apr,
       volume = {225},
        pages = {1-20},
          doi = {10.1017/S0022112091001957},
       adsurl = {https://ui.adsabs.harvard.edu/abs/1991JFM...225....1V}
}

@ARTICLE{2000Wenger,
       author = {{Wenger}, M. and {Ochsenbein}, F. and {Egret}, D. and {Dubois}, P. and {Bonnarel}, F. and {Borde}, S. and {Genova}, F. and {Jasniewicz}, G. and {Lalo{\"e}}, S. and {Lesteven}, S. and {Monier}, R.},
        title = "{The SIMBAD astronomical database. The CDS reference database for astronomical objects}",
      journal = {\aaps},
         year = 2000,
        month = apr,
       volume = {143},
        pages = {9-22},
          doi = {10.1051/aas:2000332},
archivePrefix = {arXiv},
       eprint = {astro-ph/0002110},
 primaryClass = {astro-ph},
       adsurl = {https://ui.adsabs.harvard.edu/abs/2000A&AS..143....9W}
}

@ARTICLE{1979SchneiderElmegreen,
       author = {{Schneider}, S. and {Elmegreen}, B.~G.},
        title = "{A catalog of dark globular filaments.}",
      journal = {\apjs},
         year = 1979,
        month = sep,
       volume = {41},
        pages = {87-95},
          doi = {10.1086/190609},
       adsurl = {https://ui.adsabs.harvard.edu/abs/1979ApJS...41...87S}
}

@ARTICLE{1992Schilke,
       author = {{Schilke}, P. and {Walmsley}, C.~M. and {Pineau Des Forets}, G. and {Roueff}, E. and {Flower}, D.~R. and {Guilloteau}, S.},
        title = "{A study of HCN, HNC and their isotopometers in OMC-1. I. Abundances and chemistry.}",
      journal = {\aap},
         year = 1992,
        month = mar,
       volume = {256},
        pages = {595-612},
       adsurl = {https://ui.adsabs.harvard.edu/abs/1992A&A...256..595S}
}

\appendix
\section{Additional maps}\label{sec:maps}

\begin{figure*}[ht!]
	\centering
        \includegraphics[width=0.721\textwidth]{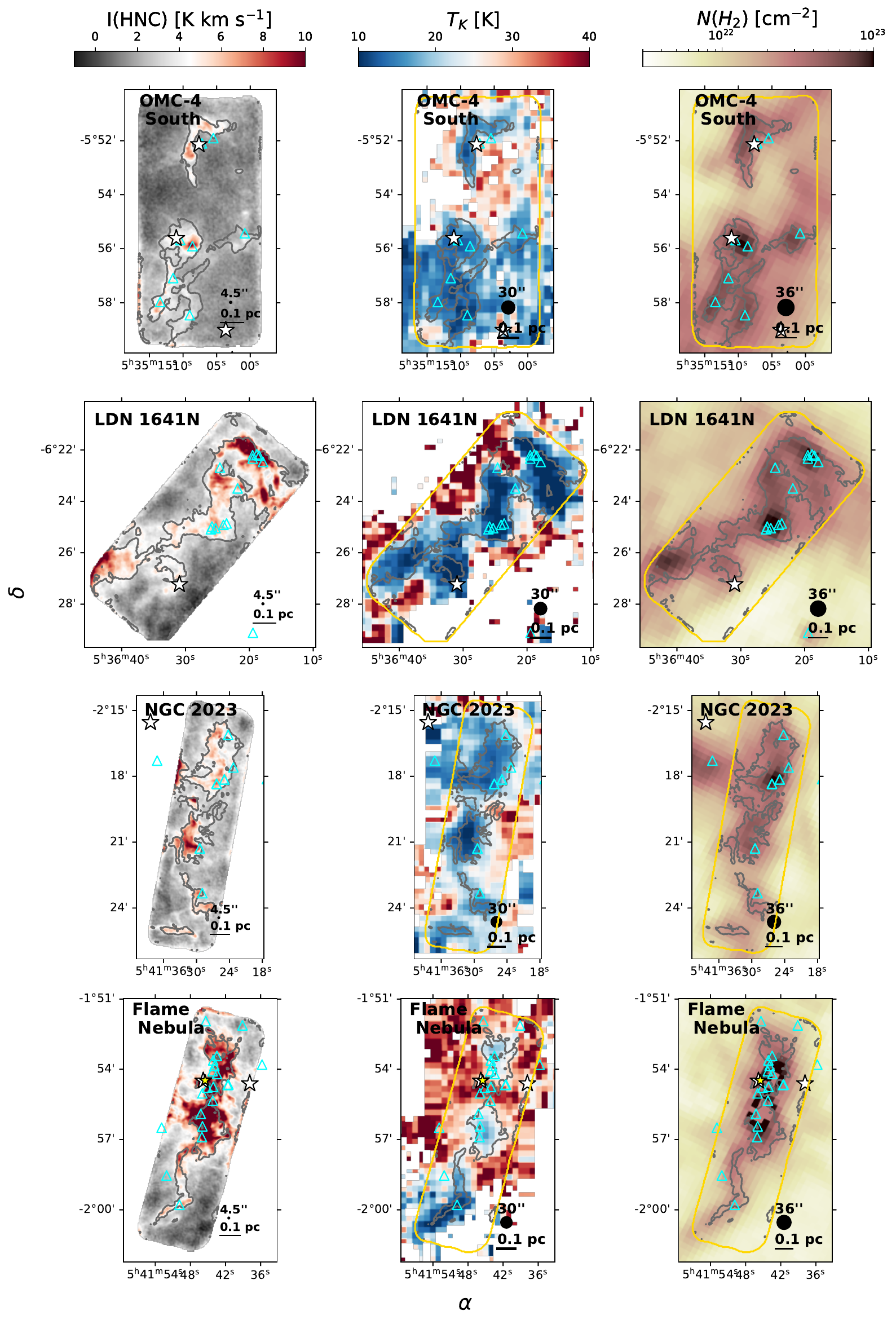}
	\caption{Similar to Fig.~\ref{fig:OMC3_maps}, maps of the four remaining star-forming regions in the EMERGE Early ALMA Survey, namely OMC-4 South, LDN~1641N, NGC~2023, and the Flame Nebula. From left to right we plot the integrated intensity I(HNC) observed at 4.5~arcsec resolution with ALMA+IRAM-30m (left panels), the gas kinetic temperature $T_K$ derived from the HCN-HNC ratio observed with IRAM-30m at 30~arcsec resolution (central panels; see Paper I), and the total column density $N(\text{H}_2)$ map as observed by Herschel at 36~arcsec resolution \citep[right panels;][]{2014Lombardi}. Beam sizes and scale bars are placed in the bottom right corner.}%The OMC-3 maps are shown in Fig.~\ref{fig:OMC3_maps}.}
	\label{fig:survey_maps} 
\end{figure*}

\begin{figure*}[ht!]
	\centering
        \includegraphics[trim={4 1cm 4 3cm},clip,width=0.85\textwidth]{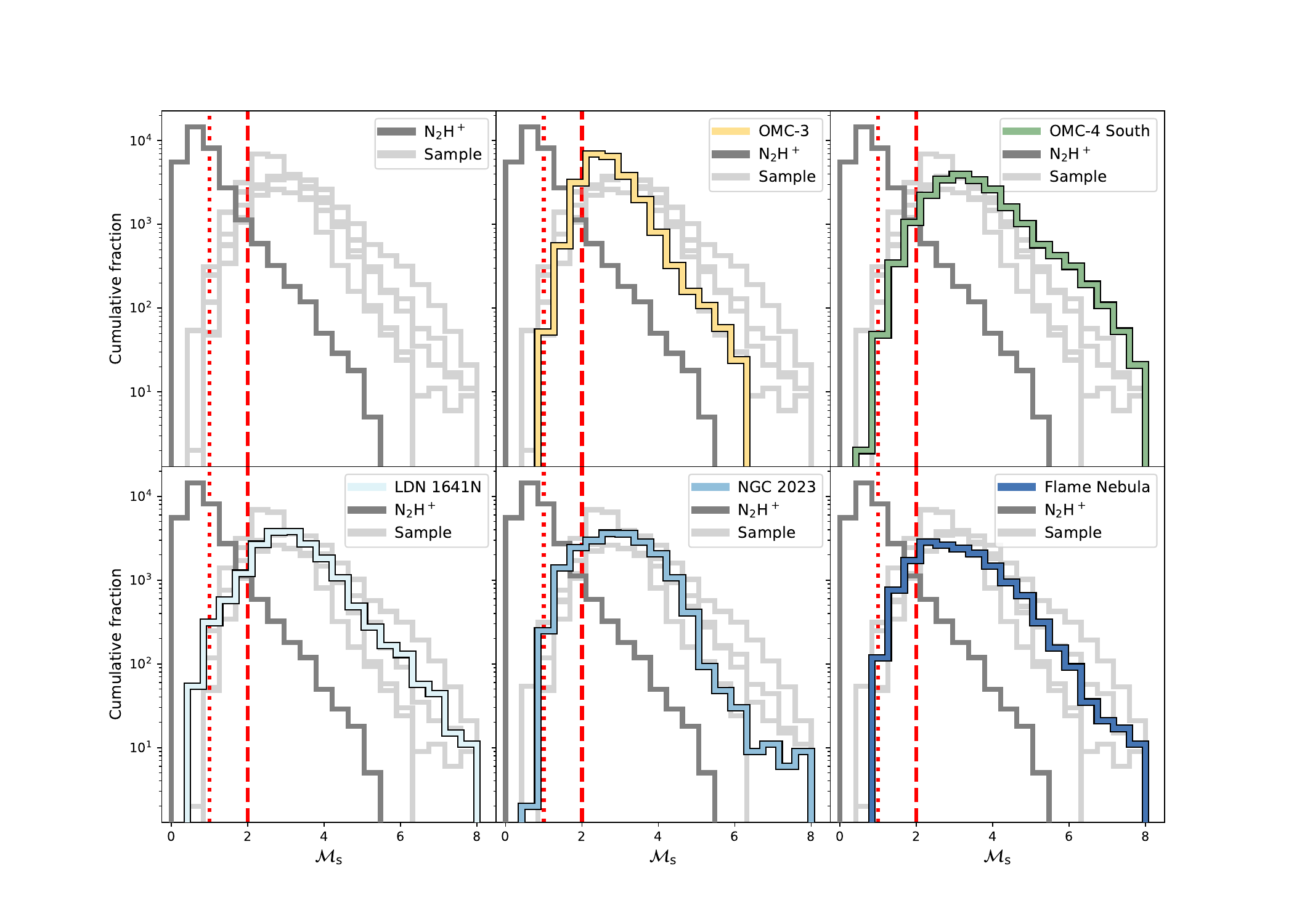}
	\caption{Individual distribution of the Mach number for the five regions in the EMERGE Early ALMA Survey (see Fig. \ref{fig:mach_number_statistics}) compared to the whole sample and the \nthp~data \citep[excluding OMC-1 and -2, see][for a full discussion]{2024Soccia,2024Soccib}. The red dotted and dashed lines at $\sigma_\text{nt}/c_\text{s}=1$ and 2 mark the transition between the sub-, tran-, and super-sonic regime, respectively.}
	\label{fig:mach_hist} 
\end{figure*}

\begin{figure*}[ht!]
	\centering
        \includegraphics[trim={4 1cm 4 3cm},clip,width=0.85\textwidth]{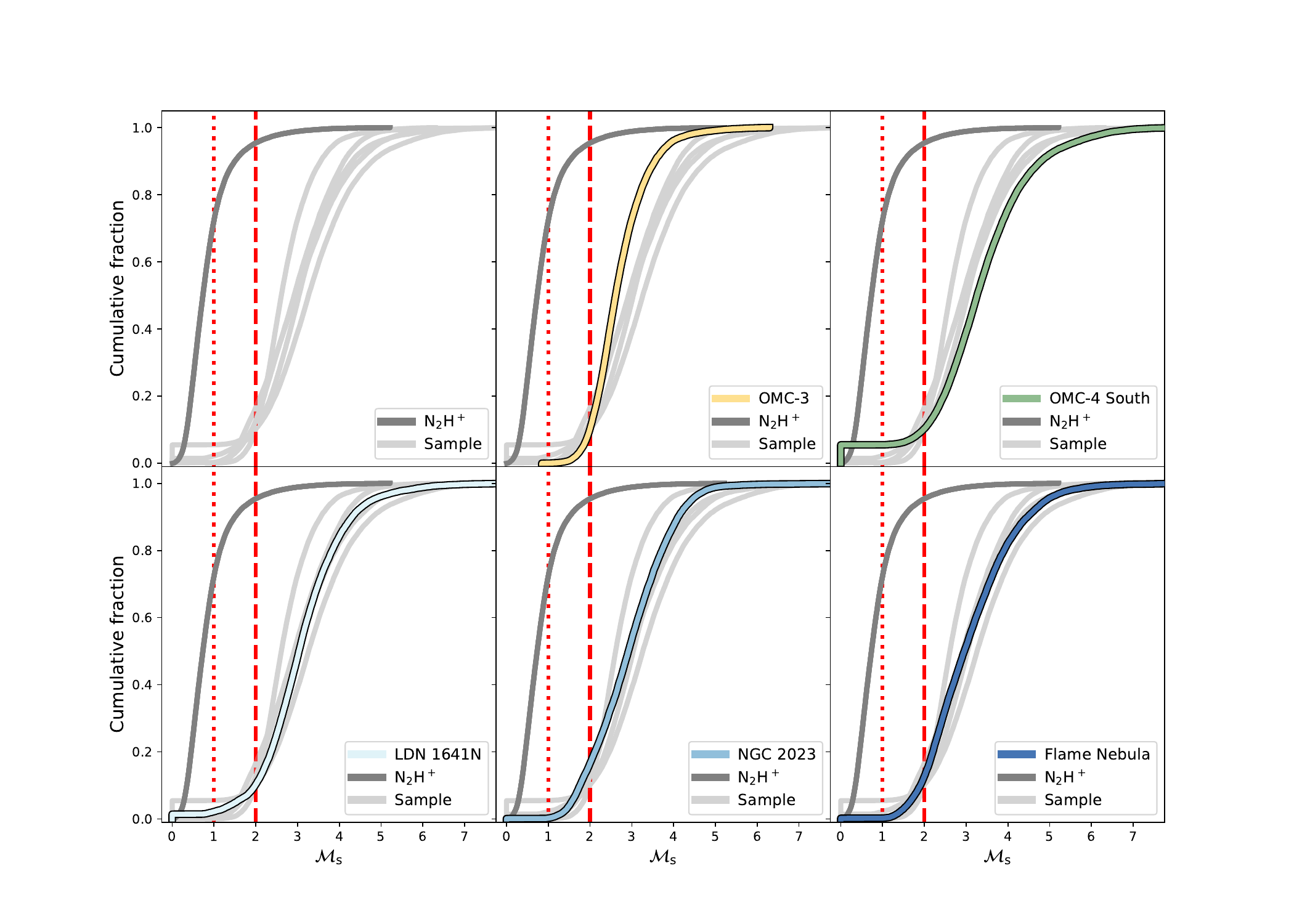}
	\caption{Individual cumulative distribution of the Mach number for the five regions in the EMERGE Early ALMA Survey (see Fig. \ref{fig:mach_number_statistics}) compared to the whole sample and the \nthp~data \citep[excluding OMC-1 and -2, see][for a full discussion]{2024Soccia,2024Soccib}. The red dotted and dashed lines at $\sigma_\text{nt}/c_\text{s}=1$ and 2 mark the transition between the sub-, tran-, and super-sonic regime, respectively.}
	\label{fig:mach_cdis} 
\end{figure*}

In this appendix, we present all the maps of our EMERGE Early ALMA survey in OMC-4 South, LDN 1641N, NGC 2023, and the Flame Nebula (see also Fig.~\ref{fig:OMC3_maps} for those in OMC-3). We show in Fig.~\ref{fig:survey_maps} the \hnc~integrated intensity I(HNC) (left) observed at 4.5 arcsec resolution with ALMA+IRAM-30m, the gas kinetic temperature $T_K$ (middle) derived from the HCN-to-\hnc~ratio observed with IRAM-30m at 30~arcsec resolution (see Paper I), and the total gas column density $N(\text{H}_2)$ maps (right) derived from \herschel~observations at 36~arcsec resolution \citep{2014Lombardi}. The gray contours show the \nthptrans~integrated intensity above 3$\sigma$ (the \nthp~maps are visible in Paper III).
We also include Figs.~\ref{fig:mach_hist} and \ref{fig:mach_cdis} to better compare the $\mathcal{M}$ distribution of the individual regions of the EMERGE Early ALMA Survey with respect to the entire sample.

\section{Multi-component fit}\label{sec:appendix_fit}

\begin{table*}[ht!]
\caption{\hnc~multi-component fit}
    \centering
    %\tiny
    \begin{tabular}{lcccccccc}
        \toprule
        \multicolumn{1}{l}{ Target} & \multicolumn{1}{c}{Fit} & \multicolumn{1}{c}{Single comp.} & \multicolumn{2}{c}{Multi-comp.} \\
        & & & 2 comp. & 3 comp. \\		     	
        \midrule
         OMC-3  & 24316 & 21538 (89$\%$)& 2758 (11$\%$)& 20 ($<1\%$)\\
         OMC-4 South    & 22743 & 20245 (89$\%$)& 2442 (11$\%$)& 56 ($<1\%$)\\
         LDN 1641N  & 19379 & 16475 (85$\%$)& 2903 (15$\%$)& 1 ($<1\%$)\\
         NGC 2023   & 20900 & 15009 (72$\%$)& 5849 ($28\%$)& 42 ($<1\%$)\\
         Flame Nebula   & 16147 & 8866 (55$\%$)& 6954 (43$\%$)& 327 (2$\%$)\\
        \midrule
         Sample &  103485 & 82133 (79$\%$)& 20906 (20$\%$)& 446 ($<1\%$)\\
        \bottomrule
    \end{tabular}
    \label{tab:fit_statistics}
\end{table*}

\begin{table}[h]
\caption{Kinematics of the diffuse gas: properties extracted from the multi-component fit}
    \centering
    \begin{tabular}{lcr}
        \toprule
        Target &  $\Delta$V    &   $\mathcal{M}_\text{s}$\\
        & (\kms)\\
        \midrule
         OMC-3  & $1.7_{-0.4}^{+0.5}$ & $2.5_{-0.6}^{+0.7}$\\
         OMC-4 South    & $1.8_{-0.6}^{+0.8}$ & $3.1_{-1.1}^{+1.1}$\\
         LDN 1641N  & $1.6_{-0.6}^{+0.5}$ & $2.8_{-1.0}^{+1.0}$\\
         NGC 2023   & $1.6_{-0.7}^{+0.7}$ & $2.5_{-1.1}^{+1.2}$\\
         Flame Nebula   & $1.5_{-0.5}^{+1.0}$ & $2.1_{-0.9}^{+1.2}$\\
         \midrule
         Sample & $1.7_{-0.6}^{+0.7}$ & $2.6_{-1.0}^{+1.1}$\\
         \bottomrule
    \end{tabular}
    \label{tab:multi_fit_properties}
\end{table}

\begin{table}[ht!]
\caption{\hnc~fit errors}
    \centering
    %\tiny
    \begin{tabular}{lcccccccc}
        \toprule
        \multicolumn{1}{l}{ Target} & \multicolumn{2}{c}{Single comp.} & \multicolumn{2}{c}{Multi-comp.} \\
        & $\sigma_{V_{lsr}}$ & $\sigma_{\Delta V}$ & $\sigma_{V_{lsr}}$  &$\sigma_{\Delta V}$  \\		     	
        \midrule
         OMC-3  & 0.02 & 0.04& 0.02 & 0.05\\
         OMC-4 South    & 0.06 & 0.2& 0.06& 0.15\\
         LDN 1641N  & 0.03 & 0.07& 0.03& 0.08\\
         NGC 2023   & 0.03 & 0.10& 0.04&0.11\\
         Flame Nebula & 0.04 & 0.08& 0.04&0.09\\
        \midrule
         Sample &  0.04 & 0.09& 0.04& 0.10\\
        \bottomrule
    \end{tabular}
    \label{tab:fit_errors}
\end{table}

In this appendix, we discuss the fitting routine mentioned in Sect.~\ref{sec:kinematics}. We performed both a single- and multi-component fitting on our data using customized GILDAS/CLASS \citep{2005Pety,2013GildasTeam} routines following similar fitting strategies than other papers in the literature \citep{2013Hacar,2018Hacar}. 
The latter shows the majority ($\sim79\%$) of the spectra fitted with only one component in \hnc, while only a marginal number of spectra with two ($\sim20\%$) or three ($<1\%$) components. We report in Tab.~\ref{tab:fit_statistics} the statistics on the number of components fitted for each target.
The distribution of single- and multi-component spectra is regular across the whole sample, with the Flame Nebula as the only outlier that presents $\sim45\%$ of multi-component spectra.
For comparison, we produced velocity dispersion maps. From these second-moment maps M$_2$ we derived an independent measure of the linewidth following the relation $\Delta V=\sqrt{2ln2\ M_2}$.

\begin{figure*}[h!]
	\centering
        \includegraphics[trim={0 2.5cm 0 2cm},clip,width=1\textwidth]{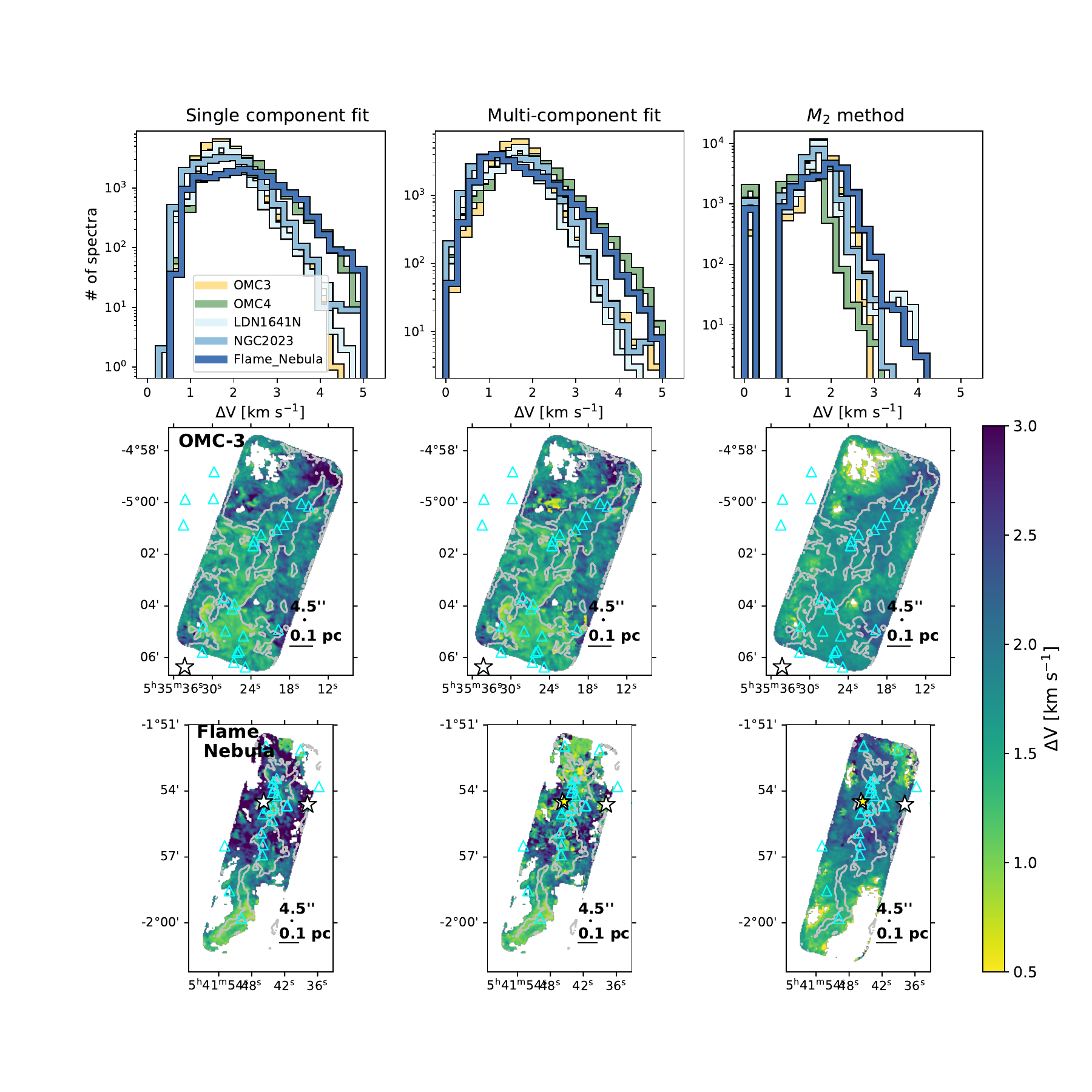}
	\caption{Linewidth estimates from different methods. From left to right: single (left panels), multi-component fit (central panels), and moment two method (right panels) maps. We display the histogram of the linewidth for all the targets (top), the linewidth maps for  OMC-3 (middle), and for the Flame Nebula (bottom). The gray contours show the \nthptrans~integrated intensity above 3$\sigma$ (the \nthp~maps are visible in Paper III). Beam sizes and scale bars are placed in the bottom right corner.}
	\label{fig:method_comparison_linewidth.} 
\end{figure*}

Since we focus on the kinematics, we compare the linewidth values and the maps obtained from these three methods in Fig.~\ref{fig:method_comparison_linewidth.}. The overall $\Delta V$ distribution (top row) is similar for single- and multi-component between 0 and 5~\kms, with systematically lower median values for the multi-component fit (all values in Tab.~\ref{tab:multi_fit_properties}). The distribution obtained from the second moment map appears to be 
much narrower and slightly skewed towards higher values. This discrepancy may be due to the moment map definition, by summing the emission along the spectra axis. Considering the positions with multiple Gaussian components, the velocity dispersion estimated by the second moment map will represent the separation between the components, not its linewidth. Following this interpretation, this difference is mostly visible in the Flame Nebula, as expected since is the region where the highest number of multiple components has been fitted. 
We also compare the linewidth maps for OMC-3 (middle row) and the Flame nebula (bottom row). For the multi-component fit we produce this linewidth maps selecting the value extracted from the brightest component. In OMC-3 we notice that the spatial distribution of $\Delta V$ is similar between the three different methods, mirroring the discrepancy observed in the histograms. This result is not surprising, since the multi-component detected in OMC-3 is only $11\%$ of the fitted spectra. The discrepancy is much more visible in the Flame Nebula, where only the $55\%$ of the spectra is fitted with a single component. Again, the overall distribution of the values is similar in the South where the values are below 2~\kms, however, in the North the values in the single component map are much higher with respect to the other methods.  

As already explained in Sect.~\ref{sec:kinematics}, our goal is investigating the kinematics of the diffuse gas and its spatial distribution. Producing linewidth and velocity gradients maps requires the selection of a single value to describe each pixel. Thus, we need to reduce the spectra fitted with multiple components either by taking (i) the brightest component among the
ones fitted, (ii) the broadest, or (iii) their average.
Considering that none of the methods is flawless and the results are in general in agreement, and that a selection bias would be introduced in case of multi-component analysis, we adopt the single-component fitting routine in our study as the simplest approach. According to this choice, the linewidth estimated following this single-component routine is to be considered as an upper limit, as multiple components may be fitted by a single broad Gaussian function. On the other hand, the velocity gradient calculation may also be affected by this choice, with high velocity gradients resulting as a jump between different components.

\begin{figure*}[h!]
	\centering
        \includegraphics[trim={0 0cm 0 0cm},clip,width=1\textwidth]{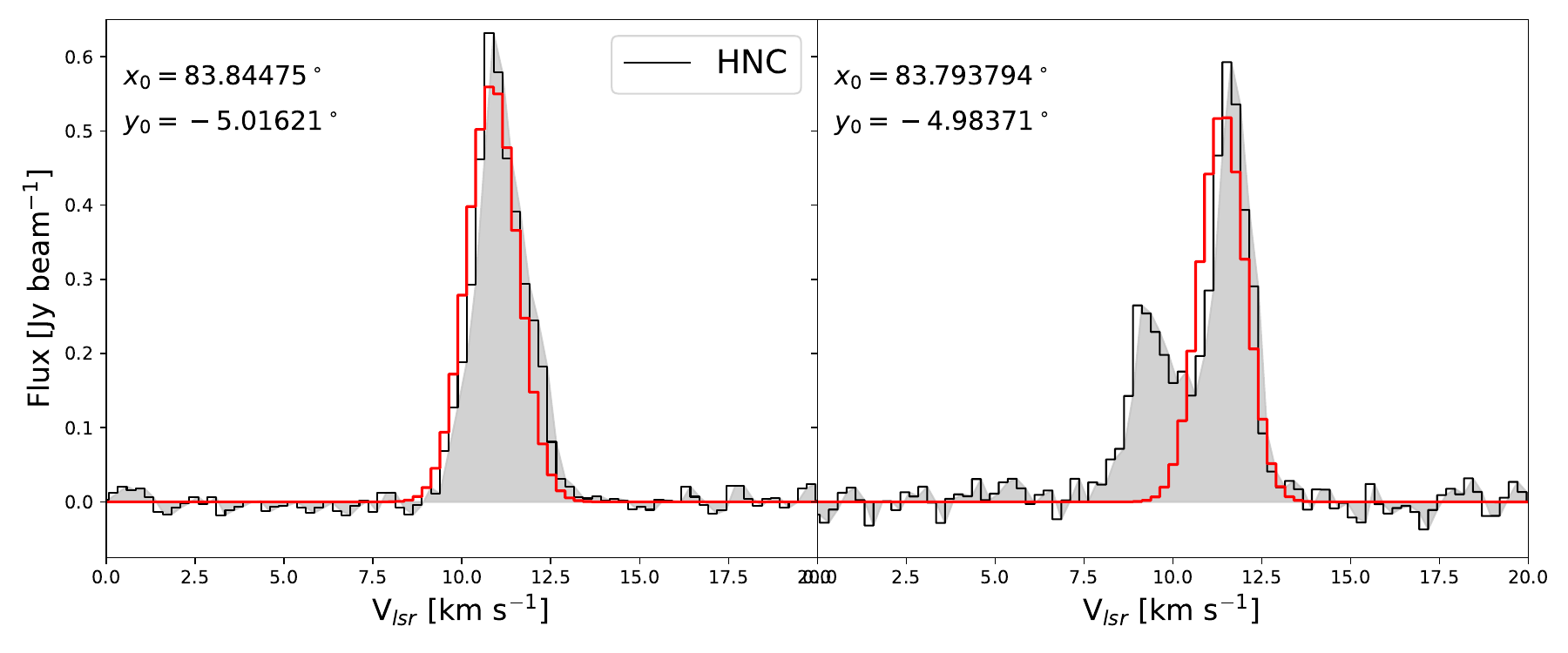}
	\caption{Examples of fitted spectra in OMC-3. The red line show the single component  Gaussian fit.}
	\label{fig:spectra_in_OMC3} 
\end{figure*}

We reported a two examples of spectra fitted with the single component routine in OMC-3 in Fig.~\ref{fig:spectra_in_OMC3}. As discussed above, the majority of the fitted spectra (89\%) are similar to the narrow lines (left), while there is only a small fraction (11\%) of multi-component spectra (right).

\begin{figure*}[h!]
	\centering
        \includegraphics[width=1\textwidth]{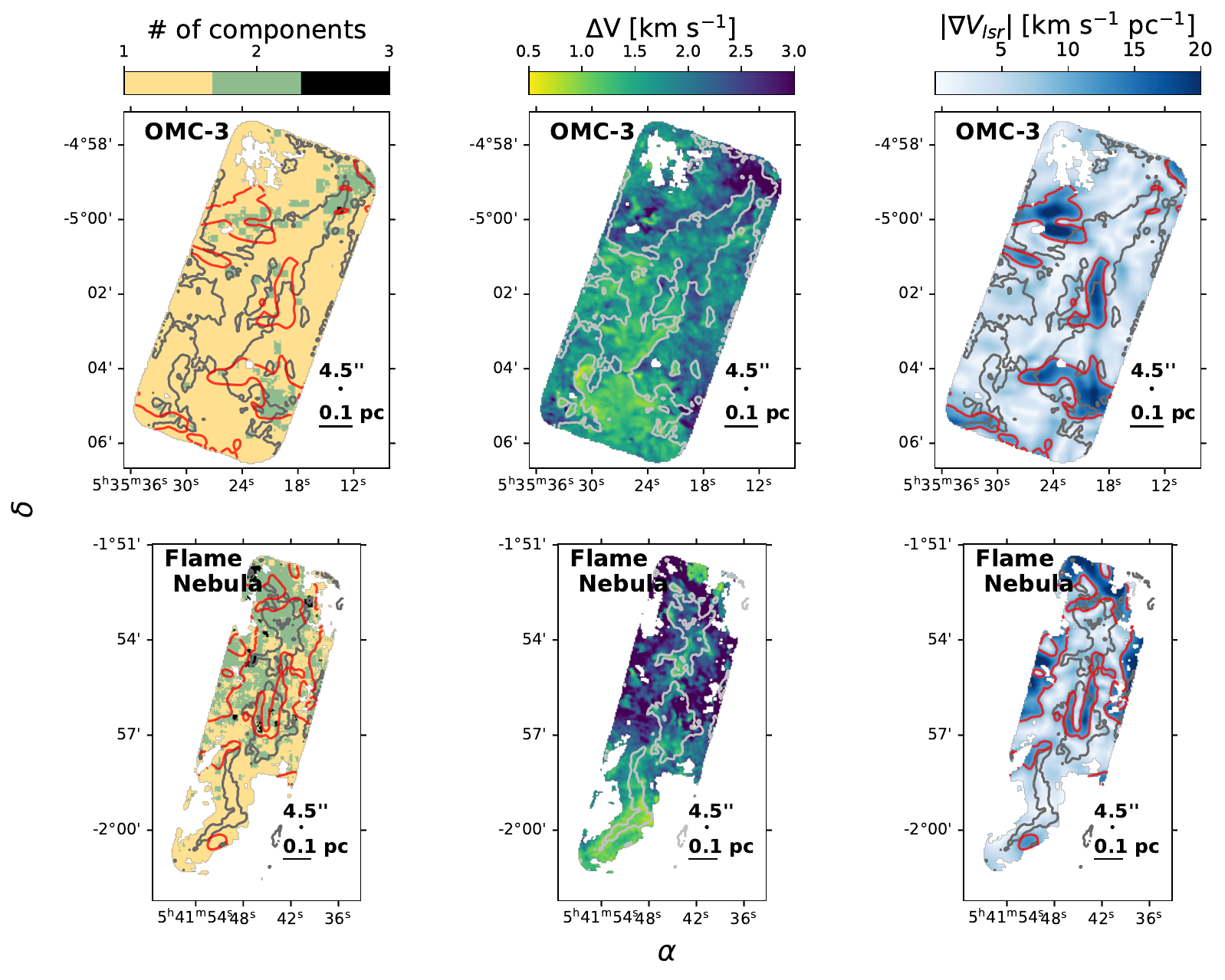}
	\caption{Influence of the number of fitted components in the analysis of the OMC-3 (top) and the Flame Nebula (bottom) regions. From left to right we display the  number of components obtained using the multi-component fit (left panel), the recovered linewidth $\Delta V$ (central panel) obtained from the single-component routine and velocity gradient $|\nabla V_\text{lsr}|$ (right panel panel) maps again from the single-component fit. The gray and red contours show the \nthptrans~integrated intensity above 3$\sigma$ (the \nthp~maps are visible in Paper III), and the high velocity gradient regions identified in Fig.~\ref{fig:grad_sel}, respectively. Beam sizes and scale bars are placed in the bottom right corner.}
	\label{fig:num_components} 
\end{figure*}

To check that our choice of a single-component would affect our analysis, we compare the spatial distribution of the fitted multi-components with our linewidth and velocity gradient maps in Fig.~\ref{fig:num_components} for OMC-3 (top) and the Flame Nebula (bottom). Considering OMC-3, we notice that there is no systematic spatial correlation between the location of multi-components and the increase of $\Delta V$ (middle panel) or high velocity gradients (right panel) regions (marked by red contours). The picture is more complex for the Flame Nebula (bottom) since the number of multi-components is larger ($\sim45\%$) and they are located mostly in the northern region that corresponds to the area severely affected by feedback (see Sect.~\ref{sec:feedback}). However, no clear correlation is observed for the high shear regions.

\section{Velocity gradients: method comparison}\label{sec:appendix_gradients}

\begin{figure*}[htbp]
	\centering
        \includegraphics[width=0.65\textwidth]{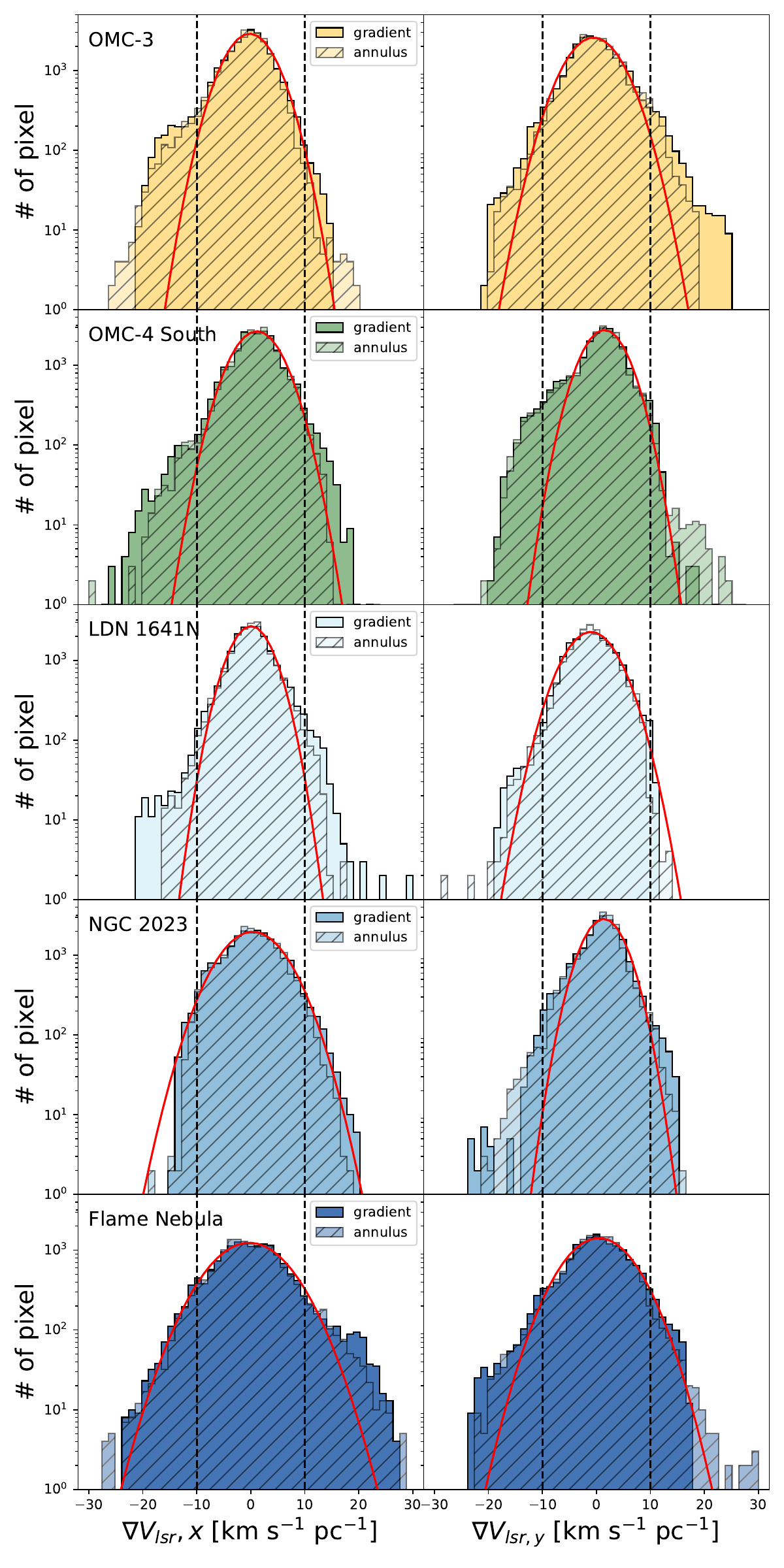}
	\caption{Histogram of the projection of the gradient along the $\alpha$ (left) and $\delta$ axes (right) in all the targets in our survey. We compare the results obtained using the standard gradient method (solid area) and the annulus method (hatched area) for a $L=0.04$~pc lag. The red lines show the Gaussian fit applied to the distribution.}
	\label{fig:grad_selection_method_comparison} 
\end{figure*}

To demonstrated that validity of velocity gradient analysis in finding high shear regions we have compared the results presented in Sect.~\ref{sec:velocity_gradients} with those obtained using (a) the velocity increments method presented in Sect.~\ref{sec:dissipation_statistics} and traditionally used in previous works of the literature, and (b) the annulus method.

Following \citet{1993Goodman}, our original calculations estimate the observed velocity gradient $\nabla V_{lsr}$ by minimizing the best fitted velocity gradient to all pixels within a box of size L. 
In the annulus method, instead, we computed the velocity gradients in an annulus $|\nabla V_{lsr}|_{annulus}$, applying the same procedure but only on the annular region at distance L, without considering the points at shorter distances \citep{1996Lis,1998Lis}. In these latter methods, we considered all the points between $L-0.5$~pix $L+ 0.5$~pix, similar to the method presented in \citet{1998Lis}. We set $L=0.04$~pc (equivalent to $20.25$~arcsec or $4.5$~beams) and a minimum number of 20 independent, Nyquist-sampled pixels considering on average 72 independent pixels per annulus. 

\begin{figure*}[htbp]
	\centering
        \includegraphics[trim={2cm 2cm 2cm 1cm},clip,width=1\textwidth]{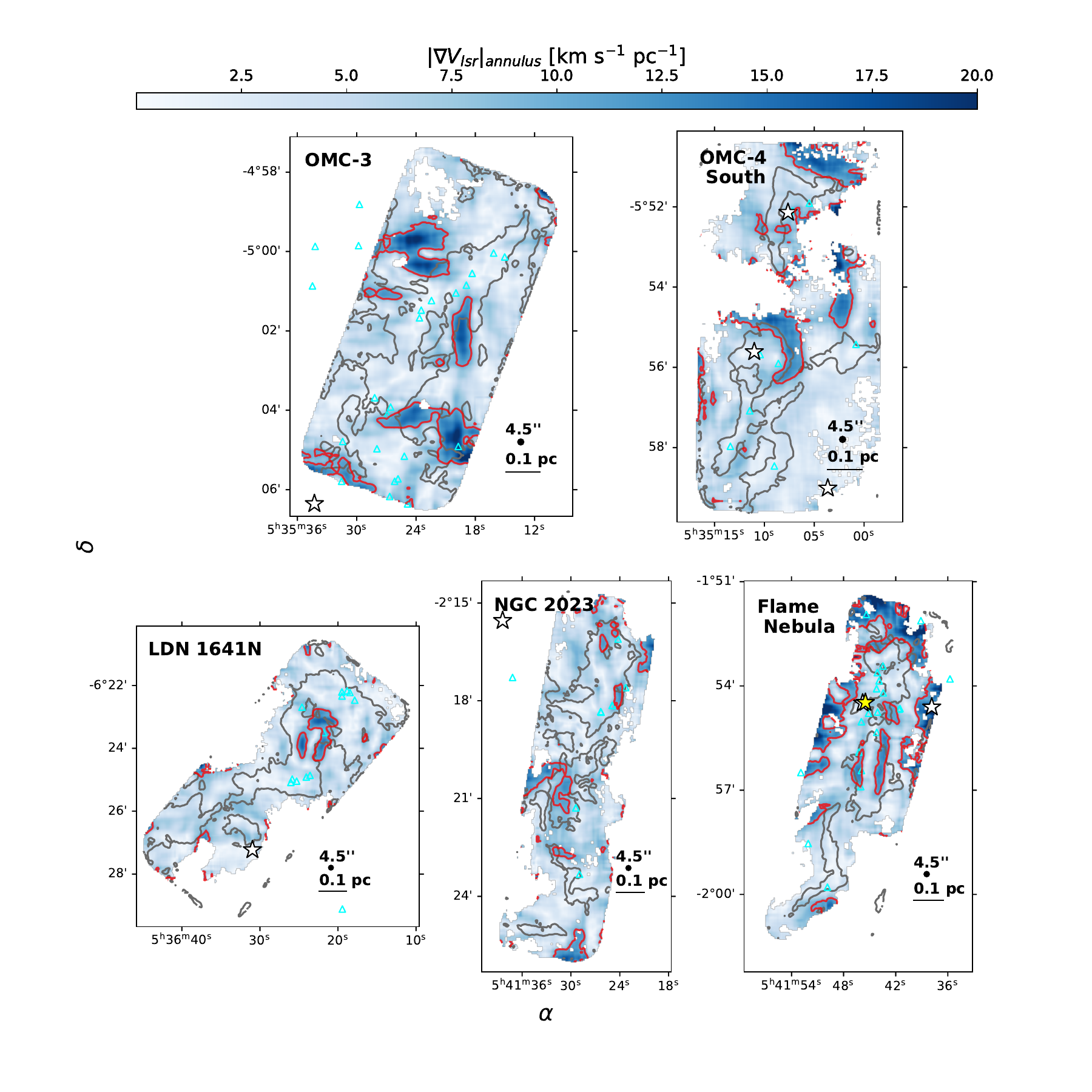}%
	\caption{\hnctrans~$V_\text{lsr}$ annulus gradient maps of the five star-forming regions in the EMERGE Early ALMA Survey at $4.5$~arcsec resolution with ALMA+IRAM-30m. The gray contours show the \nthptrans~integrated intensity above $3\sigma$ (the \nthp~maps are visible in Paper III). The red contours display the high velocity gradient regions identified in Fig.~\ref{fig:grad_selection_method_comparison}. Symbols are similar to those in Fig.~\ref{fig:OMC3_maps}. Beam sizes and scale bars are placed in the bottom right corner.}
	\label{fig:gradient_maps_annulus} 
\end{figure*}

Similar to our standard gradients, the annulus method can be used to decompose the observed gradient into each specific x and y projection, that is $\nabla V_{lsr},x|_{annulus}$ and $\nabla V_{lsr},y|_{annulus}$ gradients, respectively. Figure~\ref{fig:grad_selection_method_comparison} displays the results of the annular method (dashed area) compared to those obtained in our previous gradient calculations (solid area) in Fig.~\ref{fig:grad_sel}. The resulting distributions show a statistically indistinguishable peak, width, and skewness in all regions in our survey. Likewise, the departure from a Gaussian profile occurs at similar values of $\sim$~10~\kmspc, in close agreement to our previous results. In Fig.~\ref{fig:gradient_maps_annulus}, we display the distribution of gradients obtained by this annulus methods. Not surprisingly, both the distribution and magnitude of these gradients (blue colors) as well as the areas exhibiting high shear (enclosed by red contours) are similar to those obtained by our standard gradients (see Fig.~\ref{fig:grad_maps}). 

\begin{figure*}[htbp]
	\centering
        \includegraphics[width=0.65\textwidth]{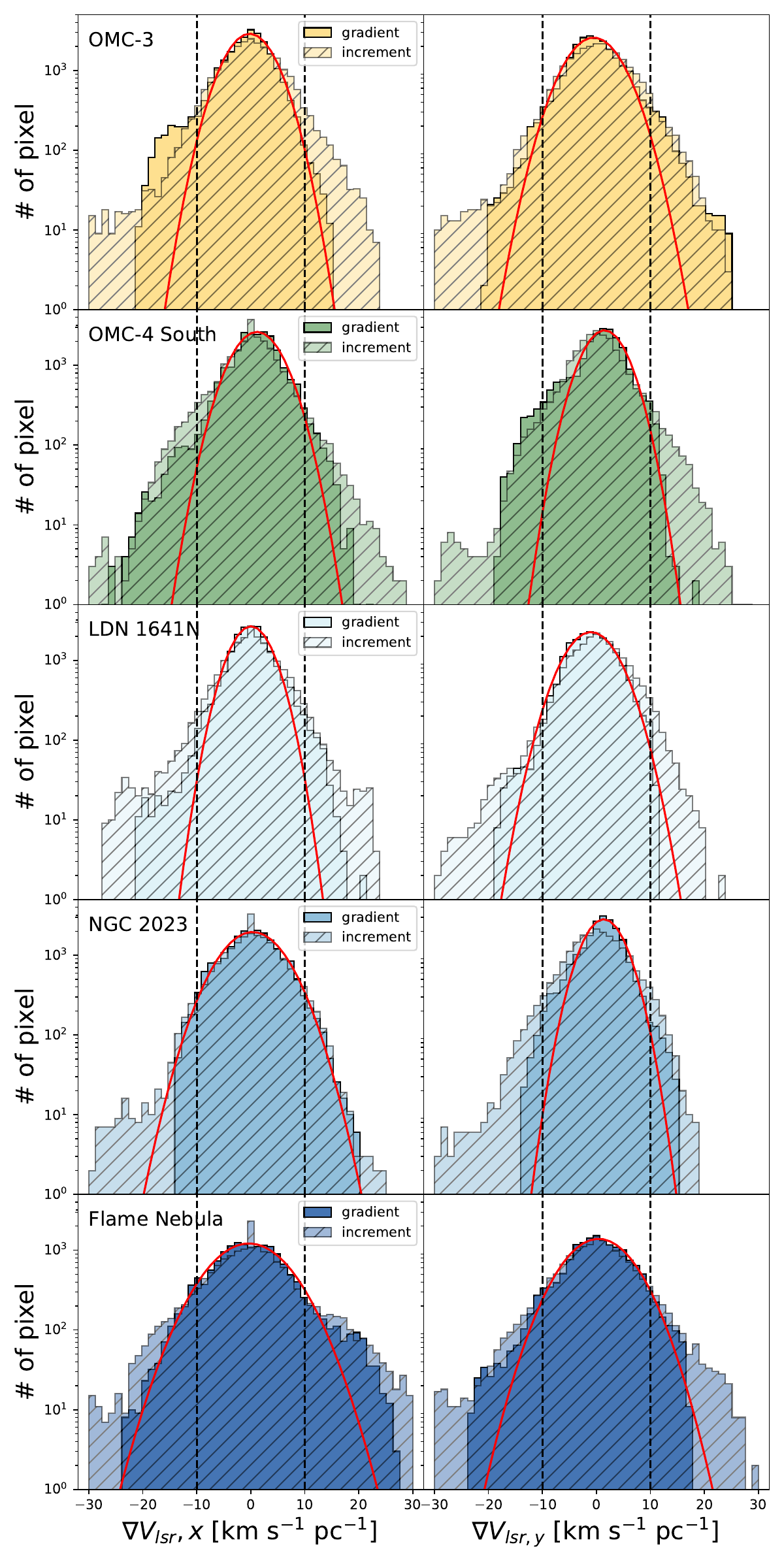}
	\caption{Histogram of the projection of the gradient along the $\alpha$ (left) and $\delta$ axes (right) in all the targets in our survey. We compare the results obtained using the standard gradient method (solid area) and the increment method (hatched area) for a $L=0.04$~pc lag. The red lines show the Gaussian fit applied to the distribution.}
	\label{fig:grad_selection_method_increment} 
\end{figure*}

\begin{figure*}[htbp]
	\centering
        \includegraphics[trim={0cm 1cm 0cm 0cm},clip,width=1\textwidth]{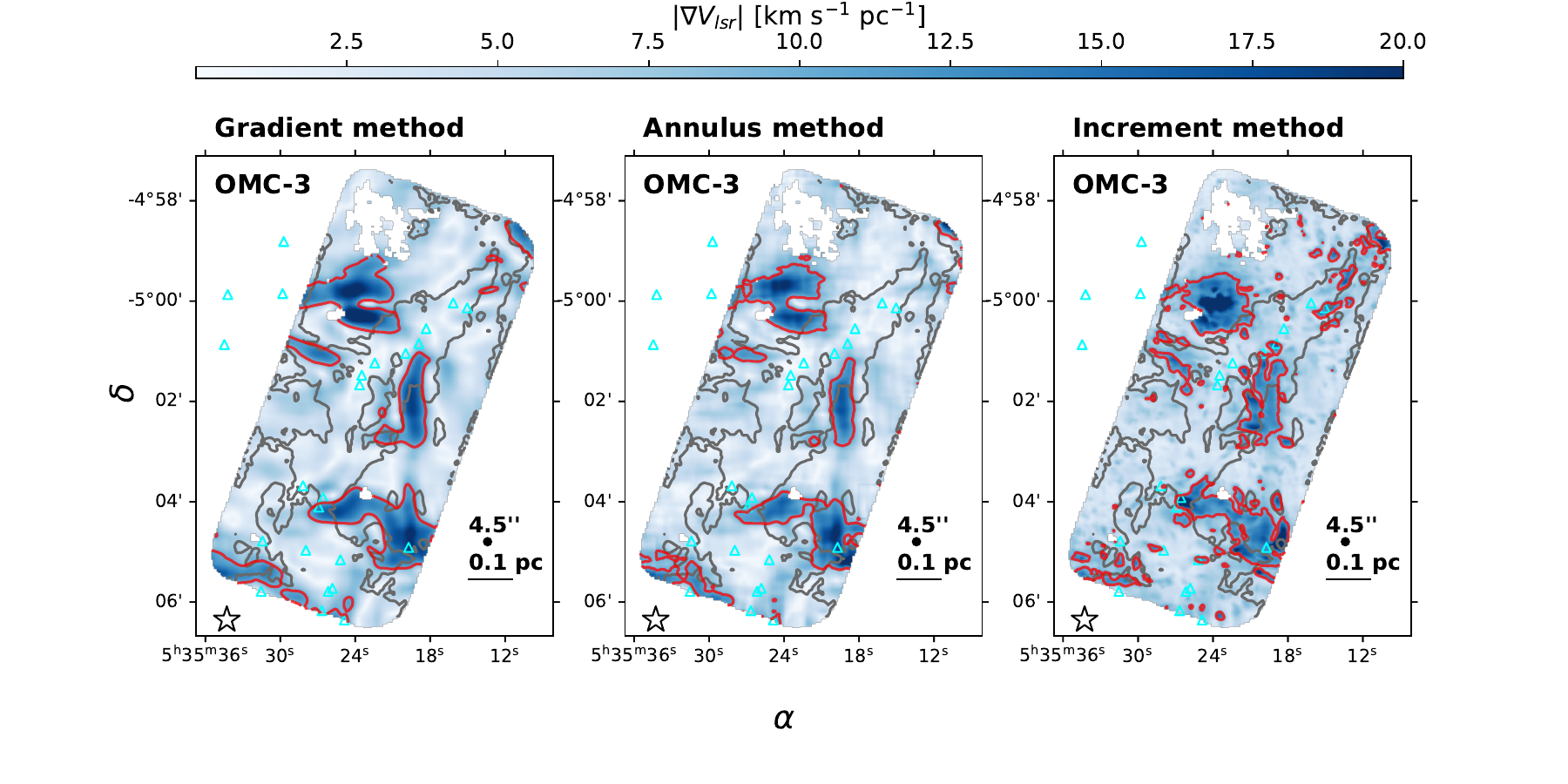}
	\caption{Comparison between the maps obtained with the gradient, annulus, and increment methods in OMC-3, all displayed in the same units and scale. The gray contours show the \nthptrans~integrated intensity above 3$\sigma$ (the \nthp~maps are visible in Paper III). The red contours display the high velocity gradient regions identified in Figs.~\ref{fig:grad_sel}, \ref{fig:grad_selection_method_comparison}, and \ref{fig:grad_selection_method_increment}. Symbols are similar to those in Fig.~\ref{fig:OMC3_maps}. Beam sizes and scale bars are placed in the bottom right corner. The maps of the remaining targets in the sample are shown in Figs.~\ref{fig:grad_maps}, \ref{fig:gradient_maps_annulus}, and \ref{fig:gradient_maps_increment}.}
	\label{fig:grad_maps_comparison} 
\end{figure*}

\begin{figure*}[htbp]
	\centering
        \includegraphics[width=1\textwidth]{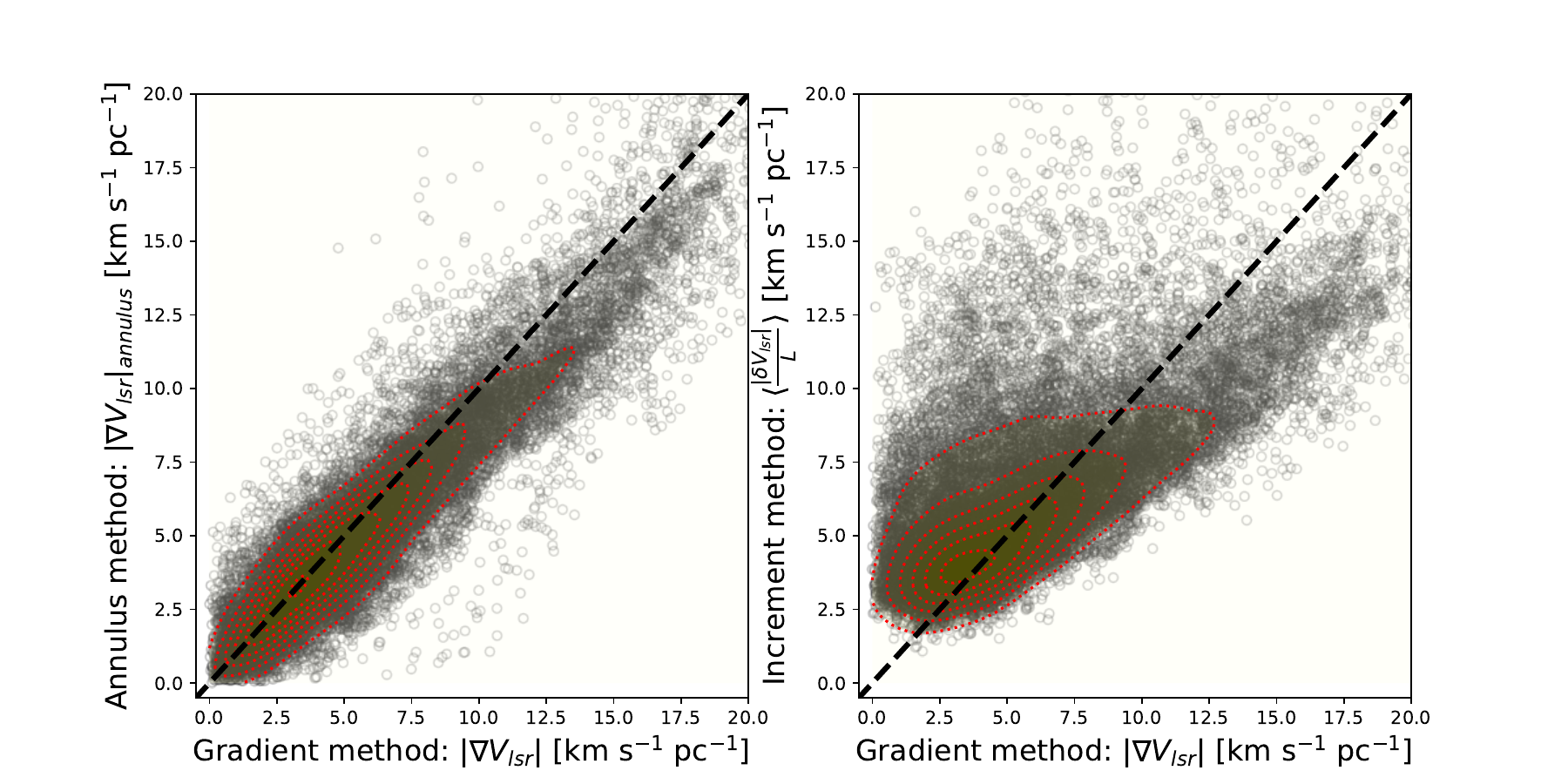}
	\caption{Comparison between the results obtained with the gradient, annulus, and increment methods. The red dashed line marks the points for which the results coincide.}%
	\label{fig:gradient_comparison} 
\end{figure*}

We display a comparison of the three methods in OMC-3 in Fig.~\ref{fig:grad_maps_comparison}, all displayed in gradient units. As seen by these comparisons the three methods produce similar results on the distribution and magnitude of velocity gradients within our targets.
To systematically quantify the similarities of these results we compared pixel-by-pixel the magnitude of the gradients obtained by the three methods, $|\nabla V_{lsr}|$ (gradient method) $|\nabla V_{lsr}|_{annulus}$ (annulus method), and  $|\nabla V_{lsr}'|=\frac{\langle \delta V_{lsr}\rangle}{L}$ (velocity increments), in Fig.\ref{fig:gradient_comparison}. As expected, the comparison of the gradient and annulus methods in Fig.~\ref{fig:gradient_comparison}~(left panel) shows a positive and almost linear (red dashed line) correlation between these two estimates and illustrates the excellent agreement of these two methods. A similar positive correlation is also observed when comparing the gradient method with the velocity increments in Fig.~\ref{fig:gradient_comparison}~(right panel). As discussed in Fig.~\ref{fig:gradient_maps_increment} however this correlation appears noisier, particularly at low gradient values. This is expected by the lower number of pixels used in this method (only those in the annulus) in comparison with our gradient calculations (also including all points within it). 
Although sub-linear, the correspondence between these two methods tightens at large gradient values, particularly in the regions of interest above $\gtrsim$~8~\kmspc.

%Our comparisons demonstrate the equivalence of these three methods (gradients, annulus, and velocity increment) as statistical tools to quantifying the velocity field in our clouds. 
Our comparisons show how these three methods (gradients, annulus, and velocity increment) produce comparable results and can be used as statistical tools to quantify the velocity field in our clouds. 
As demonstrated in this Appendix, the three methods reproduce similar gradient values both in terms of magnitude and location. %As expected, the velocity increment method systematically produces noisier results likely originated by the use of individual differences compared to the global fit used in the other two methods. 
Likewise, these comparisons allow to safely compare our gradient results with previous estimates using other techniques such as annulus and velocity increments \citep{1996Lis,1998Lis,2003Pety}. Among the three methods explored here, we decided to use our gradient estimates as the simplest and most stable approach, producing consistent estimates of both magnitude and orientation. We remark however that the results of this paper would be similar if any of the other two methods would be employed.

\end{document}